# High-magnitude, spatially and directionally programmable, and sustained strain engineering of 2D semiconductors

Boran Kumral[1*], Pedro G. Demingos[2], Shuo Yang[1], Md Akibul Islam[1, 3], Peter Serles[1], Da Bin Kim[4], Dian Yu[2], Akhil Nair[1], Akshat Rastogi[1,2], Nima Barri[1], Cristina H. Amon[1,5], Jane Howe[2], Sjoerd Hoogland[4], Edward H. Sargent[4, 6, 7], SungWoo Nam[3, 8], Chandra V. Singh[2], Tobin Filleter[1*]

[1] Department of Mechanical & Industrial Engineering, University of Toronto, Toronto, ON, Canada

[2] Department of Materials Science and Engineering, University of Toronto, Toronto, ON, Canada

[3] Department of Mechanical and Aerospace Engineering, University of California, Irvine, Irvine, CA, USA

[4] Department of Electrical and Computer Engineering, University of Toronto, Toronto, ON, Canada

[5] Department of Chemical Engineering and Applied Chemistry, University of Toronto, Toronto, ON, Canada

[6] Department of Chemistry, Northwestern University, Evanston, IL, USA

[7] Department of Electrical and Computer Engineering, Northwestern University, Evanston, IL, USA

[8] Department of Materials Science and Engineering, University of California, Irvine, Irvine, CA, USA

[*]Corresponding author emails:

Boran Kumral: boran.kumral@mail.utoronto.ca

Tobin Filleter: filleter@mie.utoronto.ca

## Abstract

Crystalline two-dimensional (2D) semiconductors often combine high elasticity and in-plane strength, making them ideal for strain-induced tuning of electronic characteristics, akin to strategies used in silicon electronics. However, existing techniques have not achieved strain in 2D materials that is simultaneously high in magnitude (>1%), stable over long periods, and spatially and directionally programmable. In this context, programmable strain means that the strain level and direction can be deterministically engineered across a single layer. Here, we apply spatially programmable biaxial strain ($\varepsilon_b$) up to ~2.2% with a spatial gradient of ~0.1 %$\varepsilon_b$ μm$^{-1}$ in monolayer $MoS_2$ via conformal transfer onto patterned substrates fabricated using two-photon lithography. The induced $\varepsilon_b$ is stable for months and enables local band gap tuning of ~0.4 eV in monolayer $MoS_2$, ~25% of its intrinsic band gap. Further, by tailoring substrate topography, we introduce uniaxial (anisotropic) strain, demonstrating control over both strain magnitude and direction. We also extend this strain engineering framework to $MoS_2$-$WS_2$ bilayer heterostructures. Overall, we introduce a distinct regime of strain-enabled control in 2D semiconductors to support the development of optoelectronics and nanoelectronics with engineered optical and electronic landscapes.

## Main

Crystalline, layered, and atomically thin two-dimensional (2D) semiconductors such as transition metal dichalcogenides (TMDs) have emerged as promising channel materials for further scaling of transistors and increasing device density[1–4]. Unlike bulk semiconductors such as silicon (Si), they retain high carrier mobilities and low leakage currents even at thicknesses below 1 nm. 2D semiconductors are also well-suited for lightweight, broad-spectrum optoelectronic devices, including high-specific-power (i.e., high power-per-weight) solar cells and high-specific-detectivity photodetectors[5–10]. Their atomically thin structure, strong light-matter interactions, and strain-tunable direct bandgaps make them particularly attractive for these applications.

A proven method to modulate the electronic and optoelectronic characteristics of 2D semiconductors is strain engineering via the introduction of in-plane lattice strain. Strain alters the lattice spacing of materials, which leads to changes in the overlap of electron orbitals and thus band structure, positioning strain engineering as an effective means to tailor electronic and optoelectronic characteristics of semiconductors. Strain engineering is routinely employed in commercial complementary metal-oxide-semiconductor (CMOS) technologies to boost performance by tuning doping and mobility of Si[11–13]. As 2D semiconductors have garnered scientific and industrial attention for use in nanoelectronics and optoelectronics, there has been considerable interest in strain engineering of 2D materials. 2D strain engineering thus far has shown promise, with studies reporting significant enhancements in electron mobility for 2D TMD-based transistors[14–16] and memristors[17] under tensile strains well below their fracture limits[18,19]. For example, it has been shown that tensile strain of only 0.1-0.2% introduced by stressor layer deposition can increase the on-state current of monolayer molybdenum disulfide ($MoS_2$) transistors by 60%[16]. Higher magnitudes of strain (>1%) can further modulate electronic characteristics, induce phase transitions (e.g., semiconducting-to-metallic crystal structure in group VI TMDs)[20], and generate pseudo-magnetic fields[21]. Additionally, 2D strain engineering has enabled proof-of-concept 2D-material-based optoelectronic devices with broad-spectrum sensing and emission capabilities[10]. For example, it has been demonstrated that by straining 2D black phosphorus by substrate deformation, the operating range of an optoelectronic sensor can be actively adjusted[10]. Strain has also been shown to guide and funnel excitons in monolayer tungsten diselenide ($WSe_2$)[22], which is promising for single-photon emission applications[23].

Many 2D materials can sustain tensile and compressive strains greater than ~10% without inelastic relaxation[21]. This high elastic strain limit is largely attributed to their crystallinity and in-plane covalent bonding. Strain engineering of 2D materials has mostly been performed through transient and non-deterministic techniques such as deforming substrates[10,24–31] and bulging[32–35]. While transient techniques have enabled experimental characterization of strained 2D materials, they are incompatible with scalable, industrial deployment. Available techniques which can introduce sustained strain in 2D materials include stressor layer deposition[16,17,36,37] and use of pre-patterned substrates[38–44], which exploits the low bending stiffness and atomically smooth surface of 2D materials to impose in-plane strain through surface topography. Stressor layer deposition enables reliable and CMOS-compatible generation of modest biaxial strain ($\varepsilon_b$) levels, typically up to ~1%, beyond which delamination typically becomes a limiting factor.

Substrates containing well-defined, pre-patterned features can enable local control of strain in 2D layers conformed on their surfaces[45]. However, demonstrations of patterned substrates to introduce strain in 2D materials has mostly been limited to scanning probe-based patterning[41,46,47], micro-electro-mechanical systems (MEMS)-based patterning[14,38], dispersed nanoparticle-based patterning[39,42], and wrinkling[44]. Although these approaches can introduce strain in 2D materials, they have not simultaneously achieved high-magnitude (>1%), spatially and directionally programmable, and long-term stable strain, likely due to limitations in the geometric complexity of patterns attainable with these techniques. Reported 2D strain engineering techniques lack the ability to introduce strain with spatial programmability (i.e., controlled variation of strain levels across different regions of a single 2D layer) or with directional programmability (i.e., controlled variation of strain direction across different regions of a single 2D layer). Spatial programmability described here differs from previous approaches that rely on patterned substrates with uniform features or uniformly thick stressor layers, which can only produce strain gradients uniform in magnitude and direction across a monolayer.

In this work, we address this limitation by using two-photon lithography (2PL), a sub-micrometer resolution additive manufacturing technique, to fabricate complex three-dimensional (3D) substrates with micrometer-scale, valley-shaped features of systematically varied aspect ratios (ARs). When 2D semiconductors are conformally transferred onto these substrates, we achieve long-term, spatially programmable $\varepsilon_b$ of up to ~2.2% with a spatial gradient of ~0.1 $\%\varepsilon_b$ $\mu m^{-1}$. We demonstrate the resulting strain-induced modulation of optoelectronic and electronic characteristics, including band gap modulation from ~1.66 eV to ~1.24 eV in monolayer $MoS_2$, and extend this approach to bilayer heterostructures. Furthermore, by designing substrate patterns that can impose predominantly uniaxial (anisotropic) deformation in conformal 2D materials, we show that the geometric versatility of 2PL allows directional control of strain. The ability to program strain directionally within different regions of a single monolayer can be leveraged as a key advance, with implications for controlled quasiparticle funneling in 2D semiconductor-based devices. A central advance of this work is a predictive framework that links designed substrate topography to resulting strain and electronic characteristics in a conformed 2D semiconductor. Specifically, we establish a predictive framework in which (i) substrate topography is deterministically designed and fabricated via 2PL, (ii) the resulting strain field in a conformal 2D semiconductor is predicted from the substrate topography using continuum-level analytical theory and finite element analysis (FEA), (iii) these predictions are validated against strain independently extracted from Raman spectroscopy, and (iv) the corresponding strain-induced modulation of electronic band gap is predicted by density functional theory (DFT) and validated against optical and electrical measurements.

This framework, to our knowledge, is the only approach that can achieve high-magnitude (>1%), spatially programmable (i.e., controlled variation of strain levels across different regions of a single 2D layer), directionally programmable (i.e., controlled variation of strain direction across different regions of a single 2D layer), and long-term sustained strain in 2D semiconductors. A benchmarking analysis in **Supplementary Note 1** compares its performance with works of 2D strain engineering reported in literature. Overall, these results demonstrate the potential of topographically engineered substrates to achieve precise and tunable strain profiles in 2D materials. Such control over strain could enable the spatial and directional programming of electronic characteristics in 2D-material-based nanoelectronics. In addition, engineered band-gap landscapes in 2D semiconductors may advance lightweight optoelectronic devices capable of broad-spectrum absorption and emission across a wide range of photon energies.

## Patterned substrate design for programmable strain engineering

We designed surfaces with micrometre-scale valleys defined by localized sinusoidal depressions (Fig. **1a**). The AR of each valley is defined by AR = h/L where h and L are the valley amplitude and period, respectively. The surfaces are extended into the third dimension to form 3D pre-patterned substrate models, which are then fabricated using 2PL (Figs. **1b-d** and **Supplementary Note 2**). Within a single substrate, the ARs of individual valleys are varied, enabling a single 2D monolayer conformed to the surface to experience spatially varying levels of biaxial tensile strain. This strain landscape can be deterministically engineered through the design of the substrate topography (Fig. **1e**). Monolayer $MoS_2$ samples were exfoliated on Au-coated $SiO_2$-Si substrates (see **Methods** and **Supplementary Notes 3, 4**), then transferred onto the pre-patterned substrate (Fig. **1c**). See **Methods** and **Supplementary Note 5** for details on the conformal transfer process. Since Raman spectroscopy can locally, rapidly, and non-destructively characterize strain in 2D materials we used an optically transparent 2PL resin, IP-Visio, to minimize background fluorescence during spectral acquisition.

We evaluated the in-plane strain generated in monolayer $MoS_2$ conformed to valleys both using continuum-level analytical theory and FEA simulations (see **Methods** and **Supplementary Notes 6, 7**). Analytical and FEA predictions of the maximum $\varepsilon_b$ imparted to $MoS_2$ for valleys of different ARs are shown in Fig **1f**. The insets in Fig **1f** show the analytical and FEA predictions of the $\varepsilon_b$ field in monolayer $MoS_2$ conformed to a valley with an AR of 0.2 (see **Supplementary Note 6** for additional analytical and FEA-based $\varepsilon_b$ field predictions). These predictions of strain guided the engineering of valley ARs to deterministically apply strain to conformed monolayers.

Our substrate design also considers the interplay between strain energy in the $MoS_2$ monolayer and its adhesion to the patterned substrate. As $MoS_2$ is conformed to the surface, it stores elastic energy, which must remain lower than the interfacial adhesion energy ($\gamma$) between the two interfaces to ensure stable conformity. Using DFT (see **Methods**), we calculated the strain energy as a function of $\varepsilon_b$ and experimentally measured adhesion via AFM using a tip made from the same 2PL resin, IP-Visio, used in substrate fabrication (Fig. **1g**). Adhesion measurements between the IP-Visio tip and monolayer $MoS_2$ revealed an $\gamma$ of $0.095 \pm 0.016$ J m$^{-2}$ (see **Supplementary Note 8**). Predicting the strain energy of $MoS_2$ as a function of $\varepsilon_b$ and evaluating the $\gamma$ of the interfaces helps ensure that the chosen ARs for the valleys can maintain monolayers in a conformed state, preventing delamination from the IP-Visio-patterned substrates.

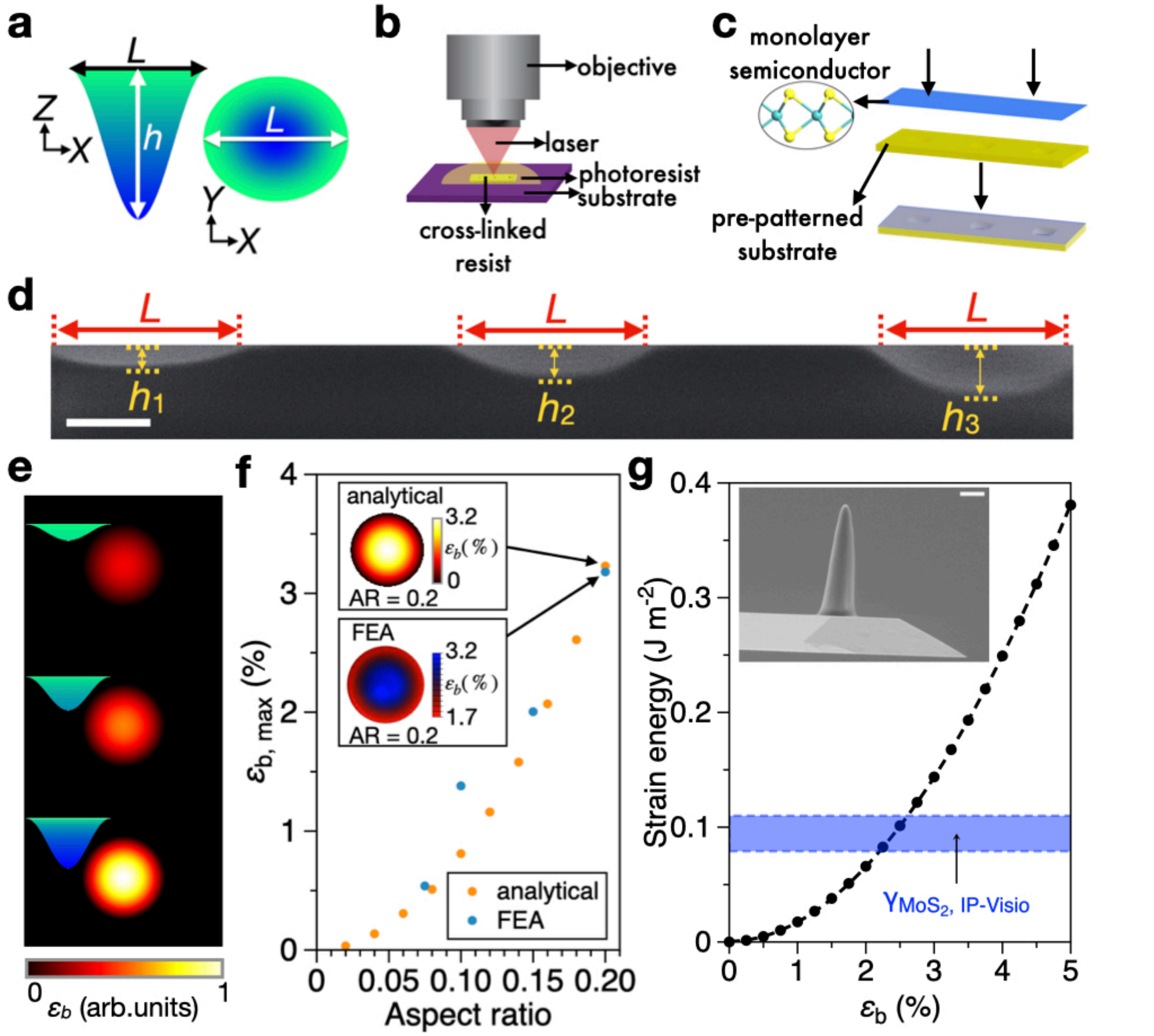

**Figure 1 | Design and fabrication of substrates for programmable strain engineering in 2D semiconductors.** **a**, Side and top-down view schematics of geometric valley profiles with aspect ratios (AR), defined by AR = $h/L$, where $L$ and $h$ are the valley period and amplitude, respectively. **b**, Illustration of two-photon lithography (2PL) printing setup. **c**, Illustration of the transfer of monolayer $MoS_2$ to the patterned substrate. **d**, Cross-sectional field emission scanning electron microscope (FE-SEM) image of a 2PL-fabricated substrate featuring periodic valleys with uniform spacing and varying amplitudes. This image shows a patterned substrate with valleys with higher ARs than those used in later sections of the manuscript, for illustrative purposes. Scale bar, 10 μm. **e,** Analytical prediction of the biaxial strain ($\varepsilon_b$) field, normalized to the maximum strain at the center of the valley with the highest aspect ratio, in monolayer $MoS_2$ conformally adhered to a pre-patterned substrate comprised of three periodically spaced valleys with uniform valley period and varying amplitudes. The center-to-center separation between

adjacent valleys is 2L. A schematic cross-section is shown at the top left of each valley, using a blue/green colormap to illustrate the corresponding valley profile and to highlight the increase in aspect ratio from the top valley to the bottom valley. **f**, Analytical and finite element analysis (FEA) predictions of the maximum strain in monolayer $MoS_2$ conformed to valleys of varying AR. Insets: the analytical (top inset) and FEA (bottom inset) predictions of the top-down view of the spatially resolved strain distribution in monolayer $MoS_2$ conformed to a valley with AR = 0.2. **g**, Density functional theory (DFT) predictions of strain energy of $MoS_2$ as a function of $\varepsilon_b$. The black dashed line is a polynomial fit, while the rectangular blue region corresponds to the measured adhesion energy ($\gamma$) between monolayer $MoS_2$ and IP-Visio. Inset: FE-SEM image of IP-Visio-fabricated tip printed on a tipless cantilever. Inset scale bar, 500 nm.

An additional important parameter in conforming a 2D monolayer on valleys is the surface roughness of the two interfaces. Low roughness is desirable to enable sufficient van der Waals (vdW) interactions between the interfaces. Exfoliated 2D materials are atomically smooth and exhibit low surface roughness ($S_{RMS} < 1$ nm). The surface roughness of the substrates was determined to be $S_{RMS} = 1.3 \pm 0.5$ nm (see **Supplementary Note 2**). We then employ a process akin to thermal molding to conform the 2D monolayers to our designed substrates (see **Methods** and **Supplementary Note 5**). The conformity of the monolayers was evaluated using AFM profiling of the valleys before and after transfer, which reveals that the monolayer topography after transfer closely aligns with the topography of the valleys (see **Supplementary Note 9**).

## Programmable, high-magnitude biaxial strain landscapes in monolayer $MoS_2$

Following successful transfer and conformity of monolayer $MoS_2$ to the patterned substrate, we characterized strain using confocal Raman spectroscopy (~1 μm spot size). This resolution is sufficient for mapping across valleys ~20 μm in diameter (Fig. **2a**). In this study, the valley diameter, also referred here as the valley period, was held at ~20 μm while the valley height was varied to control the valley AR. $\varepsilon_b$ was extracted from shifts in the E′ and $A_1'$ phonon modes, which are sensitive to strain and exhibit redshifts and blueshifts under tensile and compressive strain, respectively (see **Supplementary Note 10**). Figure **2b** shows Raman spectra of monolayer $MoS_2$ acquired from flat regions and valley centers with ARs of ≈0.044, ≈0.073, and ≈0.089 on IP-Visio-patterned substrates, normalized to the Si peak (~520.5 $cm^{-1}$; see inset of Fig. **2b**).

The intensity of the monolayer $MoS_2$ $E'$ and $A_1'$ Raman peaks at the valley regions are higher than at the flat regions, and the $E'/A_1'$ intensity ratio increases with increasing strain. The observed enhancement of both peak intensities has been observed in prior studies of biaxially strained, bulged monolayer $MoS_2$ on circular membranes, where increases in peak intensities were attributed to changes in optical interference between light scattered by the monolayer and light reflected from the substrate across strained regions[32]. Although our samples are not bulged, the curvature of the monolayers conforming to the valleys closely resembles that of bulged monolayer on a circular membrane (see **Supplementary Note 9**). Thus, the light interference effects because of the presence of curvature across the valleys may also be present in our Raman characterization. In addition, some of the variations in intensity likely arise from changes in the working distance of the confocal spectrometer as different regions are brought into focus. We rule out variations in the thickness across the underlying IP-Visio substrate as a contributing factor (see **Supplementary Note 4**). Regarding the increase in the $E'/A_1'$ intensity ratio under increasing biaxial strain, similar trends have been widely reported in studies of biaxially strained monolayer $MoS_2$. In contrast to uniaxial strain, which induces splitting behavior of the $E'$ mode (into $E'^{-}$ and $E'^{+}$) due to breaking of lattice symmetry[27,29,48], biaxial strain does not produce such splitting. Instead, both simulations[39,48] and experiments[32,36,41,49] have shown that biaxial straining of monolayer $MoS_2$ increases the intensities of its Raman modes, particularly the $E'$ mode, as well as the $E'/A_1'$ intensity ratio.

Figures **2c** and **2d** display corresponding Raman peak positions and extracted $\varepsilon_b$. Raman data from the patterned substrates were collected from three independent samples, each containing three separate valleys with ARs of 0.045 ± 0.001, 0.073 ± 0.001, and 0.089 ± 0.001. Using monolayer $MoS_2$ on $SiO_2$ as the 0% strain reference, we measured average $E'$ and $A_1'$ peak positions of 385.6 ± 0.5 $cm^{-1}$ and 404.6 ± 0.3 $cm^{-1}$, respectively, from samples either exfoliated directly onto $SiO_2$ or transferred onto $SiO_2$ following exfoliation on Au. We find that flat regions exhibit 0–0.24% biaxial tensile strain, while valley centers show progressively increasing strain with increasing AR: 0.40–0.62% for AR = 0.045 ± 0.001, 1.16–1.24% for AR = 0.073 ± 0.001, and 2.02–2.26% for AR = 0.089 ± 0.001.

We also note an instance where we recorded biaxial tensile strain of 2.87% at the valley center of a valley with AR > 0.1 (see **Supplementary Note 11**). However, this strain was not retained upon re-examination and had relaxed to an unstrained state. Notably, this high strain would have a strain energy near the limit of the interfacial adhesion energy as indicated in Fig. **1g**; therefore, delamination may be expected.

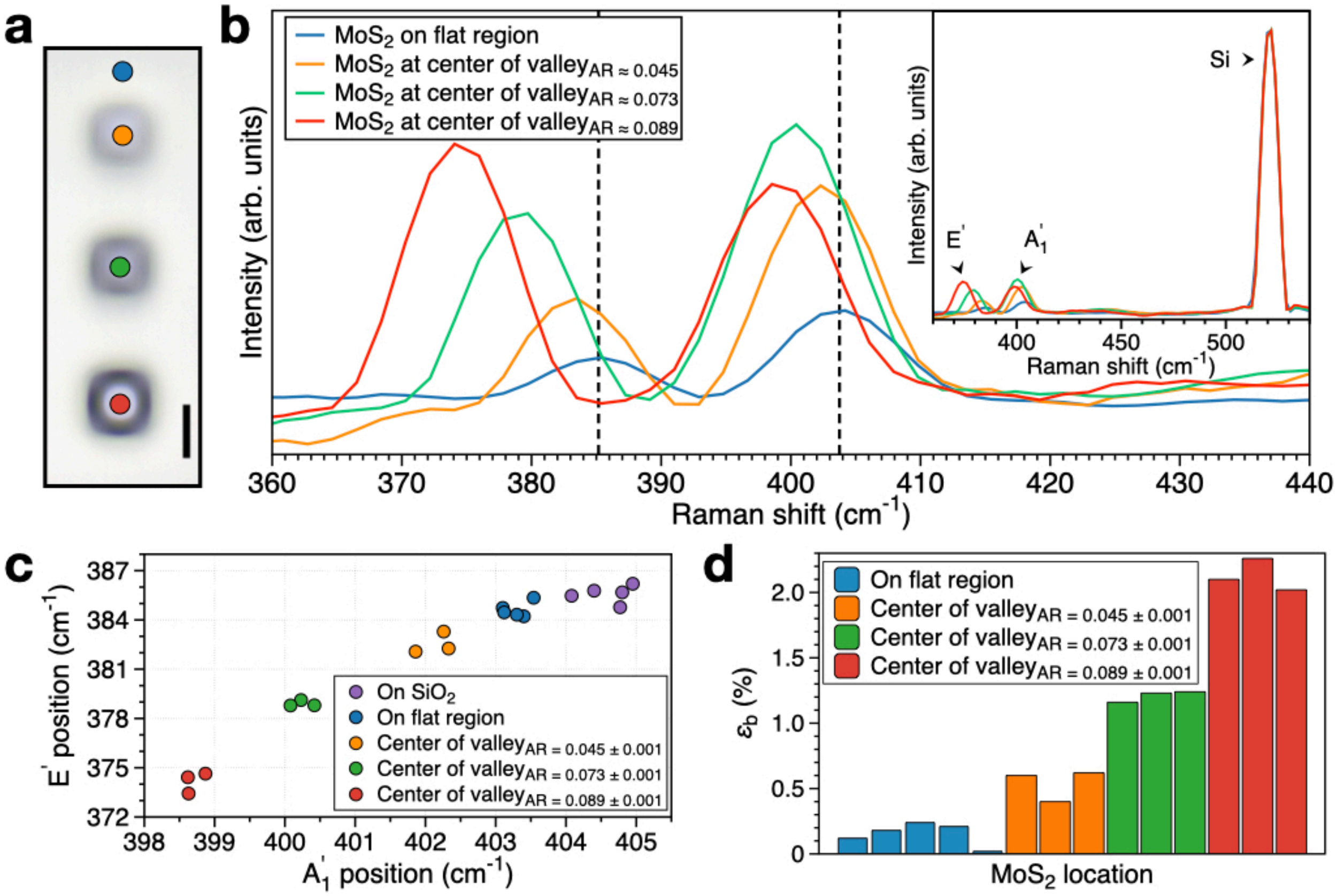

**Figure 2 | Programmable high-magnitude strain in monolayer $MoS_2$. a,** Optical microscope image of a substrate with valleys of varying aspect ratios (ARs). Colored dots (orange, green, red) indicate valley centers with AR ≈ 0.045, 0.073, and 0.089, respectively; the blue dot indicates a flat region. The distance between the centers of neighboring valleys is 2L, with L representing the valley period. Scale bar, 20 µm. **b,** Raman spectra collected from the color-coded locations in (a). As the working distance is adjusted for each region of focus, all spectra are normalized to the Si peak intensity. Vertical dashed lines indicate the E′ and $A_1$′ peak positions of monolayer $MoS_2$ on the flat region, as determined from Lorentzian fits. Inset: wide-range spectrum displaying the Si substrate peak (~520.5 $cm^{-1}$). **c,** Scatter plot of E′ versus $A_1$′ Raman peak positions for monolayer $MoS_2$ on IP-Visio valleys. Data were collected from three samples, each containing three separate valleys with ARs of 0.045 ± 0.001, 0.073 ± 0.001, and 0.089 ± 0.001. Raman peak

positions for $MoS_2$ on $SiO_2$ were obtained from two individual monolayer $MoS_2$ samples on $SiO_2$-Si substrates. **d,** Biaxial strain ($\varepsilon_b$) in $MoS_2$ extracted from Raman peak positions in (c).

Raman mapping reveals a gradient strain distribution in monolayer $MoS_2$ conformed to a valley with AR ≈ 0.089. Figure **3a** shows the AFM topography of the as-printed valley, and Fig. **3b** shows the spatial map of the extracted $\varepsilon_b$ of conformal monolayer $MoS_2$. The radial symmetry in $\varepsilon_b$ across the valley indicates uniformly biaxial strain, consistent with analytical calculations and FEA simulations. Figure **3c** shows the Raman spectra acquired along the pink arrow in Fig. **3b**, tracing a path from a flat region to the valley center. The gradual spectral shifts along this path confirm the presence of a strain gradient in monolayer $MoS_2$. We estimate a strain change rate of ~0.1 $\%\varepsilon_b$ $\mu m^{-1}$ along the pink arrow in Fig. **3b**. Large-area Raman scans of monolayer $MoS_2$ conformed to valleys with ARs of ≈0.079 and ≈0.088 (Fig. **3d**) further demonstrates spatial control of strain magnitude over extended regions.

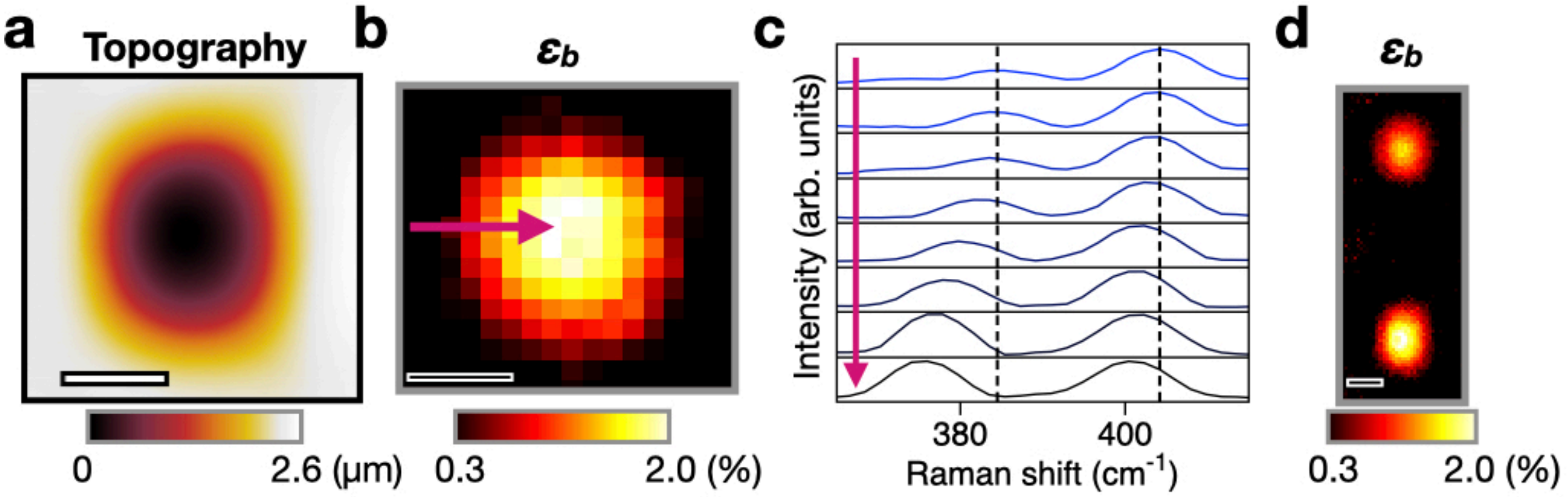

**Figure 3 | Spatially resolved strain landscapes in monolayer $MoS_2$. a,** Topographic atomic force microscope (AFM) image of a 2PL-printed substrate with a valley of an aspect ratio (AR) of ≈ 0.089. Scale bar, 10 μm. **b,** The biaxial strain ($\varepsilon_b$) map of monolayer $MoS_2$ conformed to a valley with an AR ≈ 0.089, with strain values extracted from the E′ and $A_1'$ peak position maps presented in **Supplementary Note 11**, **Supplementary Figures 15a** and **15b**. The map was acquired with 2 μm steps in both the x and y directions. The color scale range was limited to 0.3–2.0% (from −0.1 to 2.1%) strain to highlight high-magnitude spatial variations. Scale bar, 10 μm. **c,** Raman spectra of monolayer $MoS_2$ collected along the pink arrow in (b). The top spectrum corresponds to the starting point of the arrow, and the bottom spectrum corresponds to its end, with intermediate spectra sampled along the arrow path. Vertical dashed lines indicate the E′ and $A_1'$ peak positions of monolayer $MoS_2$ on the flat region (top panel), as determined from Lorentzian fits. As the working distance is adjusted for each region of focus, all Raman spectra are normalized to the Si

peak (~520.5 $cm^{-1}$). **d,** The $\varepsilon_b$ map of monolayer $MoS_2$ conformed to valleys with AR ≈ 0.079 (top valley) and AR ≈ 0.088 (bottom valley) with distance between the centers of neighboring valleys as 3L, with L representing the valley period, with strain values extracted from the E′ and $A_1'$ peak position maps presented in **Supplementary Note 11**, **Supplementary Figures 15c** and **15d**. The maps were acquired with 1 μm steps in both the x and y directions. The color scale range was limited to 0.3–2.0% (from −0.1 to 2.1%) strain to highlight high-magnitude spatial variations. Scale bar, 10 μm.

## Spatially programmable band gap modulation of monolayer $MoS_2$

We performed photoluminescence (PL) spectroscopy and conductive atomic force microscopy (C-AFM) measurements (see **Methods** and **Supplementary Notes 12** and **13**) to experimentally verify that strain modulates the band gap of monolayer $MoS_2$. Band gaps were extracted optically using PL and electronically using C-AFM.

Figure **4a** shows the PL spectra of monolayer $MoS_2$ conformed to a patterned substrate with two valleys with ARs of ≈0.051 and ≈0.087. The $\varepsilon_b$ in monolayer $MoS_2$ at the valley center is ~0.85% for AR ≈ 0.051 and ~2.00% for AR ≈ 0.087 (see **Supplementary Note 12**). The A exciton peak at ~1.84 eV on a flat region with $\varepsilon_b$ ~0.08% redshifts to ~1.74 eV at the center of the valley with AR ≈ 0.051 ($\varepsilon_b$ ~0.85%), and further redshifts to ~1.61 eV at the center of the valley with AR ≈ 0.087 ($\varepsilon_b$ ~2.00%). This corresponds to an A exciton shift rate of ~120 meV/%$\varepsilon_b$, comparable to the upper range of prior reports of strain-induced exciton shift rates in monolayer $MoS_2$ [14,29].

C-AFM, which measures out-of-plane current under applied bias, has recently been used to probe strain in 2D materials[50] and for atomic-resolution current imaging under ambient conditions[51,52]. We performed C-AFM measurements to further demonstrate that our strain-engineering approach enables localized tuning of the electronic characteristics of monolayer $MoS_2$. For these measurements, monolayer $MoS_2$ was transferred onto an Au-coated 2PL-fabricated patterned substrate containing two valleys (AR ≈ 0.044 and AR ≈ 0.070). Raman spectroscopy was used to extract the strain at the valley centers, giving $\varepsilon_b$ ~0.59% and $\varepsilon_b$ ~1.10% for AR ≈ 0.044 and AR ≈ 0.070, respectively, while the flat region exhibited $\varepsilon_b$ ~0.28% (see **Supplementary Note 13**). Differential conductance ($dI/dV$) from the flat region and from the centers of the valleys (AR ~ 0.044 and 0.070) are shown in **Fig. 4b**. The extracted band gap from these $dI/dV$ measurements decreases from ~1.66 eV in flat regions to ~1.43 eV at the AR ≈ 0.044 valley center, and to ~1.24 eV at the AR ≈ 0.070 valley center (see **Supplementary Note 13**). This corresponds to an extracted band gap shift rate of ~512 meV/%$\varepsilon_b$ for monolayer $MoS_2$.

We then compare the experimentally observed strain-induced band gap modulation, obtained from PL and C-AFM measurements, with density functional theory (DFT) predictions. The electronic band structures of monolayer $MoS_2$ were calculated using DFT as a function of $\varepsilon_b$ up to 5% (see **Supplementary Note 14**), yielding a band gap shift rate of ~206 meV/%$\varepsilon_b$. The band gaps extracted from DFT calculations are presented alongside the experimentally determined values, with all data plotted as a function of $\varepsilon_b$ from 0 to 2% in Fig. **4c**.

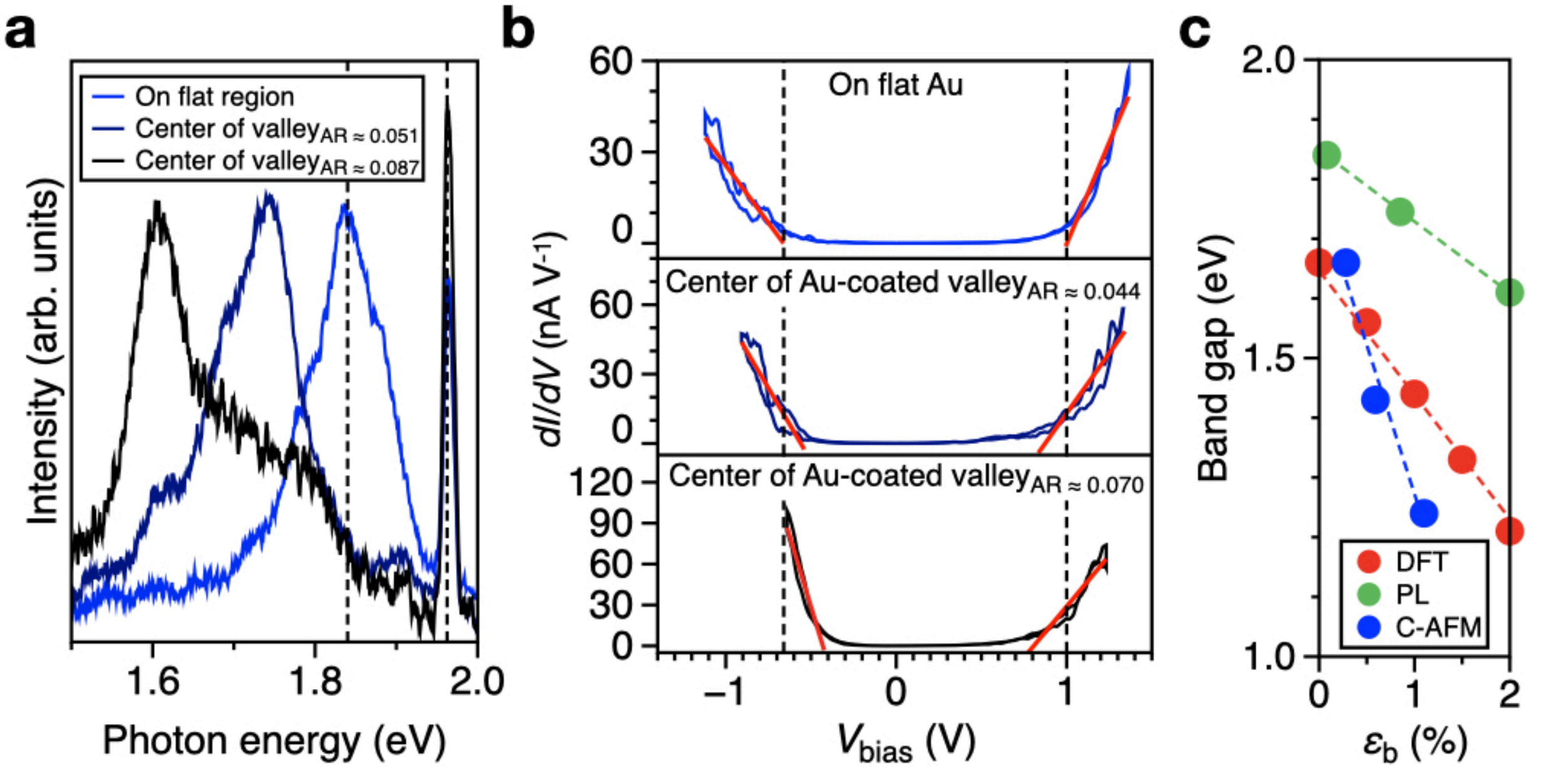

**Figure 4 | Spatially programmable band gap modulation of monolayer $MoS_2$. a,** Photoluminescence (PL) spectra of monolayer $MoS_2$ collected from a flat region, and from the centers of valleys with ARs of ≈0.051 and ≈0.087. As the working distance is adjusted for each region of focus, all PL spectra are normalized to the intensity of the $MoS_2$ PL peak on the flat region. The PL peak at ~1.96 eV originates from the IP-Visio substrate and as expected, its position remains unchanged across different locations. Vertical dashed lines indicate the A exciton position of monolayer $MoS_2$ on the flat region and the position of the IP-Visio PL peak, as determined from Lorentzian fits. **b**, Differential conductance ($dI/dV$) versus bias voltage ($V_{bias}$) curves acquired from conductive atomic force microscopy (C-AFM) measurements of monolayer $MoS_2$ conformed to an Au-coated patterned substrate containing two valleys with ARs of ≈0.044

and ≈0.070. Each panel displays d$I$/d$V$ versus $V_{bias}$ for monolayer $MoS_2$ at the center of a valley with AR ≈ 0.070 (bottom), at the center of a valley with AR ≈ 0.044 (middle), and a flat region (top). For each spectrum, linear fits (red lines) were applied to the rising edges of the conductance curves to determine the conduction and valence band edges. The electronic band gap was extracted as the voltage difference between these linear fit (red lines) zero-crossings. Black vertical dashed lines indicate the extracted positions of the valence and conduction band edges for monolayer $MoS_2$ on a flat region of the Au-coated patterned substrate. **c**, Band gap of monolayer $MoS_2$ as a function of biaxial strain ($\varepsilon_b$), determined from PL spectroscopy, C-AFM, and density functional theory (DFT). Dashed lines are linear fits to guide the eye.

We note that the PL, DFT, and C-AFM values in Fig. **4c** are not strictly equivalent measures of the band gap. The PL A exciton energy reflects the optical gap, the DFT band gap predictions are based on a computational first-principles approach, and the C-AFM-derived band gap is extracted from a transport measurement sensitive to tip-induced band bending, contact effects, and set-point force. The band-gap modulation rate measured by C-AFM, which exceeds 500 meV/%$\varepsilon_b$, is markedly larger than the shift rates obtained from PL measurements or predicted by DFT. Although previous studies have reported that monolayer $MoS_2$ exfoliated on Au can experience substantial $\varepsilon_b$ (1–1.5%)[53,54], the discrepancy observed here cannot be completely attributed to Au-induced strain. While some induced strain (~0.28%) is present in the flat regions of monolayer $MoS_2$ transferred onto an Au-coated patterned substrate (**Supplementary Note 13**), the large $\varepsilon_b$ reported for monolayers exfoliated on Au is not preserved. This loss of strain is also observed when such monolayers exfoliated on Au are subsequently transferred onto $SiO_2$ or IP-Visio (see **Supplementary Note 3**). Metallic behavior is observed on the exposed Au-coated regions of the IP-Visio substrate, indicating a low-resistance and Ohmic tip–sample contact (see **Supplementary Note 13**). It is also possible that the pronounced band-gap modulation observed in the C-AFM measurements results from the much more localized area probed by the C-AFM tip relative to the ~1 μm spot size of the Raman and PL setups. The tip radius was measured to be ~50 nm (see **Supplementary Note 13**). Previous studies have shown that C-AFM measurements can introduce strain at high set-points and that the applied set-point force strongly affects the tip–sample interface [50]. In our experiments, we consistently used a set-point of ~10 nN to minimize any force-related artifacts in the electrical measurements (see **Supplementary Note 13**).

## Long-term stability of induced strain

We assessed the long-term stability of the induced strain in monolayer $MoS_2$ conforming to valleys with ARs of ≈0.040, ≈0.075, and ≈0.085. Raman spectroscopy measurements were collected at the valley centers immediately after fabrication, four months later, and again eight months later (see **Supplementary Note 15** for the corresponding Raman spectra). Throughout this period, the sample was stored under ambient conditions (22–23 °C, 30–60% relative humidity). Figure **5a** identifies the measurement locations and timestamps for the Raman characterization, while Figs. **5b** and **5c** show the Raman peak positions and the extracted $\varepsilon_b$ values, respectively. The strain remained stable over several months, with no significant Raman peak shifts detected at either the four- or eight-month mark. The small variations of ~0.2% in $\varepsilon_b$ observed at the valley centers over time are likely attributed to manual uncertainties in the Raman system. Specifically, they may arise from slight variations in the manually adjusted working distance and lateral focus during repeated measurements.

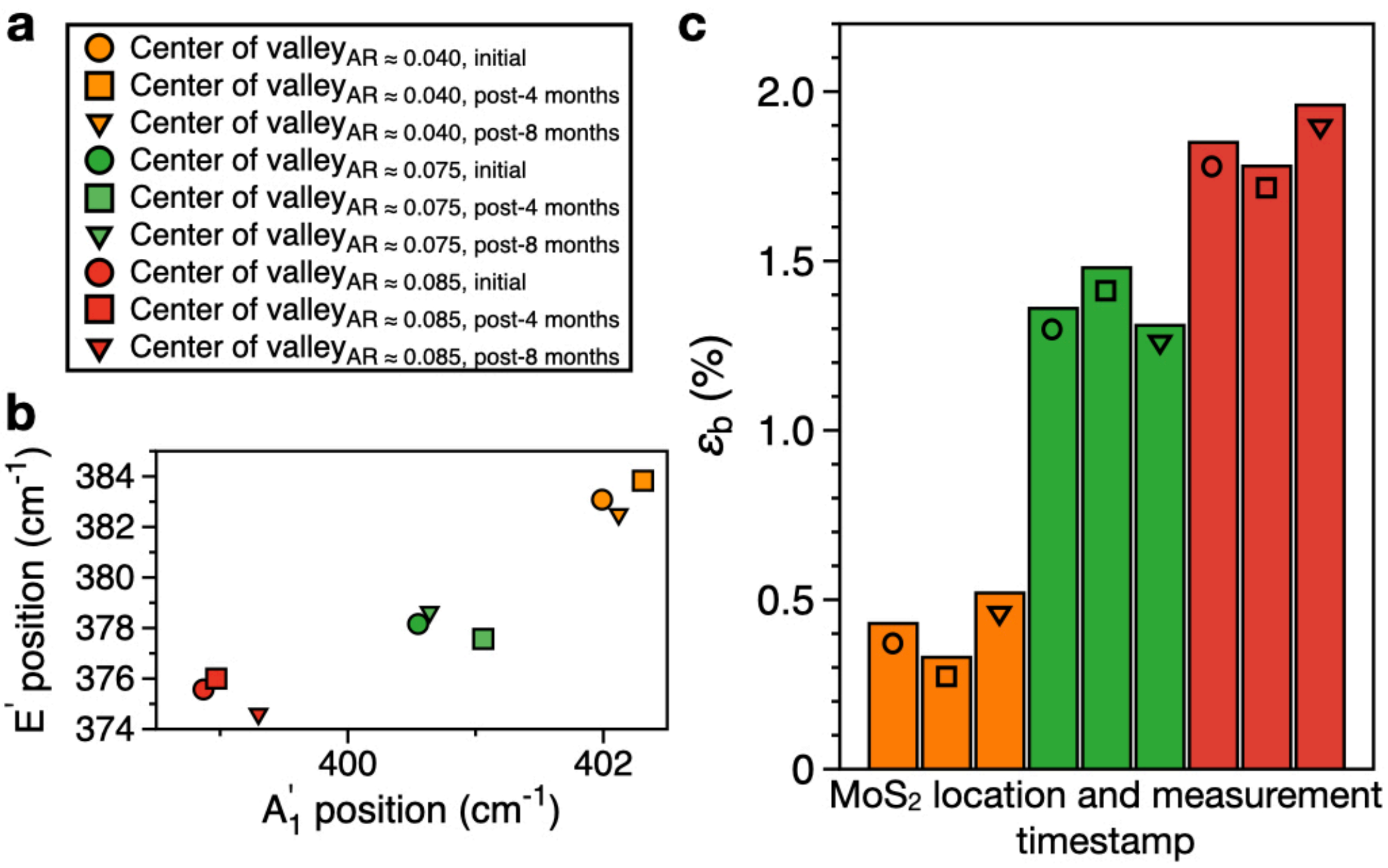

**Figure 5 | Long-term stability of induced strain in monolayer $MoS_2$. a,** Legend indicating the measurement locations and timestamps for monolayer $MoS_2$ conformed to the patterned substrate with

valleys of aspect ratios (ARs) of ≈0.040, ≈0.075, and ≈0.085. Symbols correspond to Raman measurements from each valley center performed shortly after fabrication, 4 months post-fabrication, and 8 months post-fabrication. **b,** Scatter plot of E′ versus $A_1'$ Raman peak positions collected from the corresponding monolayer $MoS_2$ regions indicated in (a), as determined from Lorentzian fits of Raman spectra presented **Supplementary Note 15, Supplementary Fig. 21**. **c,** Biaxial strain ($\varepsilon_b$) in monolayer $MoS_2$ extracted from the Raman peak positions shown in (b) for the corresponding regions indicated in (a).

## Directionally programmable strain engineering of monolayer $MoS_2$

While monolayers conforming to valley geometries experience biaxial strain, we leveraged the geometric design flexibility enabled by 2PL to fabricate structures that induce uniaxial strain in conformal monolayer $MoS_2$. To this end, we designed substrates featuring trench-like structures with cross-sectional profiles resembling half of a capsule shape (i.e., a half-cylinder with semicircular ends). Figure **6a** presents a height profile of such a trench, while Fig. **6b** shows the corresponding cross-sectional profiles along the x- and y-directions. These features exhibit different aspect ratios along the two axes, with $AR_x > AR_y$. Consequently, a monolayer conforming to this geometry is expected to experience predominately anisotropic strain, with $\varepsilon_{xx} > \varepsilon_{yy}$. We confirm this expectation using FEA simulations, which show that a monolayer conformed on a trench-like structure indeed exhibits $\varepsilon_{xx}$ and negligible $\varepsilon_{yy}$, with some presence of $\varepsilon_{yy}$ within the regions of the monolayer conformed to the semicircular ends of the trench (Fig. **6c**).

Figure **6d** shows an optical microscope image of a 2PL fabricated substrate containing a trench-like feature with an $AR_x \approx 0.081$, and AFM measurements confirm the fidelity of the fabricated structure (see **Supplementary Note 16**). Monolayer $MoS_2$ was then transferred and conformed to the substrate. In contrast to biaxial strain, uniaxial strain breaks lattice symmetry in monolayer $MoS_2$, leading to splitting of the E′ Raman mode into $E'^-$ and $E'^+$ components[27,29,48]. Figure **6e** shows Raman spectra collected along the path indicated by the pink arrow in Fig. **6d**, tracing a path along the x-direction from a flat region to the center of the trench. Signs of emerging E′ peak splitting and increased linewidth are observed for monolayer $MoS_2$ at the trench center (bottom spectra in Fig. **6e)**. Based on prior studies of uniaxial strain in monolayer $MoS_2$ [29], we estimate $\varepsilon_{xx}$ at the trench center to be 1.4–1.6% for a trench with $AR_x \approx 0.081$. We note that the Raman spectra for uniaxial strain are presented without smoothing to preserve the visibility of peak splitting. In contrast, smoothing does not significantly affect key features of biaxially strained samples, particularly peak positions which are used to extract strain (see **Supplementary Note 17**).

To directly compare the strain states and demonstrate directional programming of strain within a single monolayer, monolayer $MoS_2$ was transferred onto a substrate featuring both a trench-like structure and a valley. The valley has an AR ≈ 0.085 and the trench has an $AR_x \approx 0.080$ and $AR_y \approx 0.027$. Raman spectra were collected from three representative regions: the flat surface, the valley center (within a ±2 μm distance of the valley center) and trench y-centerline (avoiding regions near the semicircular ends), with ten spectra acquired at each location (Figs. **6g–i**). To allow direct comparison, the spectra in Figs. **6g–i** are shown without smoothing. Averaged over the ten measurements, $MoS_2$ on the flat region (Fig. **6g**) exhibit E′ peak positions at 385.2 ± 0.5 $cm^{-1}$ and 403.8 ± 0.3 $cm^{-1}$, which corresponds to $\varepsilon_b$ in the range of -0.04 to 0.15%. At the valley center (Fig. **6h**), these peaks redshift to 375.7 ± 0.5 $cm^{-1}$ and 401.5 ± 0.3 $cm^{-1}$, respectively, corresponding to $\varepsilon_b$ of ~1.89%. Along the trench y-centerline (Fig. **6i**), the E′ peaks are located at 380.6 ± 1.5 $cm^{-1}$ and 402.9 ± 0.7 $cm^{-1}$, also showing a redshift and corresponding to $\varepsilon_{xx}$ of 1.4-1.6%. The Raman spectra at the valley center display features consistent with biaxial strain, which is redshifted along with an increase in the $E'/A_1'$ intensity ratio. In addition, Raman spectra collected along the valley do not show signs of the E′ peak splitting. In contrast, most (at least 6 out of 10) spectra collected along the trench y-centerline (Fig. **6i**) exhibit the signature of uniaxial strain (i.e., signs of peak splitting or broadening). However, not all spectra from the trench y-centerline display clear signs of E′ peak splitting. This is likely because the induced uniaxial strain ($\varepsilon_{xx}$ ~1.4-1.6%) is insufficient to produce fully resolved peak separation, while the observed splitting signal is comparable to the spectral resolution and experimental noise level, particularly for the spectra in Figs. **6g–i**, which were acquired at relatively low Raman laser power to minimize laser-induced heating effects.

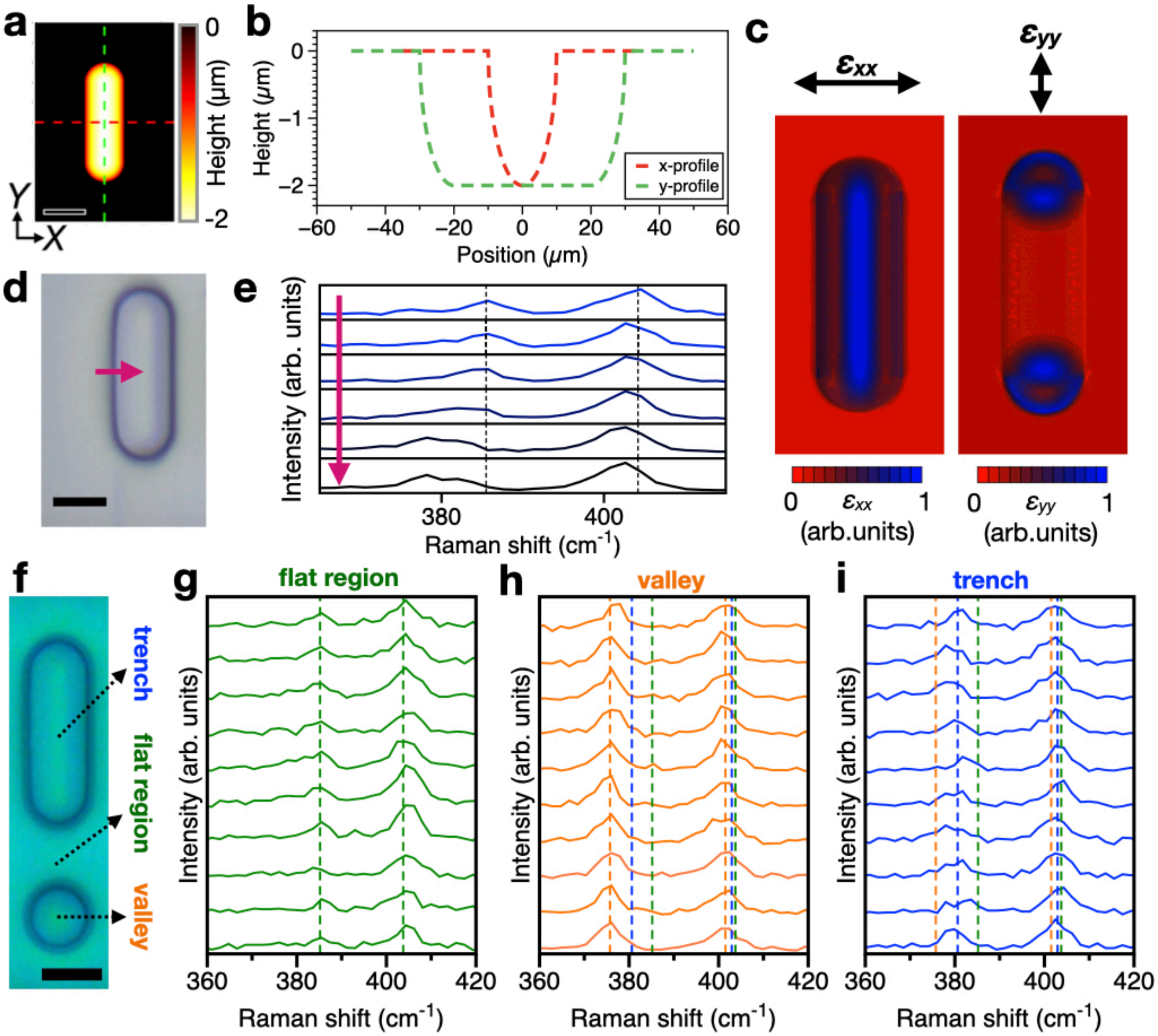

**Figure 6 | Uniaxial and biaxial strain engineering of monolayer $MoS_2$. a,** Height plot of a surface design containing a trench with $AR_x \approx 0.1$ and $AR_y \approx 0.03$. Scale bar, 20 μm. **b,** X- and Y-direction height profiles of the trench, corresponding to the red and green dashed lines in (a), respectively. **c,** Finite element analysis predictions of the top-down view of the spatially resolved uniaxial strain distributions ($\varepsilon_{xx}$ and $\varepsilon_{yy}$) in monolayer $MoS_2$ conformed to a trench with $AR_x = 0.15$ and $AR_y = 0.05$. $\varepsilon_{xx}$ and $\varepsilon_{yy}$ values are normalized to the maximum strain magnitude; non-normalized strain maps are provided in **Supplementary Note 16, Supplementary Fig. 23**. **d,** Optical microscope image of a substrate containing a trench with $AR_x \approx 0.081$ and $AR_y \approx 0.027$. Scale bar, 20 μm. **e**, Raman spectra of monolayer $MoS_2$ collected along the pink arrow in (d). The top spectrum corresponds to the starting point of the arrow, and the bottom spectrum corresponds to its end, with intermediate spectra sampled along the arrow path. **f**, Optical microscope image of monolayer $MoS_2$ conformed to a 2PL-printed substrate to a trench with $AR_x \approx 0.080$ and $AR_y \approx 0.027$, and

a valley with an AR of ≈ 0.085. Scale bar 20 μm. **g-i,** Raman spectra of monolayer $MoS_2$ collected from regions conforming to a valley and a trench shown in (f), with spectra from the flat region (g), the valley center (within a ±2 μm distance of the valley center) (h), and the trench centerline along the y-direction (avoiding regions near the semicircular ends) (i). The green, orange, and blue vertical dashed lines indicate the average E′ and $A_1'$ peak positions, as determined from Lorentzian fits, for monolayer $MoS_2$ on the flat region, at the center of a valley (AR ≈ 0.085), and along the trench centerline ($AR_x \approx 0.080$, $AR_y \approx 0.027$), respectively. Wide-range spectrum including the Si substrate peak is shown in **Supplementary Note 16**, **Supplementary Fig. 24**.

## Spatially programmable strain in bilayer heterostructures

Heterostructures composed of stacked 2D materials provide a versatile platform for engineering emergent electronic, optical, and mechanical properties that are inaccessible in the individual constituent layers[55]. We also show that our patterned substrates can strain vdW heterostructures. A monolayer tungsten disulfide ($WS_2$)–$MoS_2$ stack (see **Methods** and **Supplementary Note 18**) was transferred and conformed onto a valley with AR ≈ 0.070. Raman measurements acquired at both a flat region and the valley center spatially resolve the strain distribution within the heterostructure, while the distinct, non-overlapping Raman peaks of $WS_2$ and $MoS_2$ enable layer-specific strain characterization (Fig. **7a**). Reference peak positions on $SiO_2$ define 0% strain (see Figs. **7b, c** and **Supplementary Notes 4** and **19**). In flat regions, $WS_2$ exhibits negligible strain, while $MoS_2$ shows ~0.35% tensile strain; at the valley center, the tensile strain increases to ~0.50% for $WS_2$ and ~1.10% for $MoS_2$ (Fig. **7d**). See **Supplementary Note 20** for details on strain extraction from Raman spectra of monolayer $WS_2$**.** The lower strain in $WS_2$ indicates interlayer slippage at the 2D-2D interface, as the bottom $MoS_2$ layer, which is directly contacting the substrate, experiences greater strain. A similar trend has been observed in heterostructures strained using stressor layers (albeit at lower strain levels), where the layer interfacing with the strain-imparting material (in that case the top layer) is strained at a higher magnitude than the underlying layer[36]. In addition, it has been reported that incommensurately stacked 2D monolayers can exhibit superlubricity[56]. We also carried out DFT simulations of the sliding energy of the $WS_2$–$MoS_2$ bilayer (**Supplementary Note 18**), demonstrating the low energy required for sliding to occur at this interface.

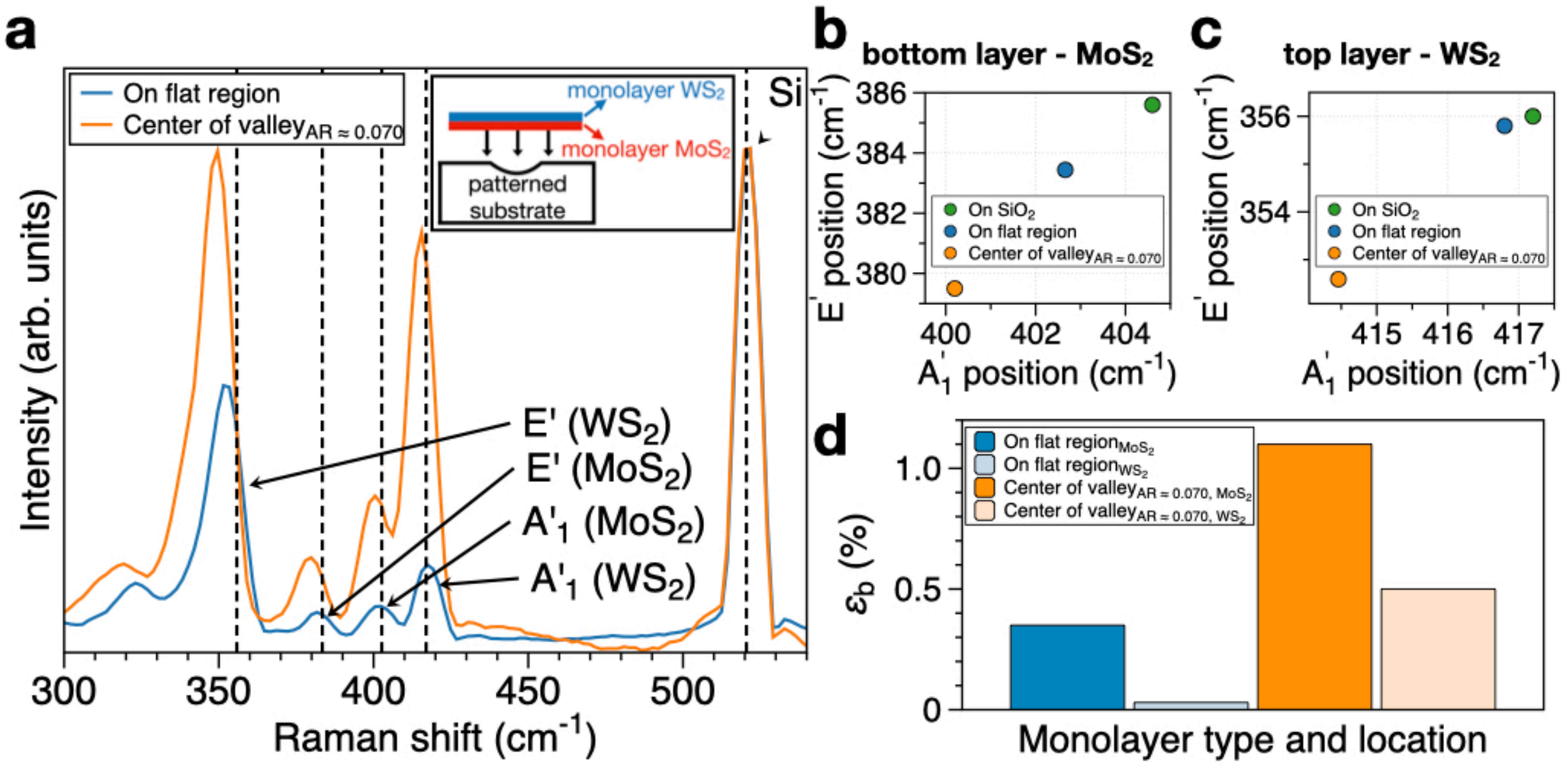

**Figure 7 | Spatially programmable strain in a $WS_2$–$MoS_2$ bilayer heterostructure. a,** Raman spectra collected from a monolayer $WS_2$-monolayer $MoS_2$ heterostructure conformed to a valley with an AR of ≈ 0.070. In the heterostructure, monolayer $WS_2$ forms the top layer and monolayer $MoS_2$ the bottom layer. Spectra shown are acquired from both a flat region and the valley center. Vertical dashed lines indicate the E′ and $A_1'$ peak positions of bottom layer $MoS_2$ and top layer $WS_2$ on the flat region, as determined from Lorentzian fits, and the Si peak at ~520.5 $cm^{-1}$. The broad feature spanning ~310–370 $cm^{-1}$ encompasses multiple $WS_2$ Raman modes, including the E’ and 2LA(M) peaks. Inset: schematic of the bilayer heterostructure on a patterned substrate. **b & c,** Scatter plots of E′ versus $A_1'$ Raman peak positions for monolayer $MoS_2$ (b) and monolayer $WS_2$ (c), obtained from the heterostructure with Raman spectra shown in (a). $SiO_2$ peak positions are extracted separately from a different heterostructure. **d,** Biaxial strain ($\varepsilon_b$) in monolayer $MoS_2$ and monolayer $WS_2$ extracted from Raman peak positions in (c).

## Conclusions

The strain engineering framework presented here establishes a distinct regime of strain control in 2D semiconductors (see **Supplementary Note 1**), enabling high-magnitude, spatially programmable, directionally tunable, and long-term stable strain in both monolayers and heterostructures. Using 2PL-fabricated patterned substrates composed of valley geometries, we achieved localized $\varepsilon_b$ of up to ~2.2%, with a spatial gradient of ~0.1 %$\varepsilon_b$ μm$^{-1}$ across monolayers through tuning of valley ARs. Raman point and mapping measurements confirmed strain magnitude and gradients. Strain remained stable for over 8 months. PL measurements verified localized modulation of optical emission, and C-AFM verified corresponding local electronic modulation. In addition, by exploiting the geometric versatility of 2PL to fabricate substrates capable of inducing simultaneous uniaxial and biaxial strain at different locations within a single monolayer, we demonstrated directional programming of strain. Finally, application to a $WS_2$-$MoS_2$ heterostructure demonstrates compatibility with complex van der Waals systems.

We envision that the presented strain engineering framework can establish a pathway toward programmable electronic and optoelectronic functionality in 2D materials through stable, spatially and directionally controllable strain engineering. The ability to locally tailor electronic and optical landscapes within a single 2D layer can enable future nanoelectronics with spatially programmed electronic characteristics, as well as lightweight broadband optoelectronic devices with broad-spectrum absorption and emission. We also expect that this framework can be extended to other 2D materials that exhibit high elasticity, low bending moduli, and can form strong interfacial interactions with the patterned substrate to maintain induced strain. The 2PL resin used in this work, IP-Visio, was specifically selected for its optical transparency, ensuring that Raman spectroscopy to extract strain could be performed without interference from background fluorescence. Although IP-Visio is a polymer, recent advances have demonstrated 2PL resins that can be converted into nanoscale optical-grade silica glass[57]. Incorporating such resins into substrate fabrication could further enhance the applicability of this strain engineering framework in device technologies. Translating this strain-engineering platform into a functional device will require further work with standard fabrication steps such as electrode and dielectric deposition, lithographic patterning, and thermal processing. In addition, we envision that holographic mask lithography [58], multi-focus 2PL[59], or metalens arrays[60] can be leveraged to scale up patterned substrates while preserving a clear separation of length scales between microscale features and the overall structure. Progress in 2D strain engineering via growth on patterned substrates may further complement our approach[61].

## Methods

**2D monolayer synthesis.** $MoS_2$ and $WS_2$ monolayers were exfoliated from a bulk crystal (2D Semiconductors) on an $SiO_2$-Si substrate containing a 10 nm Au layer. The $SiO_2$-Si substrate was exposed to glow discharge for 150 seconds and then the Au layer was sputtered (Leica EM

ACE600) at a deposition rate of 0.1 nm/s. Exfoliation was performed using heat-resistant tape (Nitto Denko) within 3-5 minutes after deposition.

**Two-photon lithography.** Surfaces were designed in MATLAB and exported as STL files, which were then extruded in Blender to generate substrates. The substrate STL files were imported into DeScribe (Nanoscribe GmbH) and printed on a silicon wafer (exposed to oxygen plasma for 3 minutes prior to resin drop casting) using Nanoscribe Photonic Professional GT2 system from IP-Visio (Nanoscribe GmbH) at the Centre for Research and Applications in Fluidic Technologies (CRAFT) facility at the University of Toronto. The Nanoscribe PPGT2 system employs a 100 fs, 80 MHz pulsed laser, with a wavelength of 780 nm focused through a 25x objective. The beam has a Lorentzian profile and is immersed in the IP-Visio resin during operation. The patterned substrates are printed using a hatching distance of 0.1 μm, an adaptive slicing distance ranging from 0.1 to 1.5 μm, a power setting of 6–8 mW, and a printing speed of 10 mm $s^{-1}$. The printed samples undergo the following development process: (1) Immerse in propylene glycol methyl ether acetate (PGMEA, also known as SU8 Developer) for 20 minutes. (2) Rinse with isopropyl alcohol (IPA) for 30 seconds. (3) After removing from IPA, gently blow-dry the back of the Si wafer using $N_2$ gas. (4) 10-minute blanket UV exposure (OAI Mask Aligner).

We note that our 2PL-fabricated patterned substrates containing valleys were initially based on a MATLAB-generated pattern that produced valleys with a slight distortion from perfect radial symmetry, likely arising from discretization of the surface (Figs. **2a**, **3a** and Supplementary Figs. **2a**, **11**, **26b**). We subsequently modified the pattern generation script to correct this, resulting in valleys with improved radial symmetry (Figure. **6f** and Supplementary Figure **16**). This change did not have a noticeable effect on the conformity of transferred monolayers or on strain realization in the conformed monolayers.

**Transfer and conforming of 2D monolayer on patterned substrate.** The transfer and conforming process of 2D monolayers outlined here is illustrated in **Supplementary Note 5**, with additional discussion and detail provided in **Supplementary Note 9**. First, PMMA A5 (MicroChem) is spin-coated on $MoS_2$–Au–$SiO_2$–Si at 1000 rpm for 60 s and then baked on a hot plate at 150 °C for 60 seconds. A thermal release tape with a target window is placed on the PMMA–$MoS_2$–Au–$SiO_2$–Si stack. This thermal tape–PMMA–$MoS_2$–$SiO_2$–Si stack is then placed in a potassium hydroxide (KOH) solution, made by dissolving 5 g of KOH pellets in 50 mL of DI water, to etch the $SiO_2$ layer and isolate the thermal tape–PMMA–$MoS_2$–Au. The thermal tape–PMMA–$MoS_2$–Au is then lifted with tweezers and placed in a potassium iodide and iodine ($KI/I_2$) solution (Transene, Gold Etch) for 2 minutes to selectively etch the Au. Afterward, the PMMA–$MoS_2$ is picked up with tweezers, rinsed in DI water for 1 minute, followed by another 5-minute rinse in fresh DI water, and then dried in ambient conditions for about 30 minutes before being placed in a vacuum desiccator overnight for further drying.

Before transfer, the patterned substrate is gently blown with nitrogen gas and heated on a hot plate at 80 °C for 5-10 minutes to remove residual contaminants. The thermal tape–PMMA–$MoS_2$ is then mounted onto a micromanipulator under an optical microscope, aligned with the patterned substrate, and brought into contact. The micromanipulator has a lateral resolution of ±5 μm. Once the PMMA and Si substrate which hosts the patterned substrate are in conformal contact, the thermal tape window is removed after cutting edges of the PMMA using a razor. The Si substrate with the PMMA–$MoS_2$ on the patterned substrate is then placed in a vacuum oven, which is gradually heated to 120 °C and maintained at that temperature for 1 hour. Finally, the PMMA is removed by exposing the substrate to acetone vapor. A beaker with 10 mL of acetone is placed on a hot plate set to 115-130 °C. The Si substrate containing PMMA–$MoS_2$–IP-Visio is attached to a glass slide using double-sided carbon tape and then placed upside down on top of the beaker so that the Si substrate faces the acetone at the bottom of the beaker. The beaker is then covered with parafilm and the sample is exposed to acetone vapor for 4-10 minutes before being removed. We note that bypassing the vacuum oven heating step increased the yield (defined as the fraction of samples exhibiting both successful transfer and strain realization confirmed by Raman spectroscopy) from ~80% to 100%.

**Bilayer heterostructure preparation.** Monolayer $MoS_2$ and monolayer $WS_2$ are individually exfoliated on Au substrates. First, monolayer $WS_2$ is transferred on top of monolayer $MoS_2$ on Au. Then, the bilayer stack is transferred and conformed to the patterned substrate. The exfoliation, transfer, and conforming procedures used for preparing the bilayer heterostructure followed the same methodologies as those used for the monolayer samples.

**Raman and PL spectroscopy.** Single point and mapping Raman measurements were performed using a Renishaw inVia Confocal Raman microspectrophotometer at a laser wavelength $\lambda$ = 532 nm, 1800 I mm$^{-1}$ grating, 20x objective, and spot size ~1 µm. Laser power was kept 1-2 mW to avoid local heating induced by the laser. Mapping was conducted with x and y steps of 1 µm (Fig. **3d**) and 2 µm (Fig. **3b**). Except for Raman spectra presented in the "Uniaxial strain engineering of monolayer $MoS_2$" section, which only presents raw spectra, all other spectra were are presented after baseline removal and smoothing. Baseline removal and smoothing were carried out using the Substract baseline and Smooth function in the Raman manufacturer's (Renishaw) WiRE software.

Single point PL measurements presented in Fig. **4a** were performed using a Nanobase XperRAM RF system at room temperature at a laser wavelength $\lambda$ = 532 nm, 300 I mm$^{-1}$ grating, and 40x objective. Laser power was kept at ~1 mW to avoid local heating induced by the laser. Each spectrum was averaged over 20 individual spectral acquisition.

Single point PL measurements presented in **Supplementary Note 12, Supplementary Figure 17a** were performed using a Renishaw inVia Confocal Raman Micro-spectrophotometer at a laser wavelength $\lambda$ = 532 nm, 1200 I mm$^{-1}$ grating, and 50x objective. Laser power was kept below 10 mW to avoid local heating induced by the laser.

**AFM.** Atomic force microscopy **(**AFM) was performed using an Asylum Cypher S and MFP-3D (Oxford Instruments). Patterned substrates before and after monolayer transfer were imaged using AFM topographical imaging. AC-mode imaging was performed using a Ti-Ir-coated ASYELEC.01-R2 cantilever and k = 2.8 ± 0.5 N m$^{-1}$ (Asylum Research).

**C-AFM.** Conductive atomic force microscopy (C-AFM) was performed using an Asylum Cypher S atomic force microscope (Oxford Instruments) with a Ti-Ir-coated ASYELEC.01-R2 cantilever and k = 4 ± 0.5 N m-1 (Asylum Research). *I-V* curves were generated by sweeping a bias voltage from -2 V to 2 V for 5 cycles and averaging all measurements. The current range of our setup is ±20 nA.

A 2.5 nm Cr adhesion layer followed by a 50 nm Au layer was deposited using electron beam evaporation (Angstrom Engineering Nexdep Electron Beam Evaporator) onto a substrate fabricated with the two-photon lithography resin IP-Visio. Deposition of the Cr and Au layers were performed at a rate of ~0.2 Å $s^{-1}$.

To calculate differential conductance and extract band gap values, *I-V* data obtained from C-AFM measurements were smoothed using a Savitzky–Golay filter. Data points with current levels near ±20 nA, corresponding to the instrument's saturation limits, were excluded from the analysis. Differential conductance (d*I*/d*V*) was calculated numerically using finite differences, with the voltage midpoints between adjacent data points used as the x-axis. To isolate the rising edges toward the band extrema, data beyond the local conductance maximum in the positive voltage region (V > 0) and data preceding the maximum in the negative voltage region (V < 0) were also excluded from analysis. Linear fits were applied to the conductance values spanning from 10% to 100% of the local maximum in each region. The valence and conduction band edges were determined from the zero-crossing points of these linear fits, and the electronic band gap was estimated as the voltage difference between the two band edge positions.

**SEM imaging.** The overall surface morphology was captured using a Hitachi SU7000 Schottky field emission scanning electron microscope (FE-SEM) at an accelerating voltage of 7 kV and a chamber pressure of 30 - 50 Pa in variable pressure mode. The micrographs were captured using the ultra-variable pressure detector (UVD).

**DFT.** Density Functional Theory (DFT) calculations were performed with the VASP software[62], using GGA/PBE exchange-correlation functional, standard PAW pseudopotentials, and a plane-wave basis set. An energy cutoff of 550 eV was used. The unit cell of monolayer $MoS_2$ was modelled with a Gamma-centered k-point mesh of 15×15×1, and a vacuum of 15 Å in the *z* direction. Calculations were performed with an energy threshold of $10^{-5}$ eV, and ionic relaxation was performed for all systems until forces were lower than $10^{-2}$ eV $Å^{-1}$. The initial optimization of the system included unit cell relaxation. Following this, the system was biaxially strained by manually increasing the size of the cell and allowing atomic positions to relax for each strain value.

For $MoS_2$-$WS_2$ friction calculations, a bilayer was generated with a single unit cell of both materials which share a lattice constant of 3.18 Å. These calculations included a 15 Å vacuum layer on top of the bilayer, and van der Waals interactions were taken into account via Grimme's D3 method with zero-damping function. The *x* and *y* coordinates of the Mo and W atoms were controlled directly to mimic sliding along the zigzag and armchair directions, while their *z* coordinates were optimized in every step alongside the position of S atoms.

**FEA.** Finite element analysis (FEA) simulations were performed using Abaqus to estimate biaxial strain in monolayer $MoS_2$ conformed to valleys of varying aspect ratios. The constitutive stress-strain relationship for the $MoS_2$ monolayer was derived from DFT calculations. In the FEA model the monolayer was defined as a hyperelastic material, allowing for accurate modeling of the nonlinear mechanical response, including large deformations. The FEA model employed the M3D4R element, a 4-node quadrilateral membrane element, for the monolayer, while the valleys with different aspect ratios were defined as rigid bodies. The monolayer was initially positioned above the valley substrate with its perimeter nodes fixed, and a uniform downwards pressure was applied enabling it to conform to the valley surface, consistent with experiments. The in-plane strain distribution was extracted from the conformal monolayer.

## Acknowledgements

The authors thank the technical staff and acknowledge the use of shared facilities at the Centre for Research and Applications in Fluidic Technologies (CRAFT), the Ontario Centre for the Characterization of Advanced Materials (OCCAM), and the Toronto Nanofabrication Centre (TNFC) at the University of Toronto, as well as the Advanced Optical Microscopy Facility at the University Health Network. The technical support of A. Wasay and D. Voicu are recognized, as are D. Stratkov's discussions on the FEM simulations and R. Aguiar's discussions on polymer-based 2D material transfer. This work was supported by funding from the University of Toronto, the Digital Research Alliance of Canada, the Natural Sciences and Engineering Research Council of Canada (NSERC), and the Canada Foundation for Innovation (CFI). B.K. acknowledges support of the Fonds de recherche du Québec (FRQ) Doctoral Research Scholarship. P.S. and P.G.D. acknowledge the support of the Vanier Canada Graduate Scholarship.

## Data availability

The data supporting the plots and findings of this study are available from the corresponding author upon reasonable request.

## Ethics declarations

The authors declare no competing interests.

# Author contributions

Conceptualization: B.K.

Sample preparation: B.K., N.B.

Two-photon lithography: B.K., P.S.

Spectroscopy: B.K., M.A.I., D.B.K.

SEM: D.Y., B.K.

AFM and C-AFM: B.K., M.A.I.

Analysis of experimental data: B.K.

FEA: S.Y., B.K.

DFT: P.G.D., A.R., A.N.

Visualization: B.K.

Supervision, funding acquisition, and project administration: T.F., C.V.S., E.H.S., C.H.A., J.H., S.H., S.N.

Writing – original draft: B.K., T.F.

Writing – review & editing: All Authors

# Supplementary Information

# High-magnitude, spatially and directionally programmable, and sustained strain engineering of 2D semiconductors

Boran Kumral[1*], Pedro G. Demingos[2], Shuo Yang[1], Md Akibul Islam[1, 3], Peter Serles[1], Da Bin Kim[4], Dian Yu[2], Akhil Nair[1], Akshat Rastogi[1,2], Nima Barri[1], Cristina H. Amon[1,5], Jane Howe[2], Sjoerd Hoogland[4], Edward H. Sargent[4, 6, 7], SungWoo Nam[3, 8], Chandra V. Singh[2], Tobin Filleter[1*]

[1] Department of Mechanical & Industrial Engineering, University of Toronto, Toronto, ON, Canada

[2] Department of Materials Science and Engineering, University of Toronto, Toronto, ON, Canada

[3] Department of Mechanical and Aerospace Engineering, University of California, Irvine, Irvine, CA, USA

[4] Department of Electrical and Computer Engineering, University of Toronto, Toronto, ON, Canada

[5] Department of Chemical Engineering and Applied Chemistry, University of Toronto, Toronto, ON, Canada

[6] Department of Chemistry, Northwestern University, Evanston, IL, USA

[7] Department of Electrical and Computer Engineering, Northwestern University, Evanston, IL, USA

[8] Department of Materials Science and Engineering, University of California, Irvine, Irvine, CA, USA

[*]Corresponding author emails:

Boran Kumral: boran.kumral@mail.utoronto.ca

Tobin Filleter: filleter@mie.utoronto.ca

## Table of Contents

## Supplementary Note 1. Benchmarking of strain engineering techniques for 2D materials

**Supplementary Table 1:** Benchmarking strain engineering of 2D materials. References are ordered by their reported maximum strain magnitude. Whether each method can achieve spatial programmability, directional programmability, retains the applied strain, or has scalability potential is presented. Here, scalability potential refers to the prospective feasibility of integrating a given strain-engineering method into established electronic and optoelectronic fabrication processes. Rather than requiring full industrial readiness, this assessment emphasizes deterministic strain control, reasonable thermal requirements, and the potential for scale-up to larger substrates or higher-throughput production. 'NA' indicates that the corresponding metric was not reported. Methods are categorized as: (1) pressure or bulging, (2) mechanical substrate deformation, (3) wrinkling and buckling instabilities, (4) scanning probe–induced deformation, (5) topographic substrate patterning, (6) lattice mismatch, (7) electrostatics, (8) growth.

| Reference | Maximum strain (%) | Spatial programmability | Directional programmability | Retention of strain | Scalability potential | Method |
|---|---|---|---|---|---|---|
| [1] | 5.6 | X | X | X | X | 1 |
| [2] | 5.6 | X | X | NA | ✓ | 5 |
| [3] | 5 | X | X | X | X | 2 |
| [4] | 4.7 | X | X | X | X | 3 |
| [5] | 4.2 | X | X | ✓ | X | 2 |
| [6] | 3.7 | X | X | X | X | 2 |
| [7] | 3.5 | X | X | ✓ | ✓ | 5 |
| [8] | 3.4 | X | X | X | ✓ | 4 |
| [9] | 3 | X | X | X | X | 2 |
| [10] | 3 | X | X | NA | ✓ | 5 |
| [11] | 3 | X | X | X | X | 7 |
| [12] | 2.92 | X | X | X | X | 2 |
| [13] | 2.85 | X | X | NA | X | 5 |
| [14] | 2.8 | X | X | X | X | 2 |
| [15] | 2.5 | X | X | NA | X | 3 |
| [16] | 2.4 | X | X | NA | X | 3 |
| This work | 2.2 | ✓ | ✓ | ✓ | ✓ | 5 |
| [17] | 2 | X | X | NA | X | 5 |
| [18] | 2 | X | X | NA | ✓ | 6 |
| 19 | 1.97 | X | X | ✓ | X | 1 |
| 20 | 1.6 | X | X | X | X | 2 |
| 21 | 1.5 | X | X | X | X | 2 |
| 22 | 1.35 | X | X | NA | X | 5 |
| 23 | 1.3 | X | X | NA | X | 5 |
| 24 | 1.3 | X | X | ✓ | X | 3 |
| 25 | 1.3 | ✓ | NA | NA | ✓ | 4 |
| 26 | 1.2 | X | X | X | X | 2 |
| 27 | 1 | X | X | ✓ | ✓ | 6 |
| 28 | 1 | ✓ | NA | NA | ✓ | 5 |
| 29 | 1 | X | X | NA | X | 4 |
| 30 | 1 | ✓ | ✓ | NA | ✓ | 8 |
| 31 | 0.85 | X | X | NA | X | 3 |
| 32 | 0.85 | X | X | ✓ | ✓ | 6 |
| 33 | 0.8 | X | X | NA | ✓ | 6 |
| 34 | 0.74 | X | X | NA | ✓ | 5 |
| 35 | 0.7 | X | X | NA | ✓ | 6 |
| 36 | 0.7 | X | X | X | X | 2 |
| 37 | 0.7 | X | X | NA | ✓ | 6 |
| 38 | 0.7 | X | X | X | X | 2 |
| 39 | 0.64 | X | X | X | X | 2 |
| 40 | 0.63 | X | X | NA | X | 5 |
| 41 | 0.6 | X | X | NA | ✓ | 5 |
| 42 | 0.6 | X | X | ✓ | ✓ | 6 |
| 43 | 0.6 | X | X | X | X | 2 |
| 44 | 0.52 | X | X | X | X | 2 |
| 45 | 0.47 | X | X | NA | X | 4 |
| 46 | 0.3 | ✓ | X | NA | ✓ | 5 |
| 47 | 0.3 | X | X | NA | X | 2 |
| 48 | 0.23 | X | X | ✓ | ✓ | 6 |
| 49 | 0.07 | X | X | NA | X | 5 |

**Supplementary Note 2. IP-Visio substrates**

After parameterized sweeps of printing speed and power, a printing speed of 10 mm $s^{-1}$ and a power range of 6-8 mW was identified to be suitable for two-photon lithography of IP-Visio on an Si substrate. The fabricated IP-Visio substrates have low surface roughness of $S_{RMS} = 1.3 \pm 0.5$ nm (**Supplementary Figure 1**).

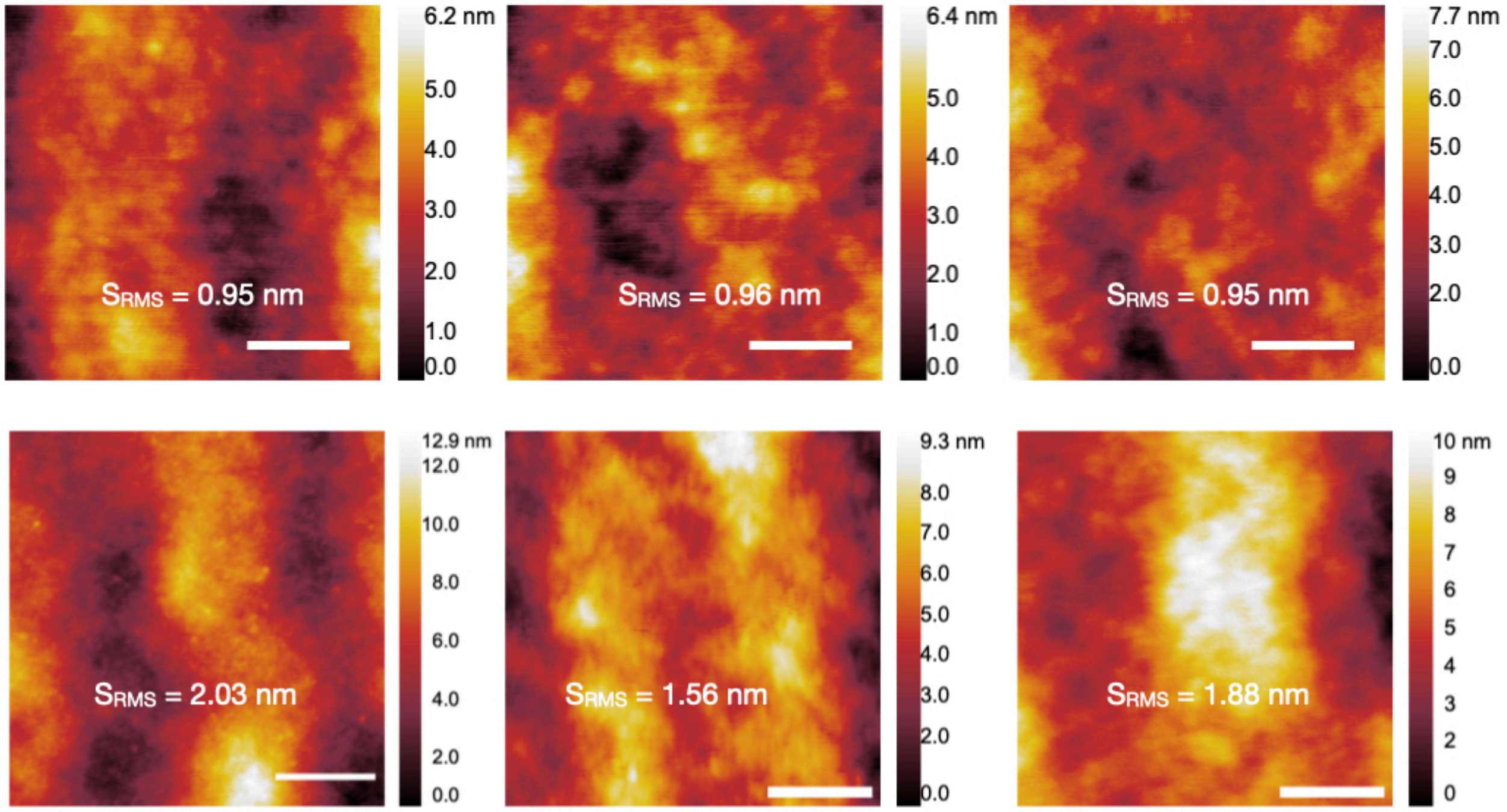

**Supplementary Figure 1 | 2PL-fabricated IP-Visio substrates.** Six 2 x 2 um topographic AFM images of IP-Visio substrates, along with their root-mean-square roughness ($S_{RMS}$) values. Scale bars, 500 nm.

Control over the valley aspect ratio (AR), defined as AR = h/L where L and h are the valley period and amplitude, respectively, is demonstrated by printing valleys with varying amplitudes. **Supplementary Fig. 2a** presents topographic AFM images of five valleys that share a uniform period but exhibit increasing amplitudes. Corresponding line profiles taken along the centerline of each valley are shown in **Supplementary Fig. 2b**. Valley period and height were extracted by fitting these profiles (**Supplementary Fig. 2c**). The transition to the valley from the flat region (left intersection in **Supplementary Fig. 2c**) and transition to the flat region from the valley (right intersection in **Supplementary Fig. 2c**) are manually selected, and then a linear fit is applied between the selected points. The valley period is taken as the distance between the left and right intersection points, while the valley amplitude is defined as the halfway distance between the height of the two intersection points. The resulting ARs for the samples, demonstrating AR control with steps of ~0.01, are provided in **Supplementary Fig. 2d**. In this work, we limited the range of the AR of fabricated valleys to ~0.02-0.15, based on the predictions of maximum biaxial strain in monolayer $MoS_2$ conformed to valleys of different ARs (Fig. **1f**).

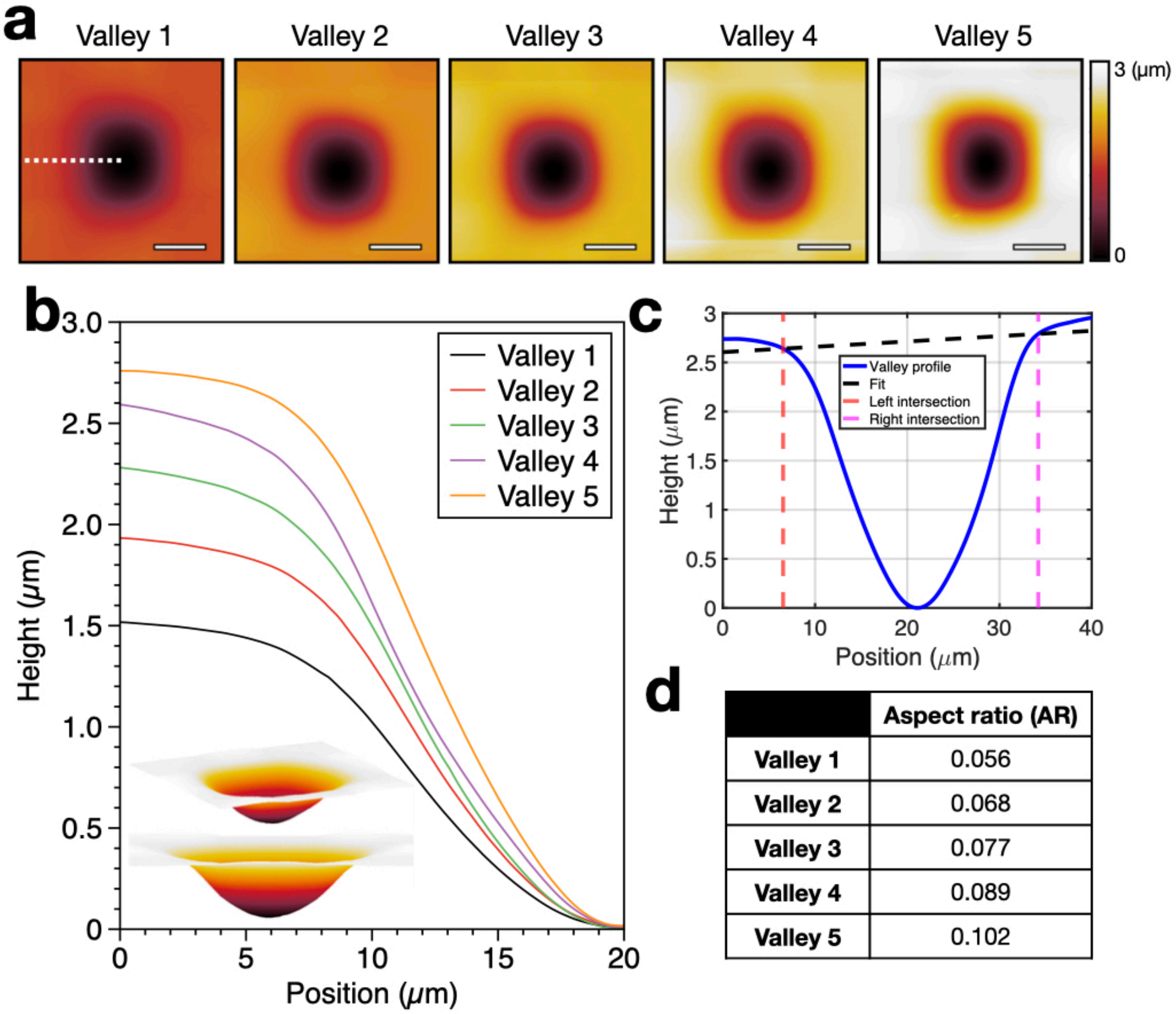

**Supplementary Figure 2 | Aspect-ratio control of valleys in substrates fabricated by two-photon lithography. a,** Topographic atomic force microscope (AFM) images of two-photon lithography (2PL) fabricated valleys of varying amplitudes. Scale bars, 10 µm. **b,** Line profiles from topographic AFM showing the half of the cross-sectional profile along the respective centreline of each valley shown in (a), with the white dashed line in (a) marking the example profile for valley 1. Inset shows 3D views of the topographic AFM data of a valley, which is not to scale. **c,** Fitting of a valley line profile from topographic AFM to extract valley height and period (not to scale). **d,** Determined aspect ratios for all valleys shown in (a).

Repeatability of AR control in 2PL-fabricated IP-Visio substrates was evaluated by printing substrates each containing three valleys with low, medium, and high aspect ratios. Two substrates were fabricated using identical printing conditions (6 mW power and 10 mm $s^{-1}$ scan speed), and a third was printed with a slightly higher dose (8 mW power and 10 mm $s^{-1}$). Under identical printing parameters, the resulting valley profiles are nearly indistinguishable, while increasing the dose introduces some variation in the profiles (**Supplementary Fig. 3a**). Extracted aspect ratios for the printed valleys are shown in **Supplementary Fig. 3b**, demonstrating that valleys produced using the same dose exhibit highly consistent aspect ratios, with sample-to-sample variation of no more than ~0.001, while a change in dose leads to slight shifts in AR values.

**a**

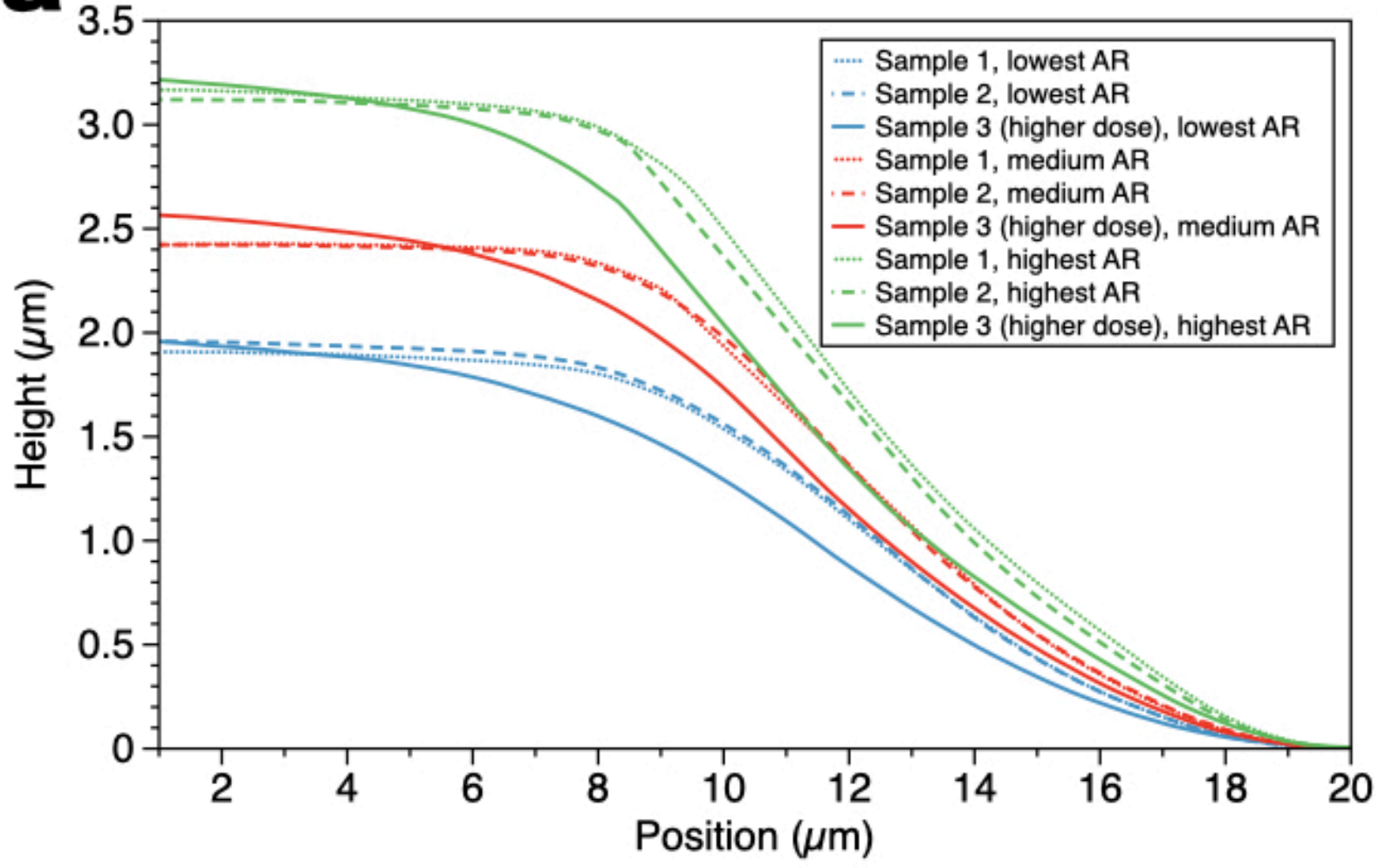

**b**

| | Aspect ratio (AR) |
|---|---|
| Sample 1, lowest AR | 0.070 |
| Sample 2, lowest AR | 0.069 |
| Sample 3 (higher dose), lowest AR | 0.065 |
| Sample 1, medium AR | 0.094 |
| Sample 2, medium AR | 0.093 |
| Sample 3 (higher dose), medium AR | 0.091 |
| Sample 1, highest AR | 0.116 |
| Sample 2, highest AR | 0.118 |
| Sample 3 (higher dose), highest AR | 0.112 |

**Supplementary Figure 3 | Repeatability of aspect-ratio control of valleys in substrates fabricated by two-photon lithography. a,** Topographic AFM line profiles showing the half cross-sectional shapes along the centrelines of the valleys. The data include three samples, each containing three valleys with increasing aspect ratios. Samples 1 and 2 were fabricated using a Nanoscribe power setting of ~6 mW, while sample 3 was fabricated at ~8 mW and all other fabrication parameters were kept the same. **b,** Extracted aspect ratios for all valleys shown in (a).

## Supplementary Note 3. Thickness characterization of monolayer $MoS_2$

The single-layer structure of the exfoliated monolayers were verified using AFM-based thickness characterization.

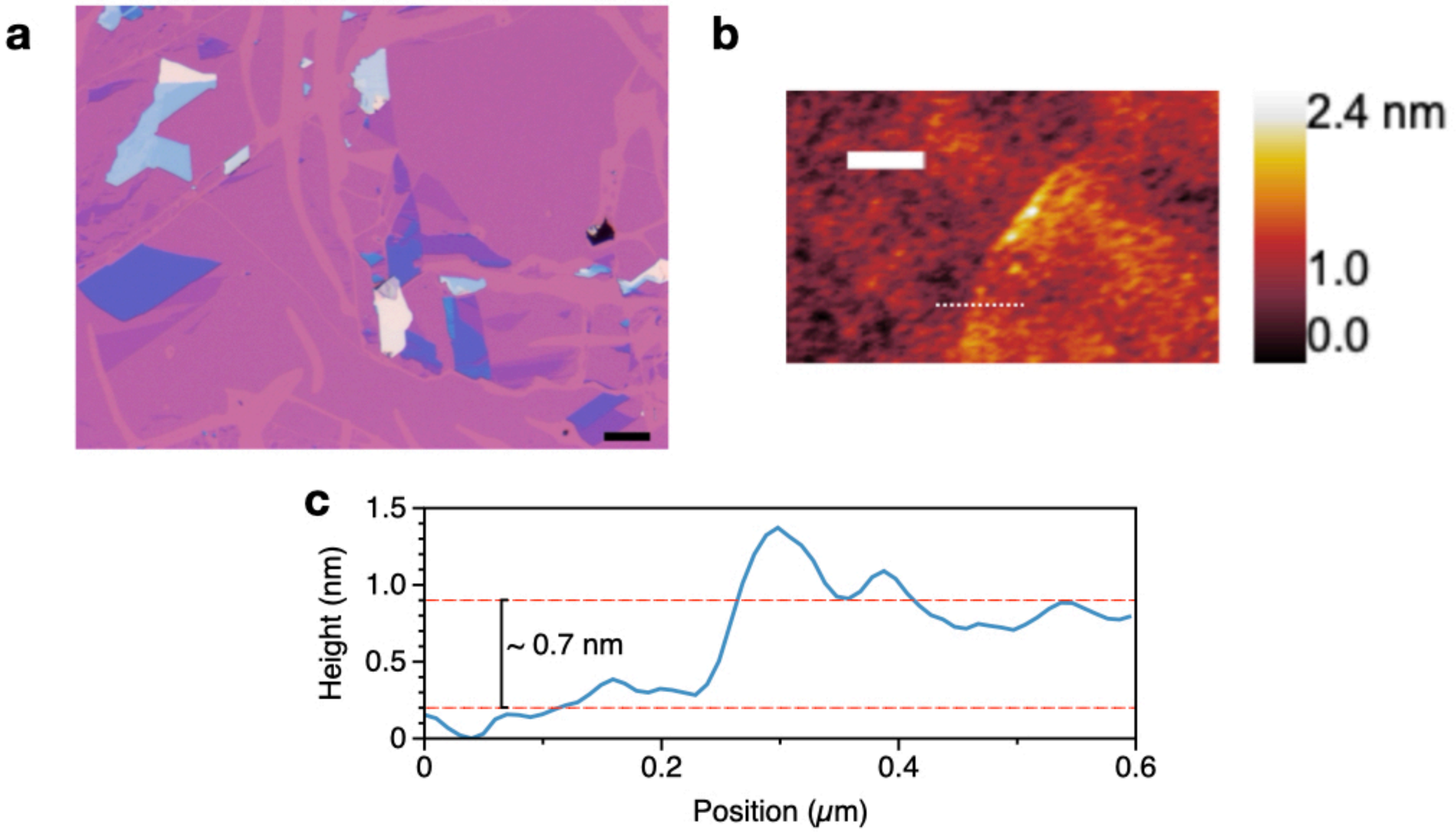

**Supplementary Figure 4 | Monolayer $MoS_2$. a,** Optical image of large, exfoliated monolayer $MoS_2$ flakes on a pre-patterned, Au-coated $SiO_2$–Si substrate. The image also contains few-layer and bulk $MoS_2$ flakes, which appear as varying shades of blue and white. Scale bar, 100 μm. **b,** AFM topographical image of monolayer $MoS_2$ on Au and **c,** profile along white dashed line shown in (b). Scale bar (b), 500 nm.

## Supplementary Note 4. Raman spectra of monolayer $MoS_2$ on different substrates

The single-layer structure of the monolayers was verified using Raman spectroscopy. Raman spectra of monolayer $MoS_2$ exfoliated on Au and on $SiO_2$ are shown in **Supplementary Figure 5a**. The Raman spectra of monolayer $MoS_2$ on $SiO_2$ reveals a peak position difference of ~19.3 $cm^{-1}$ between the E' and $A_1$' peaks, which is in the range of reported values for this peak position difference[50,51]. The Raman spectra of monolayer $MoS_2$ transferred to $SiO_2$ and to IP-Visio substrates are shown in **Supplementary Figure 5b**.

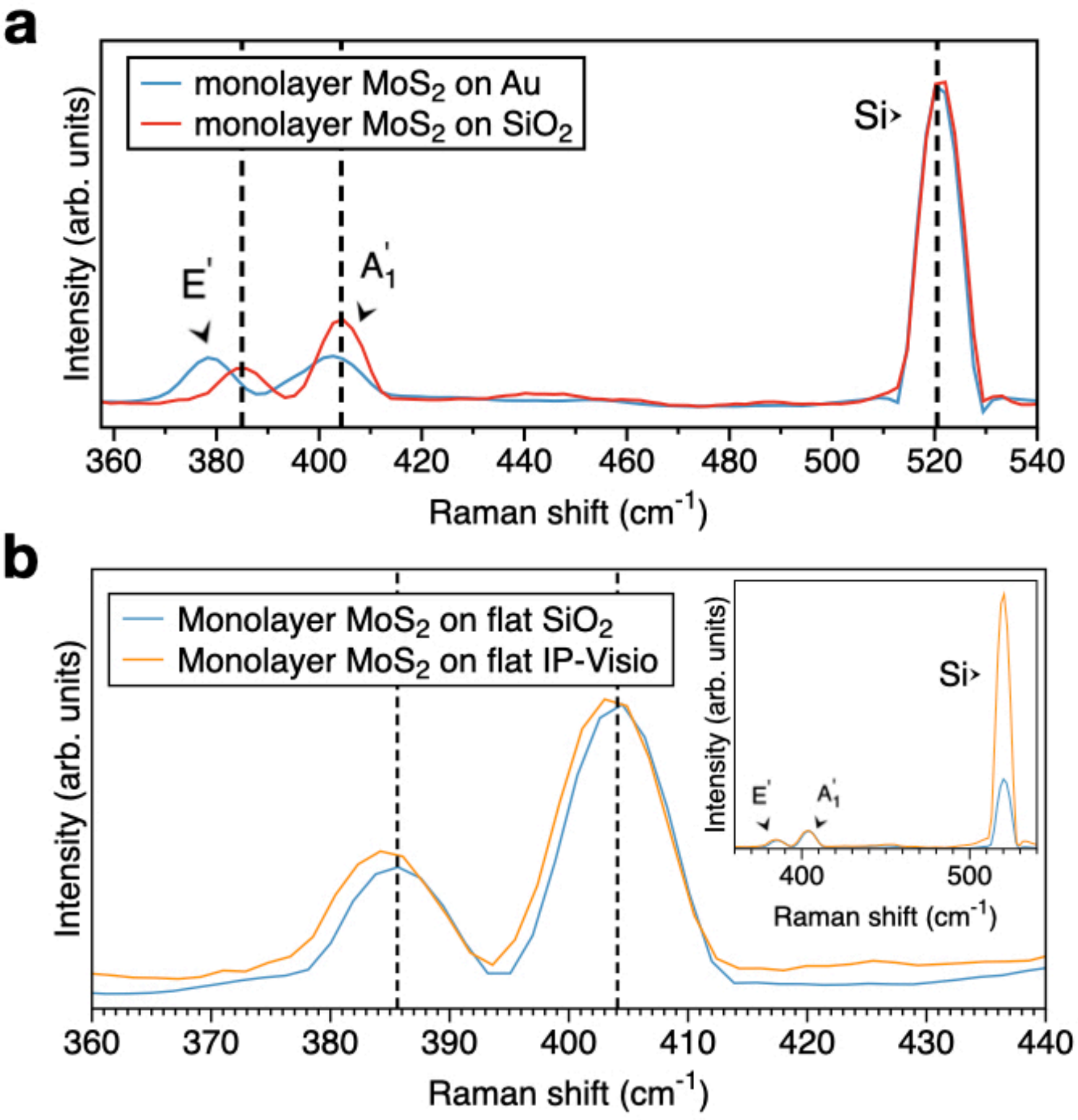

**Supplementary Figure 5 | Raman spectra of monolayer $MoS_2$ on different substrates. a,** Raman spectra of monolayer $MoS_2$ exfoliated on Au and on $SiO_2$. Spectra are normalized to the Si peak intensity. Vertical dashed lines indicate the E′ and $A_1$′ peak positions of monolayer $MoS_2$ on $SiO_2$, as determined from Gaussian fits, and the Si peak at ~520.5 $cm^{-1}$. **b,** Raman spectra of monolayer $MoS_2$ transferred to $SiO_2$ and to a flat IP-Visio substrate. Spectra are normalized to the $A_1$′ peak intensity. Vertical dashed lines indicate the E′ and $A_1$′ peak positions of monolayer $MoS_2$ on $SiO_2$, as determined from Gaussian fits. Inset: wide-range spectrum displaying the Si substrate peak (~520.5 $cm^{-1}$).

Monolayer $MoS_2$ was transferred to flat IP-Visio substrates of 2 µm and 20 µm (**Supplementary Figs. 6a and 6b**) thicknesses. It was observed that both the peak positions and the intensities of the Raman peaks of monolayer MoS2 on these substrates of varying thicknesses were not altered considerably (**Supplementary Figure 6c**).

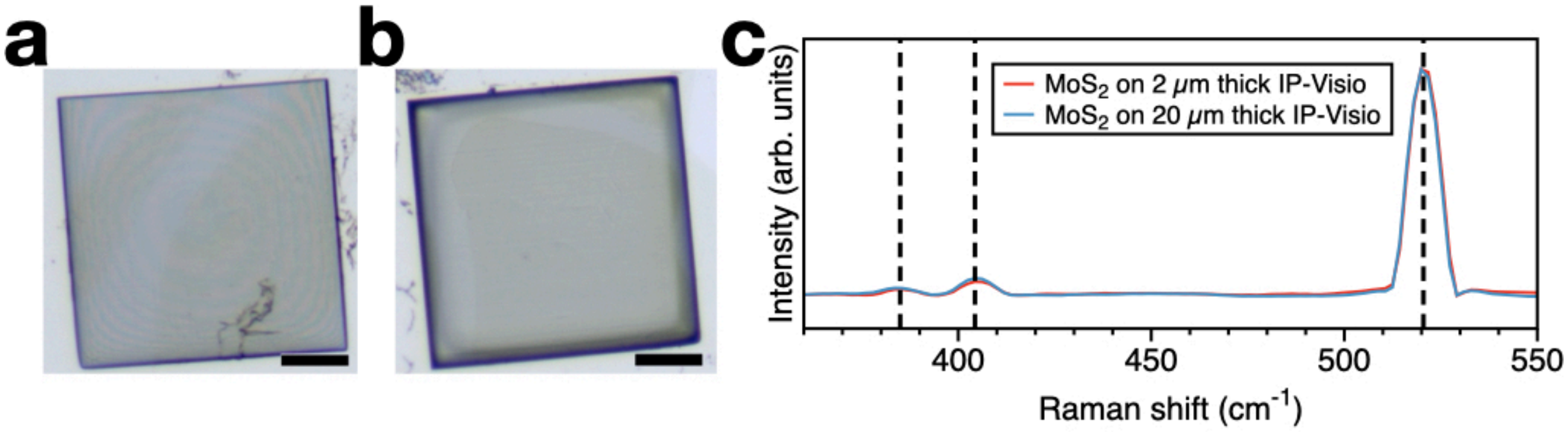

**Supplementary Figure 6 | Raman spectra of monolayer $MoS_2$ on flat IP-Visio substrates of varying thicknesses. a & b,** Optical microscope images of (a) 2 µm thick and (b) 20 µm thick IP-Visio substrates after transfer of monolayer $MoS_2$. Scale bars, 50 µm. **c,** Raman spectra of monolayer $MoS_2$ on 2 and 20 µm thick IP-Visio substrates shown in (a) & (b).

**Supplementary Note 5. Transfer and conforming of 2D semiconductor**

The transfer and conforming process for placing monolayer $MoS_2$ onto 2PL-fabricated patterned substrates is shown in **Supplementary Figure 7**. This is a multi-step procedure involving several parameters that can influence the outcome. Under consistent fabrication and transfer conditions, we estimate an ~80% success rate (based on 10 independent experiments) for achieving Raman-verified, spatially programmable strain in monolayer $MoS_2$ transferred onto 2PL-fabricated IP-Visio patterned substrates containing three valleys with different aspect ratios. We note that certain steps in the process shown in **Supplementary Figure 7** are particularly susceptible to introducing variability. For example, during PMMA removal, submerging the sample directly in acetone rather than using acetone vapor, or handling the sample too aggressively, can fracture the monolayer in the valley regions and lead to strain relaxation. It is plausible that through the development of automated and controlled transfer systems, such sources of variability could be avoided. We also note that bypassing the vacuum oven heating step (step L in **Supplementary Figure 7**) increased the yield (based on 5 independent experiments), defined as the fraction of samples exhibiting both successful transfer and strain realization confirmed by Raman spectroscopy, from ~80% to 100%. Additional procedural details are included in the Methods section of the main text.

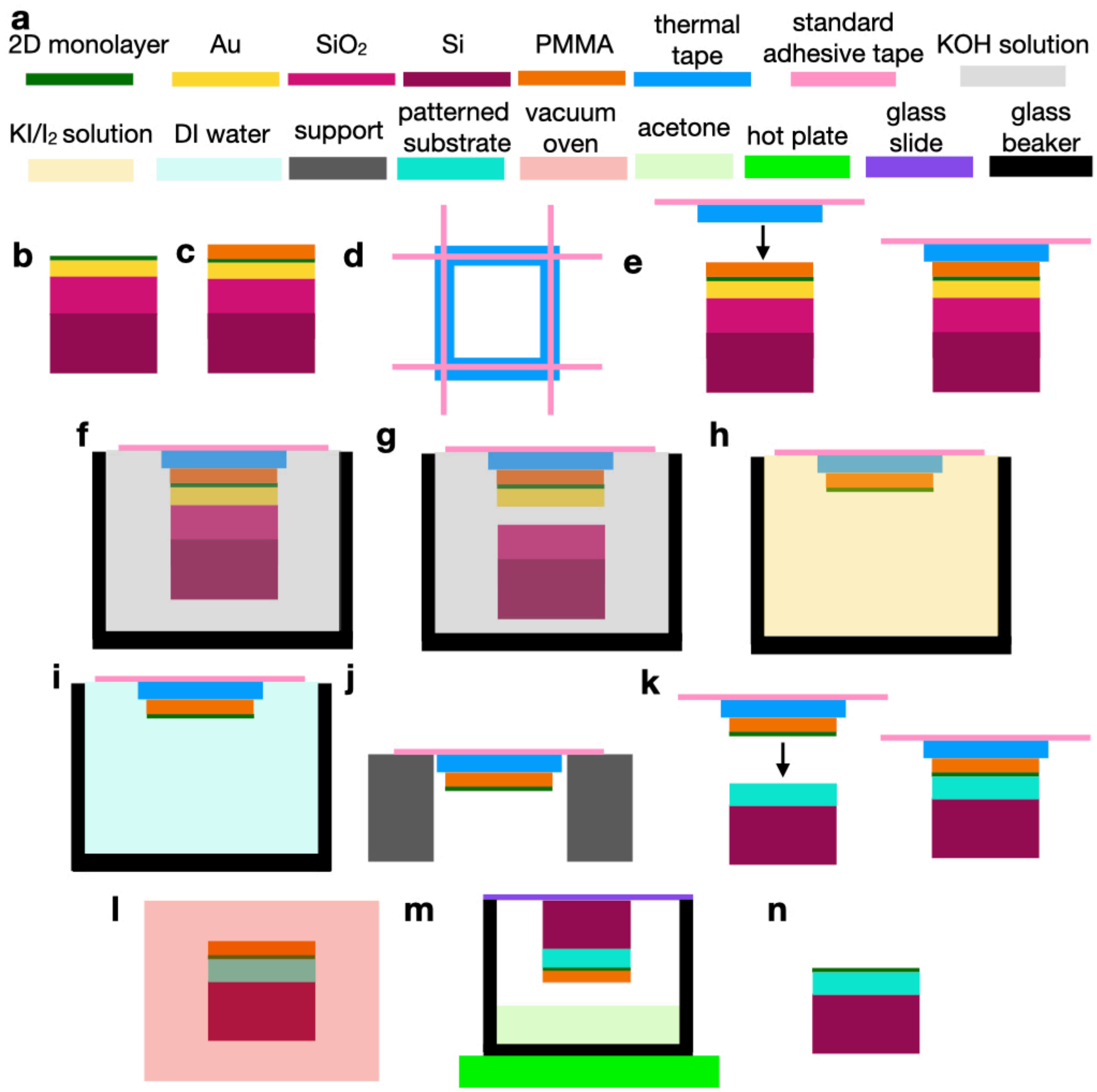

**Supplementary Figure 7 | Transfer and conforming of monolayer to patterned substrate. a,** Color codes for items outlined in the transfer process. **b,** Monolayer $MoS_2$ is exfoliated on an Au substrate. **c,** PMMA is spin-coated on the Au substrate. **d,** A thermal tape target window with lines of adhesive tape on its sides is prepared. **e,** The thermal tape window is placed on the target region containing monolayers. **f,** The prepared structure is placed in KOH solution. **g,** KOH etches the structure at the Au–$SiO_2$ interface. **h,** The structure is picked up with tweezers and placed in KI/$I_2$ solution to etch the Au. **i,** The remaining structure is placed in DI water. **j,** The sample is dried overnight by hanging over its ends. **k,** The monolayer–PMMA–thermal tape structure is placed on a patterned substrate. Alignment is performed under an optical microscope. **l,** The PMMA is cut using a razor along the thermal tape window and the thermal tape is removed. The remaining structure is placed in a vacuum oven. **m,** Acetone vapor is used to remove the PMMA. **n,** Conformed monolayer sample is prepared.

## Supplementary Note 6. Analytical and FEA predicted strain fields

Biaxial strain ($\varepsilon_b$) is imparted on a 2D material conformed to a sinusoidal valley, and for monolayer molybdenum disulfide ($MoS_2$) the biaxial strain gauge factor is 2.3 times higher than the uniaxial strain gauge factor[52]. The analytical prediction is obtained by solving the Föppl-van Kármán equation for a sinusoidal valley[53,54] (see **Supplementary Note 7** and **Supplementary Figs. 8a-d**). The analytical model assumes that the monolayer is fully relaxed to minimize its elastic energy while conforming to the substrate topography. This relaxation implies the absence of external forces at the monolayer's perimeter and no frictional interaction with the substrate. Physically, this corresponds to a scenario where the 2D material is stamped onto a frictionless substrate, allowing the layer to undergo lateral contraction. In the FEA simulations, the substrate is defined as a rigid body and monolayer $MoS_2$ is conformed to the substrate by application of uniform downward pressure on the monolayer (see **Methods** and **Supplementary Figs. 8e-h**).

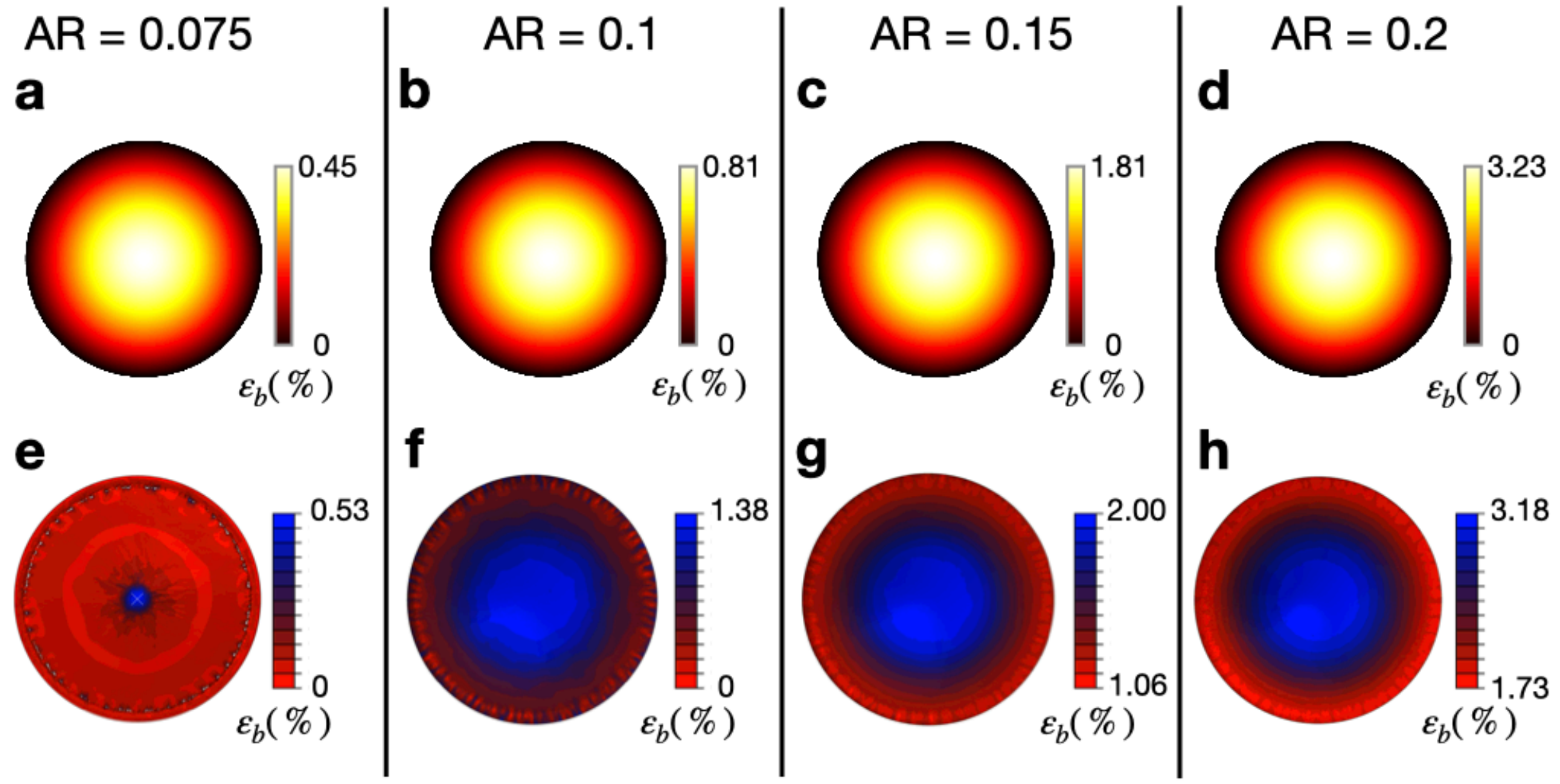

**Supplementary Figure 8 | Biaxial strain ($\varepsilon_b$) fields of monolayer $MoS_2$ conformed onto valleys of varying aspect ratios. a-h,** Top-down views of the predicted $\varepsilon_b$ strain fields in monolayer $MoS_2$ conformed onto valleys of varying aspect ratios (ARs) based on analytical theory (a–d) and finite element analysis (FEA) (e–h).

As expected, the monolayer experiences radially symmetric, biaxial tensile strain. The highest level of strain is always at the center of the valley, which decreases continuously from the center towards the edges, and reaches its minimum value near the edges. Although the maximum strain values predicted by the analytical and FEA models for each valley AR are similar, the radial strain distribution does differ. In the FEA predictions, the $\varepsilon_b$ near the valley edges is relatively high, whereas in the analytical model, the $\varepsilon_b$ near the valley edges is near zero.

Fig. **1e** in the main text is constructed directly from the continuum analytical theory presented in **Supplementary Note 7**. This analytical model assumes that the 2D material fully relaxes to minimize its elastic energy while conforming to the substrate topography. This corresponds to a frictionless "stamping" process, where the layer can contract laterally without experiencing forces at its remote boundaries. Under these conditions, the strain in the flat regions of the surface is predicted to be zero. Accordingly, Fig. **1e** is generated by assigning zero strain to the flat regions and using the analytical strain predictions for the circular valleys.

**Supplementary Note 7. Continuum level analytical theory displacement and strain fields**

The Föppl-von-Kármán equations[53] are a set of nonlinear partial differential equations describing the large deflection of linear elastic plates:

$$\Delta^2\chi = -Y\,(f_{xx}f_{yy} - {f_{xy}}^2) \quad (1)$$

Where $\chi, Y, and\ f$ are the Airy stress function, Young's modulus, and surface shape function respectively. The valley surface shape function can be defined by:

$$f(x, y, t) = h * sin(ax) * \sin(By) \quad (2)$$

h is the sinusoid height, and $a = 2\pi/L_x$ and $B = 2\pi/L_y$ where $L_x\ and\ L_y$ are the lateral periodicity in x and y directions.

Sub in derivatives into FvK and Integrate 4 times to solve for $\chi$:

$$\chi = \frac{Y*h^2}{32}(\left(\frac{B}{a}\right)^2 cos(2ax) + \left(\frac{a}{B}\right)^2 \cos(2By)) \quad (3)$$

From the Airy stress function $\chi$, the components of the strain field can be obtained where ν is the Poisson's ratio, $\varepsilon_{ij}$ is a Levi-Civita tensor, and $u_{ij}$ and is the strain tensor:

$$u_{ij} = (\tfrac{1}{Y})(\varepsilon_{ik}\varepsilon_{jl} - v\delta_{ik}\delta_{jl})\partial_k\partial_l\chi \quad (4)$$

$$u_{xx} = \left(\frac{1}{Y}\right)\left(\chi_{xx} - v\chi_{yy}\right) \quad (5)$$

$$u_{yy} = \left(\frac{1}{Y}\right)\left(\chi_{yy} - v\chi_{xx}\right) \quad (6)$$

$$u_{xx} = -\frac{h^2}{8}(B^2 cos(2ax) - va^2 cos(2By)) \quad (7)$$

$$u_{yy} = -\frac{h^2}{8}(a^2cos(2By) - vB^2cos(2ax)) \quad (8)$$

Biaxial strain ($\varepsilon_b$) in % is extracted from:

$$u_{ii} = \frac{u_{xx}+u_{yy}}{2} \quad (9)$$

$$\varepsilon_b = 100u_{ii} \quad (10)$$

For clarity, we note that $\varepsilon_{ij}$ does not represent strain, it is a mathematical tool to express the strain components $u_{ij}$ in terms of the Airy stress function $\chi$. The tensor $u_{ij}$ represents strain and is used to extract the reported biaxial strain $\varepsilon_b$. Values in the range of 0.25-0.44 have been reported for the Poisson's ratio of monolayer $MoS_2$ [55]. We use the average of the upper and lower ends of the range, 0.345.

**Supplementary Note 8. Adhesion energy of the monolayer $MoS_2$ and IP-Visio interface**

The relationship between the pull-off force ($P_c$) recorded by the AFM and the adhesion ($\gamma$) is:

$$\gamma = \frac{-P_C}{\chi \pi R} \tag{11}$$

where R is the tip radius, and $\chi$ ranges monotonically from 1.5 for the Johnson, Kendall and Roberts (JKR) limit to 2 for the Derjaguin, Muller and Toporov (DMT) limit. The tip radius (R) is determined via SEM to be ~400 nm (see **Supplementary Figure 9**). A process outlined in Grierson et al. was used to determine which regime our case corresponds to[56].

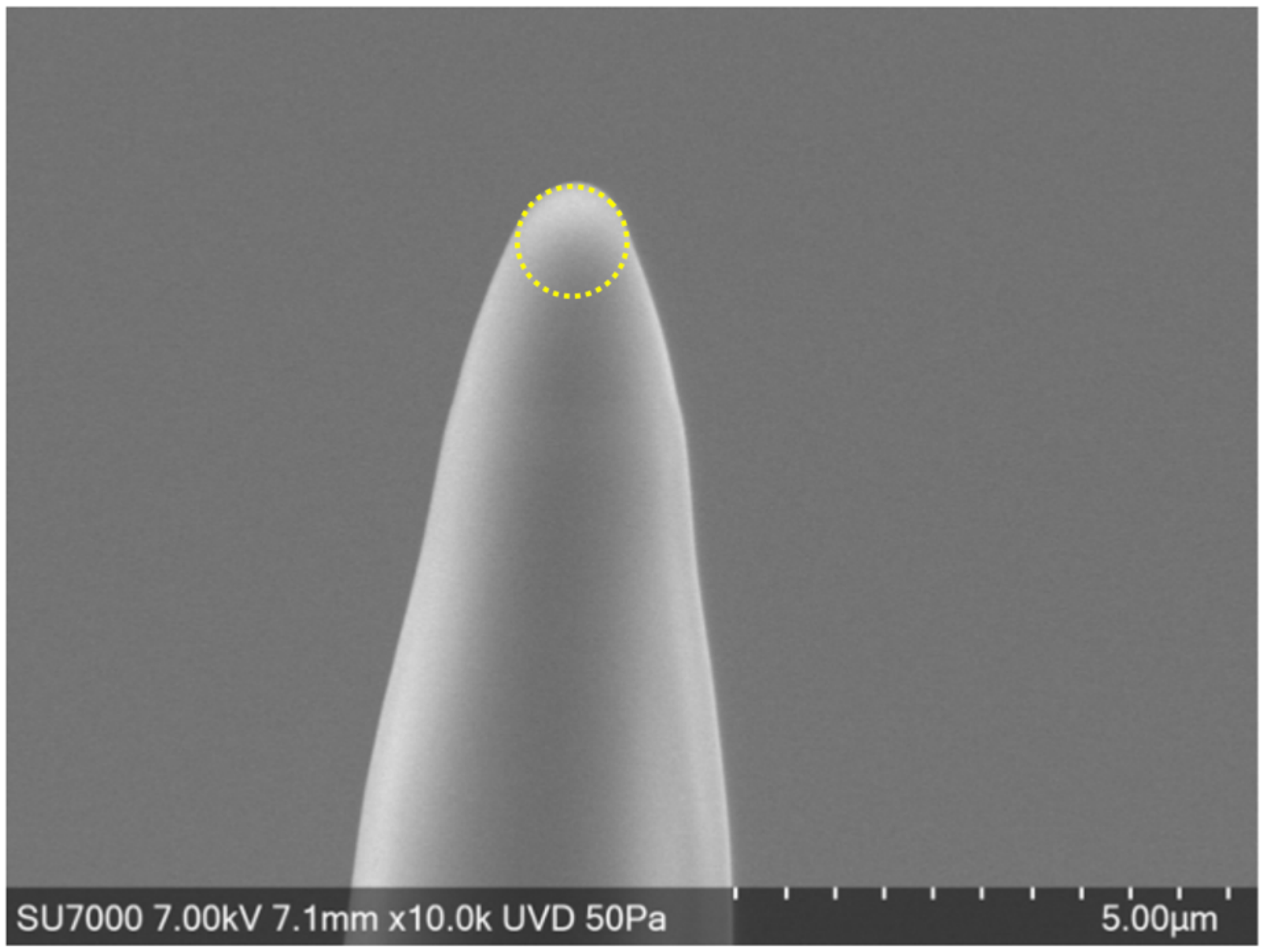

**Supplementary Figure 9 | 2PL-fabricated IP-Visio tip**. FE-SEM image of IP-Visio tip fabricated on a tipless cantilever. The dashed yellow circle marks the region from which the tip radius was measured.

The $\lambda$ parameter is given by the expression:

$$\lambda = 2\sigma_0 \left(\frac{R}{\pi \gamma K^2}\right)^{1/3} \tag{12}$$

where $\sigma_0$ is the minimum adhesion stress for a Lennard–Jones potential (with equilibrium separation $z_0$) and K is obtained from the contact mechanics-based relationship[57] valid for a sphere and a flat plane:

$$K = \frac{4}{3}\left(\frac{(1-\nu_1^2)}{E_1} + \frac{(1-\nu_2^2)}{E_2}\right) \tag{13}$$

where $E_1$ and $E_2$ are the Young's modulus and $\nu_1$ and $\nu_2$ are the Poisson's ratio of the tip and flat plane, respectively. Since there is only a single layer of $MoS_2$ on the $SiO_2$ substrate, we used the Young's modulus and Poisson's ratio of $SiO_2$ as $E_2$ and $\nu_2$ [58]. The elastic properties of the contact materials[59–61] are provided in **Supplementary Table 2**.

**Supplementary Table 2:** Young's modulus and Poisson's ratio of contact materials.

| | IP-Visio | $SiO_2$ |
|---|---|---|
| **Young's modulus (GPa)** | 2.8 | 70 |
| **Poisson's ratio** | 0.3 | 0.17 |

If $\lambda > 5$, the JKR model applies and if $\lambda < 0.1$ the DMT model applies. Values between 0.1 and 5 correspond to the 'transition regime' between JKR and DMT models. $\lambda$ is related to Tabor's parameter $\mu_T$ through the relationship $\lambda = 1.157\ \mu_T$. Tabor's parameter is given by:

$$\mu_T = \left(\frac{16R\gamma^2}{9K^2 z_0^3}\right)^{1/3} \tag{14}$$

First, we assume that our case is in the DMT regime. To test this assumption, a lower bound value is assumed for $z_0$ and smallest possible $\chi$. Thus, we use the Mo-S bond length of 2.4 Å [62] for $z_0$ and 1.5 for $\chi$. These assumptions yield a $\mu_T$ of 3.33 and $\lambda$ of 3.86, which corresponds to the transition regime. This assumption yields an upper bound $\gamma$ of 0.11 J m$^{-2}$.

Then, we assume that our case is in the JKR regime. To test this assumption, a higher bound value is assumed for $z_0$ and highest possible $\chi$. Thus, we use snap-in distance of the AFM tip, 70 nm, for $z_0$ and 2 for $\chi$. The snap-in distance is determined from the maximum snap in force (**Supplementary Figure 10**) and tip stiffness of 2.8 N/m. These assumptions yield a $\mu_T$ of 0.01 and $\lambda$ of 0.011. In the JKR regime $\mu_T$ is expected to exceed 5. Thus, again, we are not in the assumed regime. This assumption yields a lower bound $\gamma$ of 0.079 J $m^{-2}$.

We conclude that we are in the transition regime, and we determine the adhesion value between monolayer $MoS_2$ and IP-Visio to be an average of the upper and lower bound of $\gamma$, and the error as the half the difference between them. Thus $\gamma_{ML\ MoS2\text{-}IP\text{-}Visio} = 0.095 \pm 0.016$ J $m^{-2}$.

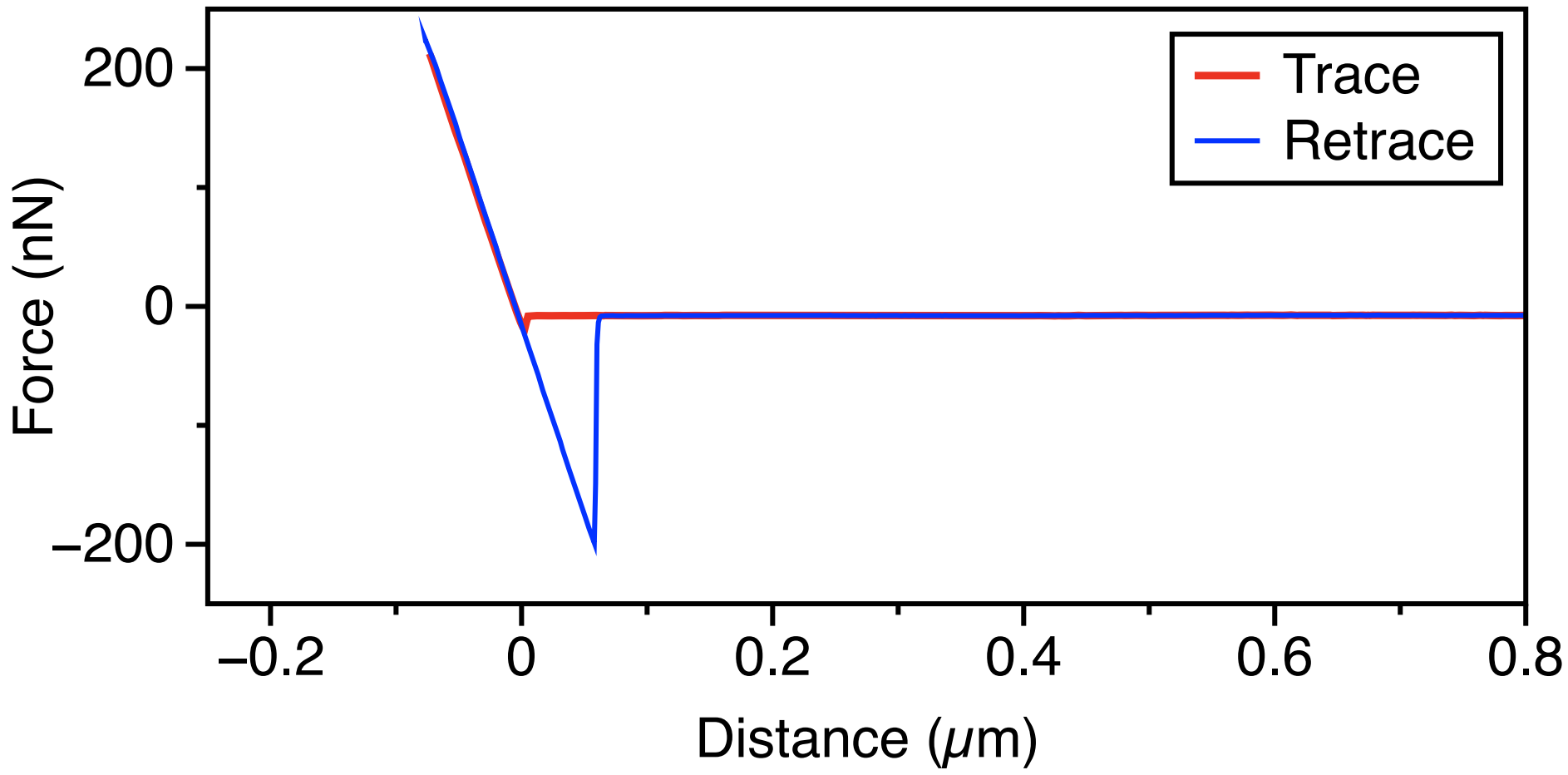

**Supplementary Figure 10 | Force-distance curve**. A representative force-distance curve measured on monolayer $MoS_2$ on $SiO_2$ using a spherical IP-Visio tip.

**Supplementary Note 9. Conformity of transferred monolayers**

**Supplementary Figure 11** provides AFM analysis of monolayer $MoS_2$ conforming to patterned substrates with valleys, including line profiles (vertical and horizontal) before and after transfer of monolayer $MoS_2$, topographic images of valleys, and 3D views of a topographic AFM image of monolayer $MoS_2$ conformed to a valley.

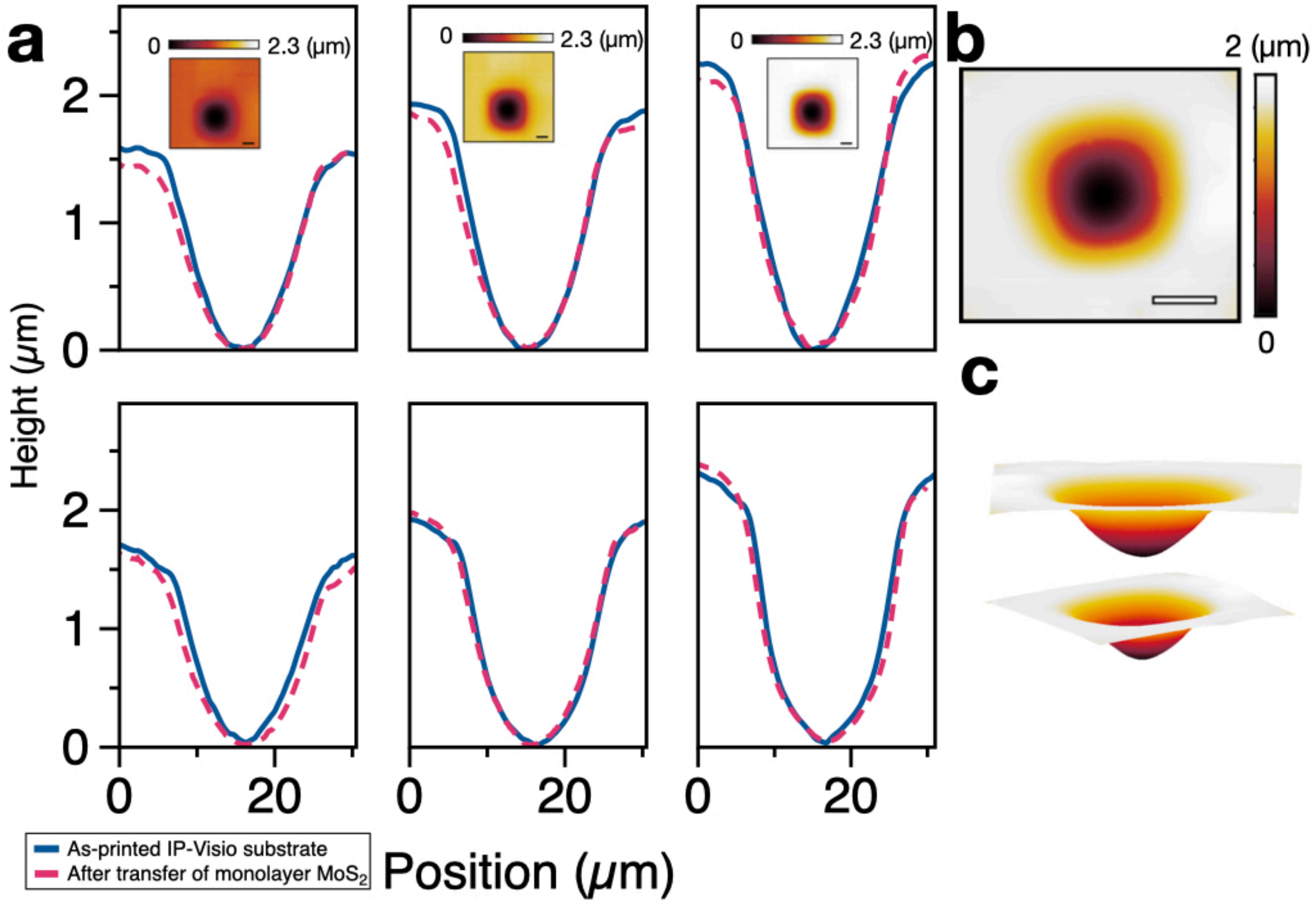

**Supplementary Figure 11 | Conformed monolayer $MoS_2$ on valleys. a,** AFM profiles of valleys before and after transfer of monolayer $MoS_2$. Insets show AFM topographic image of valleys with different aspect ratios. The top-row profiles are taken horizontally across the valley centerline, and the bottom-row profiles are taken vertically across the centerline. Scale bars, 5 µm. **b & c,** Topographic AFM image of monolayer $MoS_2$ conformed to a valley (b) and 3D views (c), which are not to scale, of (b). Scale bar, b, 10 µm.

Optical microscope images of the as-printed IP-Visio substrate, the substrate after contacting it with a monolayer $MoS_2$–PMMA stack, and the substrate following controlled heating at 120 °C in a vacuum oven are shown in **Supplementary Figs. 12a-c**. The thickness of the PMMA used to pick up the monolayers is characterized by AFM to be ~430 nm (**Supplementary Figs. 12d** and **12e**). Given that the glass transition temperature of PMMA is ~105 °C [63] and although PMMA is a much less stiff material than monolayer $MoS_2$, with a thickness of ~600 times the monolayer, the PMMA may induce interactions in its viscous state to conform the monolayer into the underlying surface valleys. In addition, the conforming of monolayer $MoS_2$ on to the patterned substrate may also be enabled by pressure applied during dissolving PMMA with acetone vapor. A complete understanding of the strain realization mechanism requires further studies. It is also noted to dissolve PMMA, submerging the sample in acetone rather than use of acetone vapor and rough handling can both cause fracture of the monolayer at the valley areas.

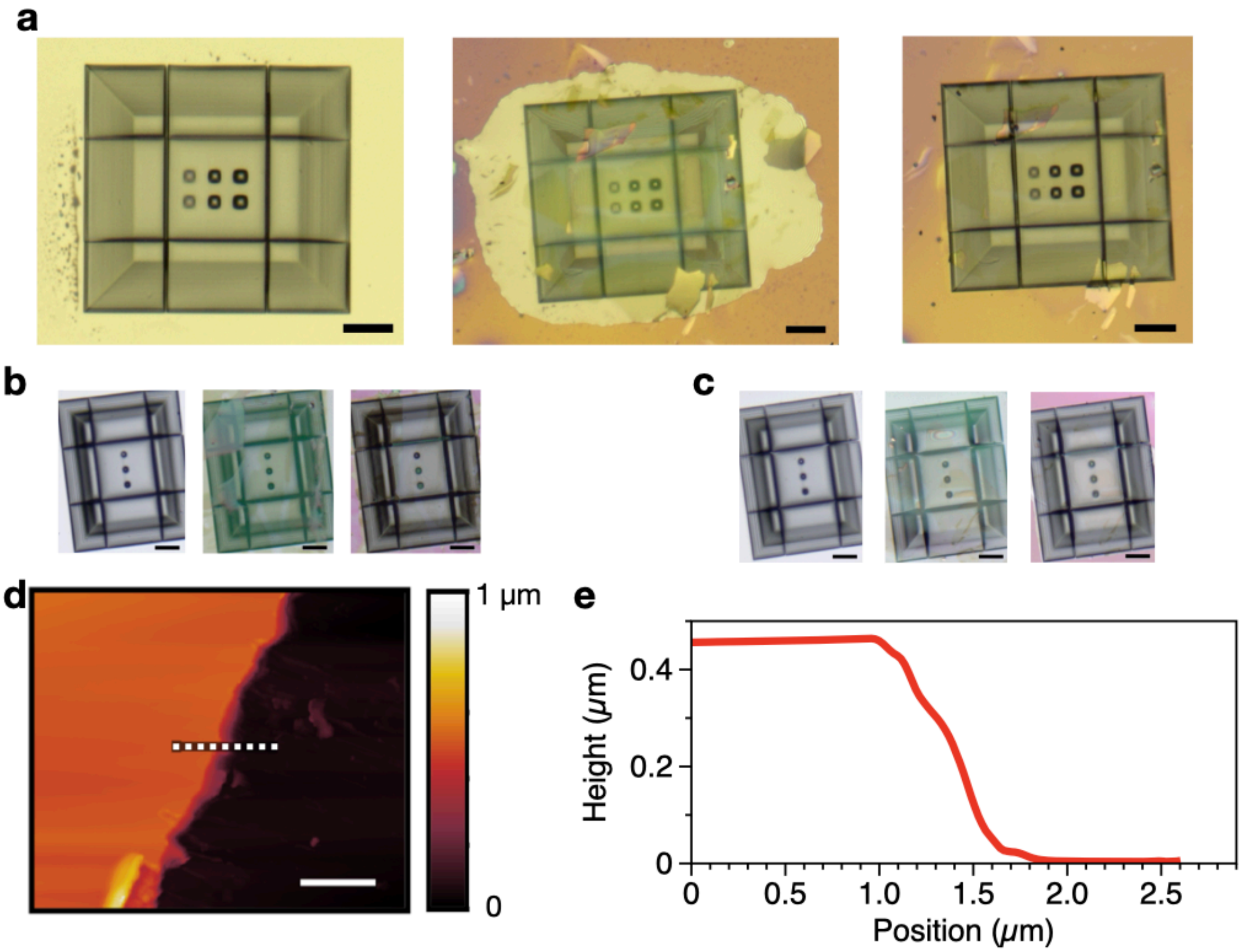

**Supplementary Figure 12 | PMMA transfer of monolayer $MoS_2$. a-c,** Optical microscope image of the as-printed patterned substrate (left image in **a-c**), image after the patterned substrate is brought into contact with the monolayer $MoS_2$–PMMA stack (middle image in **a-c**), and image after controlled heating in a vacuum oven (right image in **a-c**). Scale bars, 100 µm. **d,** Topographic AFM image of a representative

PMMA layer used in 2D material transfer that is spin-coated on $SiO_2$, showing PMMA on the left and bare $SiO_2$ on the right. Scale bar, 2 μm. **e,** Height profile along the white dashed line in (d).

**Supplementary Figure 13** shows optical microscope images of patterned substrates after successful transfer and conforming of monolayer $MoS_2$.

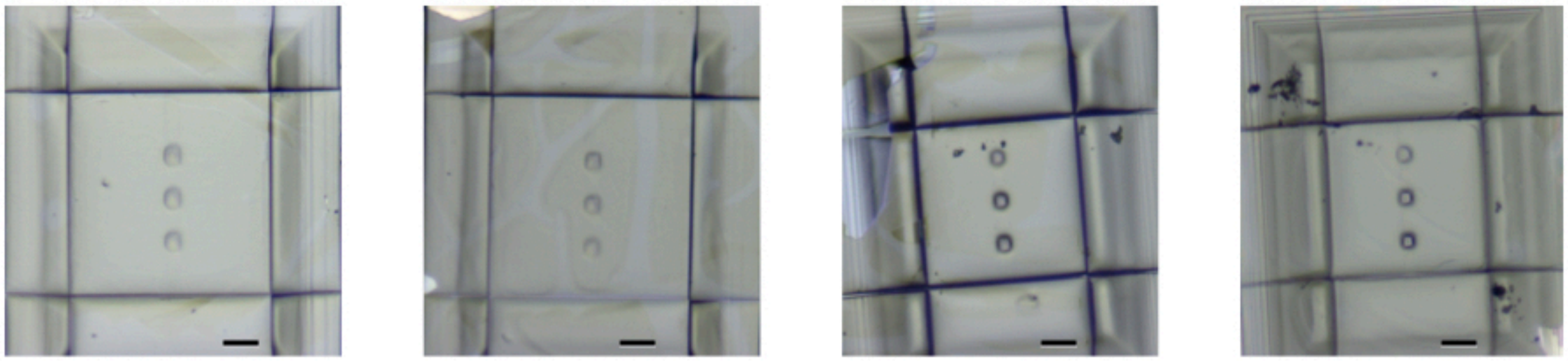

**Supplementary Figure 13 | Optical microscope images of monolayer $MoS_2$ conformed to patterned substrates.** Scale bars, 50 μm.

**Supplementary Note 10. Extraction of biaxial strain from phonon mode shifts for monolayer $MoS_2$**

It should be noted that for encapsulated 2D materials Raman can overestimate strain as revealed by grazing x-ray diffraction measurements capable of directly probing lattice spacing unlike Raman which correlates phonon-mode vibrations to strain[64]. However, for exposed monolayers Raman has been shown to accurately estimate strain within ~0.02% [64], and monolayers investigated herein are not encapsulated.

In monolayer $MoS_2$, the positions of the characteristic in-plane E′ and out-of-plane and $A_1'$ modes are sensitive to biaxial strain ($\varepsilon$) and doping ($n$). They are related to strain and doping through the following relation:

$$\begin{pmatrix} \Delta Pos\ E' \\ \Delta Pos\ A_1' \end{pmatrix} = \begin{pmatrix} -2\gamma_{E'} PosE' & k_{n,E'} \\ -2\gamma_{A_1'} PosA_1' & k_{n,A_1'} \end{pmatrix} \begin{pmatrix} \varepsilon \\ n \end{pmatrix} \qquad (15)$$

Where $\gamma_{E'}$ and $\gamma_{A_1'}$ are the Grünesian parameters, and $k_{n,E'}$ are the charge doping shift $k_{n,A_{1g}}$ coefficients.

The values of the Grünesian parameters and charge doping shift coefficients are extracted from values are extracted from Michail et al.[65], Lloyd et al.[1], and Chakraborty et al.[66] as $\gamma_{E'} = 0.68, \gamma_{A_1'} = 0.21, k_{n,E'} = \frac{0.33}{10^{13}} cm, k_{n,A_1'} = \frac{2.22}{10^{13}} cm.$

**Supplementary Note 11. Monolayer $MoS_2$ conformed to valley with an aspect ratio >0.1 and Raman peak position maps of strain programmed monolayer $MoS_2$**

**Supplementary Fig. 14a** presents the Raman spectra of monolayer $MoS_2$ acquired from the center of a valley with AR > 0.1. **Supplementary Figure 14b** shows the corresponding Raman E′ and $A_1$′ peak positions.

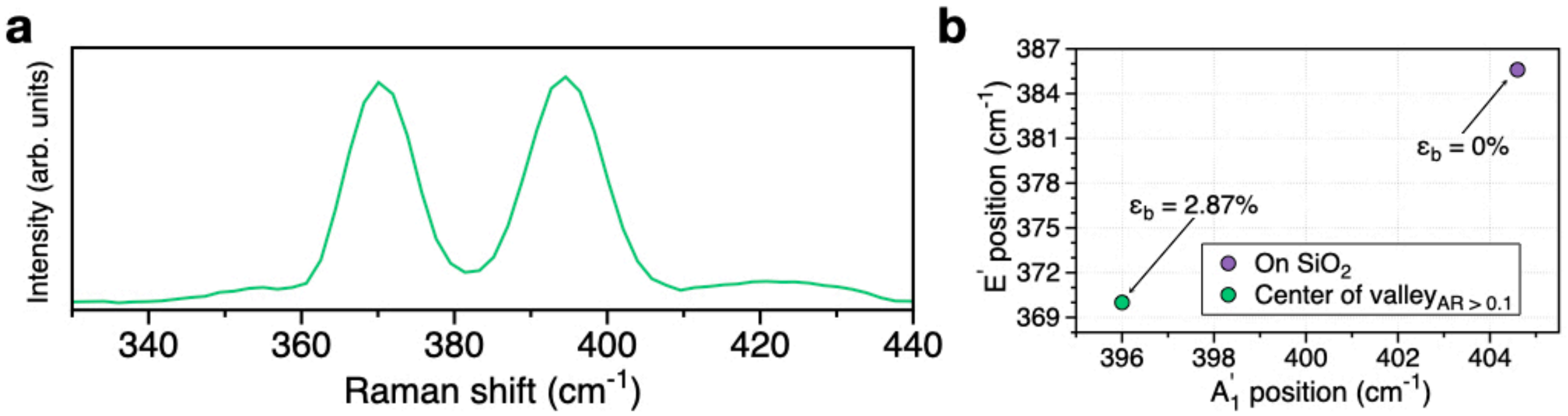

**Supplementary Figure 14 | Strain in monolayer $MoS_2$ conformed to valley with an aspect ratio of 0.15. a,** Raman spectrum of monolayer $MoS_2$ conformed to a valley with an AR > 0.1. **b,** Scatter plots of E′ versus $A_1$′ Raman peak positions for monolayer $MoS_2$, obtained from the sample with Raman spectra shown in (a) and peak positions of monolayer $MoS_2$ on $SiO_2$. In the sample shown in (a), the E′ peak position is ~370 $cm^{-1}$ and the $A_1$′ peak position is ~396 $cm^{-1}$, corresponding to a biaxial strain ($\varepsilon_b$) of 2.87%.

**Supplementary Fig. 15a** and **15b** present the E′ and $A_1$′ peak position maps, respectively, which were used to generate the $\varepsilon_b$ map of monolayer $MoS_2$ conformed to a valley with an AR ≈ 0.089 (Fig. **3b** in main text). **Supplementary Fig. 15c** and **15d** present the E′ and $A_1$′ peak position maps, respectively, which were used to generate the $\varepsilon_b$ map of monolayer $MoS_2$ conformed to a valleys with AR ≈ 0.078 (top valley) and AR ≈ 0.088 (bottom valley) (Fig. **3d** in main text).

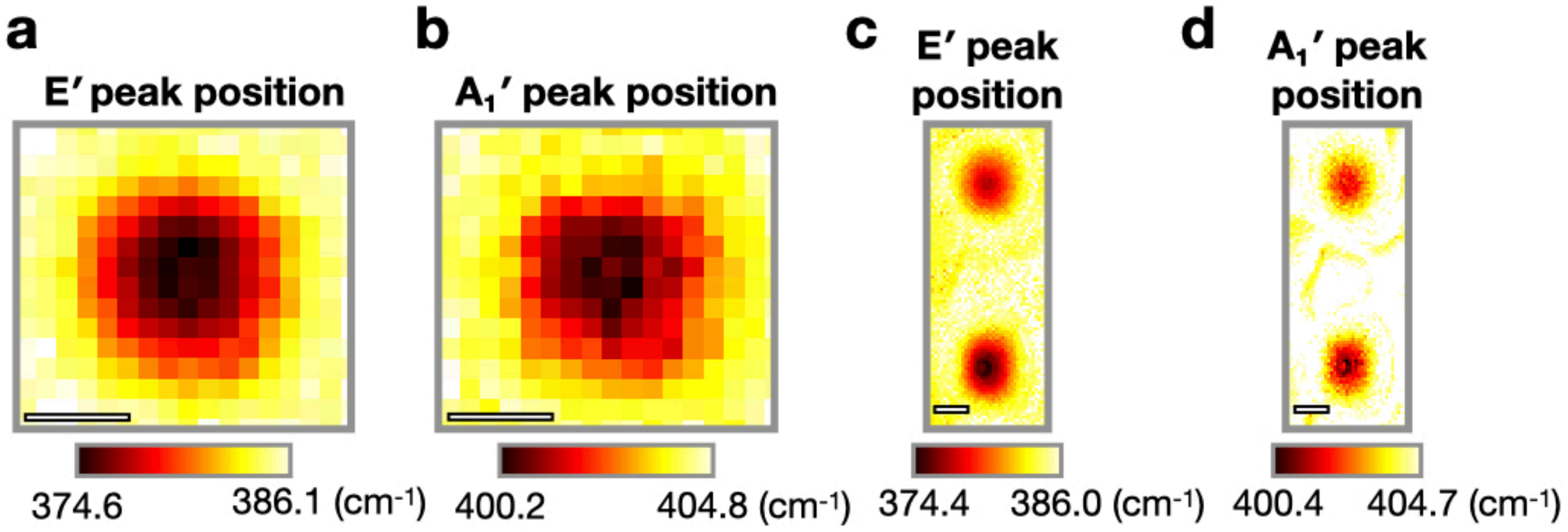

**Supplementary Figure 15 | Strain distributions in strain-engineered monolayer $MoS_2$. a & b,** Scanning Raman maps showing spatial distributions of E′ (a) and $A_1$′ (b) peak positions across monolayer $MoS_2$ conformed to a valley with an aspect ratio (AR) ≈ 0.089. The maps were acquired with 2 μm steps in both the x and y directions. Scale bars, 10 μm. **c & d,** Scanning Raman maps showing spatial distributions of E′ (c) and $A_1$′ (d) peak positions of monolayer $MoS_2$ conformed to valleys with AR ≈ 0.079 (top valley) and 0.088 (bottom valley) with distance between the centers of neighboring valleys as 3L, with L representing the valley period. The maps were acquired with 1 μm steps in both the x and y directions. Scale bars, 10 μm.

**Supplementary Note 12. PL measurements of strain engineered monolayer $MoS_2$**

**Supplementary Figure 16** presents the optical microscope image of a patterned substrate containing two valleys, along with Raman spectra, Raman peak position plots, and $\varepsilon_b$ plots of monolayer $MoS_2$ conformed to this patterned substrate, whose PL spectra are shown in Fig. **4a** of the main text, as well as AFM images of the valleys and the corresponding valley height profiles.

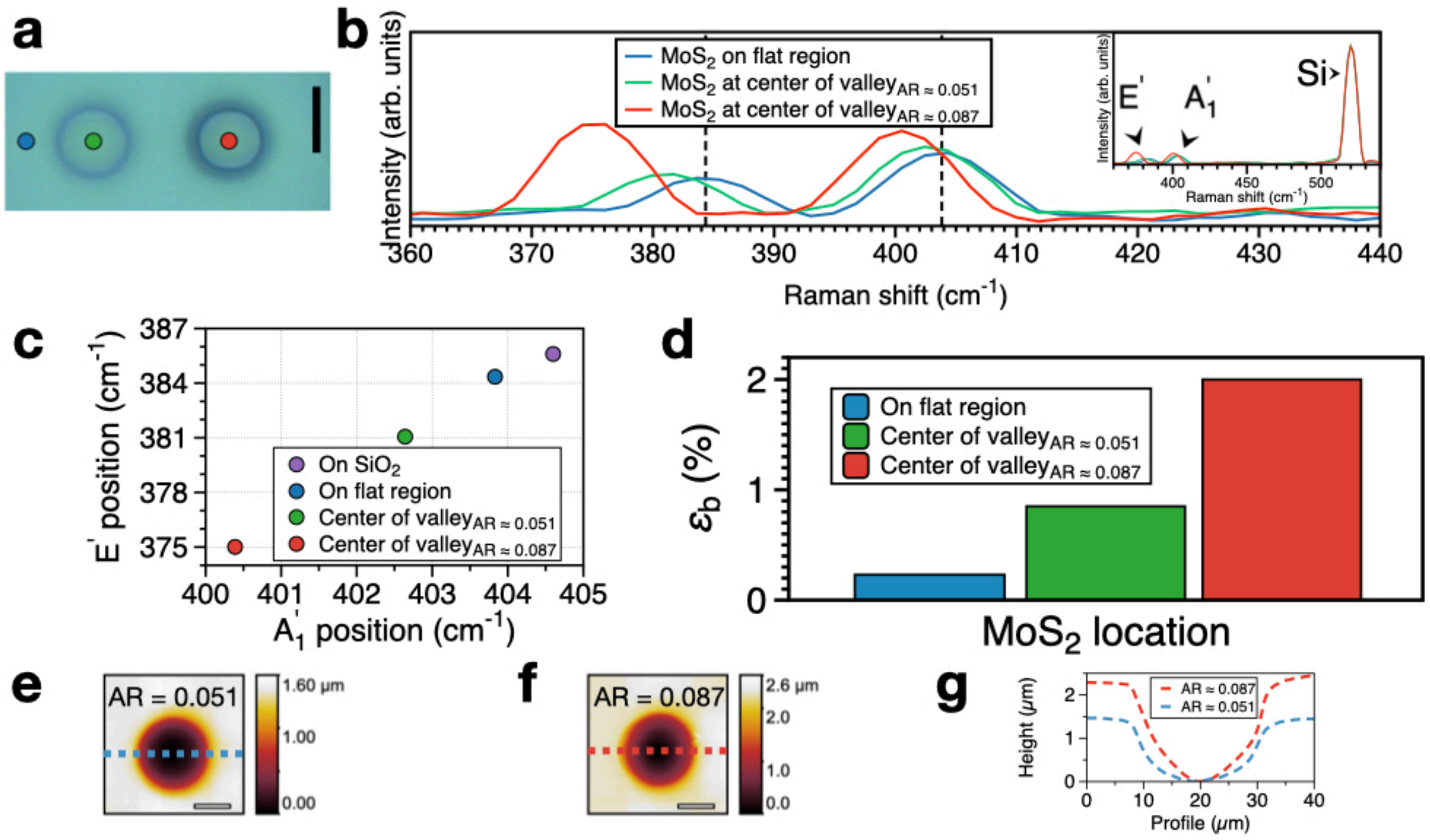

**Supplementary Figure 16 | Strain engineered monolayer $MoS_2$. a,** Optical microscope image of a substrate with valleys of varying aspect ratios (ARs). Colored dots (green and red) indicate valley centers with AR ≈ 0.051, and 0.087, respectively; the blue dot indicates a flat region. The distance between the centers of neighboring valleys is 2L, with L representing the valley period. Scale bar, 20 µm. **b,** Raman spectra collected from the color-coded locations in (a). All spectra are normalized to the Si peak intensity. Vertical dashed lines indicate the E′ and $A_1$′ peak positions of monolayer $MoS_2$ on the flat region, as determined from Lorentzian fits. Inset: wide-range spectrum displaying the Si substrate peak (~520.5 $cm^{-1}$). **c,** Scatter plots of E′ versus $A_1$′ Raman peak positions for monolayer $MoS_2$, obtained from the sample with Raman spectra shown in (b). **d,** Biaxial strain ($\varepsilon_b$) in $MoS_2$ extracted from Raman peak positions in (c). **e,** Topographic atomic force microscope (AFM) image of a 2PL-printed substrate with a valley of an aspect ratio (AR) of ≈ 0.051. Scale bar, 10 µm. **f,** Topographic AFM image of a 2PL-printed substrate with a valley of an aspect ratio (AR) of ≈ 0.087. Scale bar, 10 µm. **g,** Line profiles from topographic AFM showing height profiles of valleys with AR ≈ 0.051 (dashed blue) and AR ≈ 0.087 (dashed red).

We also present additional PL spectra demonstrating the gradual redshift of the monolayer $MoS_2$ PL peak with increasing strain. **Supplementary Figure 17a** shows PL spectra acquired along the green arrow indicated in the $\varepsilon_b$ map in **Supplementary Figure 17b**. The top spectrum in **Supplementary Figure 17a** approximately corresponds to the starting point of the arrow, while the bottom spectrum corresponds to its endpoint. Owing to uncertainties in the laser position within the PL setup, the exact start and end points may lie roughly anywhere within the green shaded region shown in **Supplementary Figure 17b**. Across these spectra, the A-exciton peak redshifts from ~1.82 eV in a flat, low strained region to ~1.72 eV near the center of the valley with AR ≈ 0.079. We further note that the PL data presented in **Supplementary Figure 17a** were acquired using a different experimental setup from the PL measurements shown in Fig. **4a** of the main text, which were collected from the sample presented in **Supplementary Figure 16**.

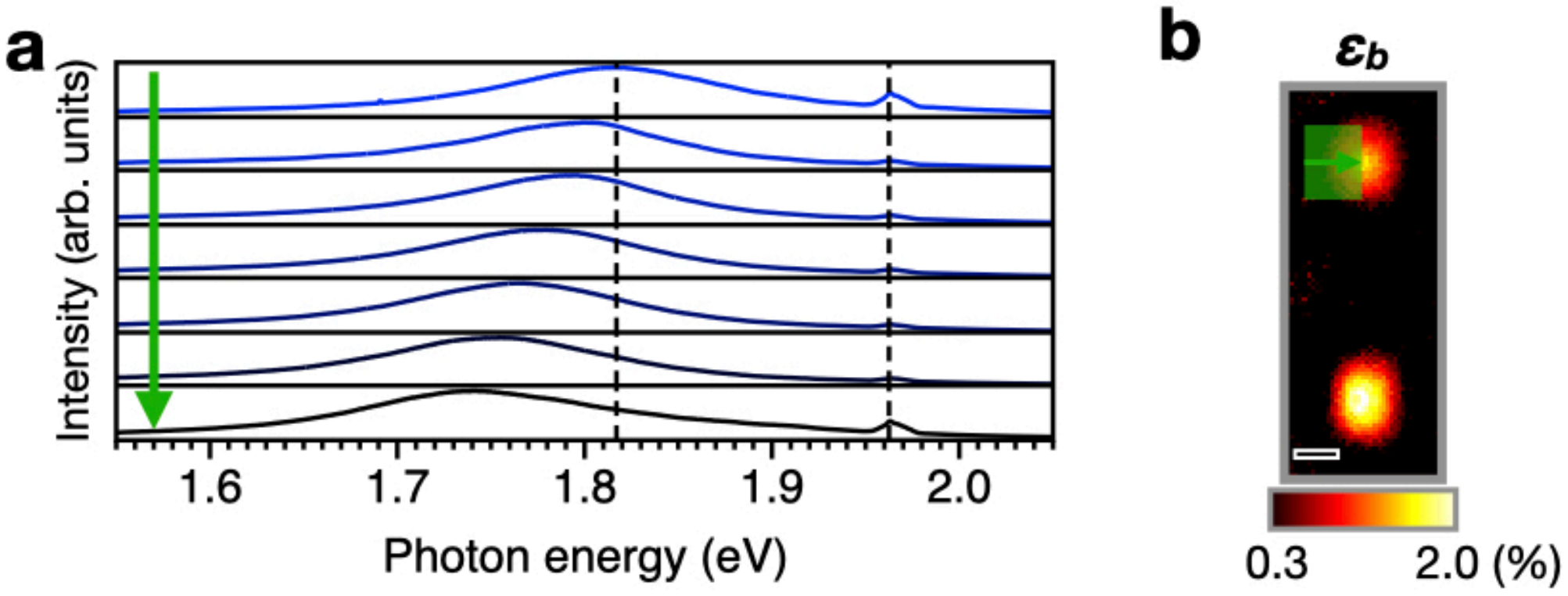

**Supplementary Figure 17 | a,** Photoluminescence (PL) spectra of monolayer $MoS_2$ collected along the green arrow in (b). The top spectrum roughly corresponds to the starting point of the arrow, and the bottom spectrum roughly corresponds to its end, with intermediate spectra sampled along the arrow path. The PL peak at ~1.96 eV originates from the IP-Visio substrate and as expected its position does not shift across the valley. As the working distance is adjusted for each region of focus, all PL spectra are normalized to the intensity of the $MoS_2$ PL peak on the flat region (top panel). Vertical dashed lines indicate the A exciton position of monolayer $MoS_2$ on the flat region and the position of the IP-Visio PL peak in the same spectrum (top panel), as determined from Lorentzian fits. **b,** The $\varepsilon_b$ map of monolayer $MoS_2$ conformed to valleys with AR ≈ 0.079 (top valley) and AR ≈ 0.088 (bottom valley) with distance between the centers of neighboring valleys as 3L, with L representing the valley period, with strain values extracted from the E′ and $A_1$′ peak position maps presented in **Supplementary Figures 15c** and **15d**. The maps were acquired with 1 µm steps in both the x and y directions. The color scale range was limited to 0.3–2.0% (from −0.1 to 2.1%) strain to highlight high-magnitude spatial variations. This $\varepsilon_b$ map of monolayer $MoS_2$ is also presented in Fig. **3d** of the main text without the green shaded region and green arrow. Scale bar, 10 µm.

We note that the PL spectrum presented for the monolayer $MoS_2$ on a flat region in Fig. **4a** was collected from a location which exhibited $\varepsilon_b$ ~0.08%, which differs from the flat region whose Raman spectra is presented in **Supplementary Figure 16**, which exhibits $\varepsilon_b$ ~0.23%. $\varepsilon_b$ values of 0.08% and 0.23% are both are in the range of -0.1% to 0.3% of $\varepsilon_b$ observed in this study for monolayer $MoS_2$ on flat IP-Visio substrates.

**Supplementary Note 13. Extraction of band gap value from C-AFM measurements**

To perform C-AFM measurements, which require a conductive path between the sample and the AFM tip, we deposited 2.5 nm of Cr followed by 50 nm of Au onto a 2PL-fabricated patterned substrate prior to transfer and conforming the monolayer $MoS_2$. Monolayer $MoS_2$ strongly adheres to Au via covalent-like quasi-bonding (adhesion energy of ~0.6 J $m^{-2}$), a property commonly used to exfoliate large-area monolayers[67–70].

The presence of strain was confirmed using Raman spectroscopy (**Supplementary Figs**. **18a–c**). **Supplementary Fig. 18a** presents the Raman spectra of monolayer $MoS_2$ acquired from a flat region of an Au-coated patterned substrate, as well as from the centers of valleys with ARs ≈ 0.044 and 0.070. **Supplementary Figures 18b** and **16c** show the corresponding Raman E′ and $A_1'$ peak positions (b) and the extracted $\varepsilon_b$ (c). A schematic of the C-AFM setup is shown in **Supplementary Fig**. **18d**, along with an FE-SEM image of the tip used which is shown in **Supplementary Fig**. **18e**. Current-voltage (*I-V*) sweeps obtained from various regions of the sample are presented in **Supplementary Fig**. **18f**.

We also performed exclusion experiments to examine the role of tip-induced artifacts in our I–V measurements. In all C-AFM experiments, the tip set-point force was maintained at 10 nN. **Supplementary Fig. 18g** shows I–V curves acquired from a flat region of monolayer $MoS_2$ on an Au-coated patterned substrate for set-points ranging from 10 to 500 nN. The I–V characteristics remain mostly unchanged up to ~100 nN, while deviations become apparent at higher forces, indicating the onset of tip-induced modulation. Since no measurable influence is observed below ~100 nN, we conclude that our operating set-point of 10 nN effectively minimizes tip-induced effects.

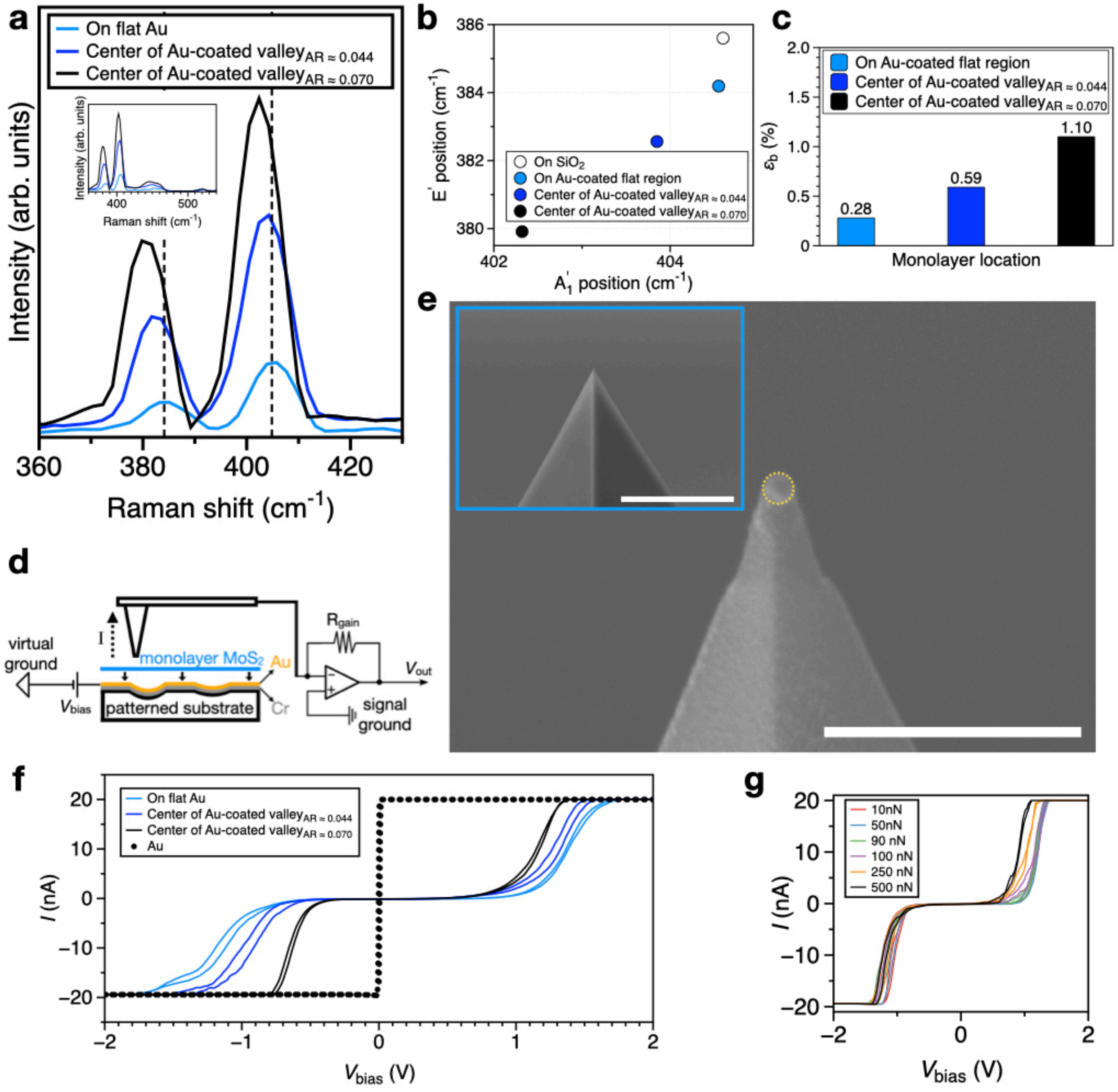

**Supplementary Figure 18 | Conductive atomic force microscopy (C-AFM) of monolayer $MoS_2$ acquired using across different topographic regions. a,** Raman spectra collected from monolayer $MoS_2$ conformed to an Au-coated patterned substrate, from monolayer $MoS_2$ on a flat region, and from monolayer $MoS_2$ at valley centers with aspect ratios (ARs) ≈ 0.044 and 0.070. Vertical dashed lines mark the E′ and $A_1'$ peak positions of monolayer $MoS_2$ on the flat region, extracted using Lorentzian fits. Inset: wide-range Raman spectrum with the Si substrate peak (~520.5 cm$^{-1}$) suppressed. **b,** Scatter plots of E′ versus $A_1'$ Raman peak positions for monolayer $MoS_2$, obtained from the sample with Raman spectra shown in (a). $SiO_2$ peak positions were extracted separately from a different sample. **c,** $\varepsilon_b$ in $MoS_2$ extracted from Raman peak positions in (b). **d,** Schematic of the sample prepared for C-AFM measurements and C-AFM setup. A

patterned substrate (IP-Visio) is coated with 2.5 nm of Cr and 50 nm of Au, followed by the transfer of monolayer $MoS_2$ conforming to the substrate topography. **e,** FE-SEM image of the AFM tip used in the C-AFM measurements. The dashed yellow circle highlights the region used to determine the tip radius. Inset: zoomed-out image. Scale bars, 500 nm. **f,** C-AFM *I-V* sweeps (±2V) collected from an exposed region of the Au-coated substrate, monolayer $MoS_2$ on a flat region and monolayer $MoS_2$ at valley centers with ARs ≈ 0.044 and 0.070. **g,** C-AFM *I–V* sweeps (±2 V) collected from monolayer $MoS_2$ on a flat region of the Au-coated substrate under increasing tip set points.

The spring constant of the tip was calibrated by using the Thermal method[71]. Each reported I-V curve corresponds to the average of five individual I–V spectra (**Supplementary Fig. 19a**). The averaged current–voltage dataset was smoothed using a Savitzky–Golay polynomial filter, which preserves local slope information while suppressing noise (**Supplementary Fig. 19b**). Current plateaus near the instrumental saturation limits (±20 nA) were identified and excluded (**Supplementary Fig. 19c**). Exclusion of these regions ensures that the derivative reflects intrinsic material transport rather than instrument limitations. The differential conductance ($dI/dV$) was computed numerically as the first derivative of the smoothed current with respect to voltage:

$$\frac{dI}{dV} = \frac{\Delta I}{\Delta V} \tag{16}$$

A second smoothing step was applied to reduce derivative noise (**Supplementary Fig. 19d**). The resulting $dI/dV$ spectrum serves as a local probe of the density of electronic states. To identify the conduction and valence band edges, the $dI/dV$ curve was divided into two regions: (1) Negative bias ($V < 0$): corresponds to electron tunneling from the tip into occupied sample states (valence band). (2) Positive bias ($V > 0$): corresponds to tunneling into unoccupied states (conduction band). Within each region, the maximum in conductance was located, and only the monotonic region up to that maximum was retained to isolate the quasi-linear portions of the conductance onset (**Supplementary Fig. 19e**). Each side of the $dI/dV$ spectrum (negative and positive bias) was linearly fitted within the range from 10% to 100% of the maximum conductance value. This fitting captures the most linear region near the onset of significant tunneling, representing the transition from the bandgap to band conduction. Extrapolating each linear fit to zero conductance provided the valence band maximum (VBM) and conduction band minimum (CBM) edges, respectively (**Supplementary Fig. 19f**): extracted band gap (EBG) = CBM − VBM. This method assumes that the onset of measurable conductance in a C-AFM I–V curve reflects the transition of the tip–sample junction from the bandgap region into conduction through available electronic states. The linear extrapolation of d$I/dV$ avoids fitting the highly nonlinear regions near saturation.

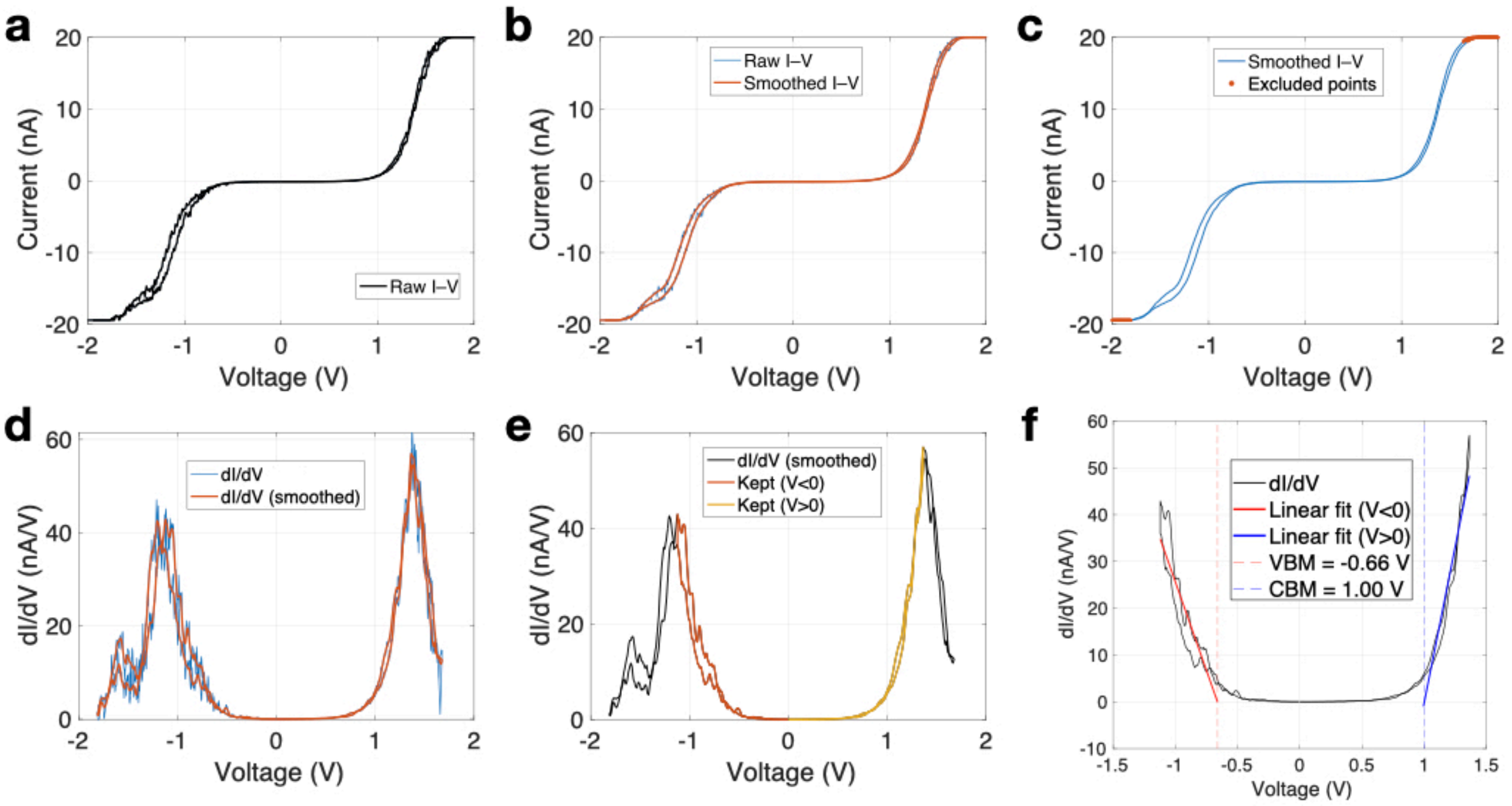

**Supplementary Figure 19 | Extraction of band gap from conductive atomic foce microscopy measurements. a**, Raw current–voltage (I–V) curve. **b**, Smoothed I–V curve. **c**, Smoothed I–V curve with excluded data points highlighted in orange. **d**, Differential conductance and its smoothed counterpart. **e**, Segments of the differential conductance curve retained for fitting to determine the band gap. **f**, Two linear fits used to identify the valence band maximum (VBM) and conduction band minimum (CBM) edges, where the energy difference between the intersection points with the x-axis define the extracted band gap.

## Supplementary Note 14. Electronic band structure under biaxial strain

**Supplementary Figure 20a** presents density functional theory (DFT)–calculated electronic band structures of monolayer $MoS_2$ under increasing biaxial strain up to $\varepsilon_b$ = 5%. **Supplementary Figure 20b** presents the extracted monolayer $MoS_2$ band gap versus $\varepsilon_b$.

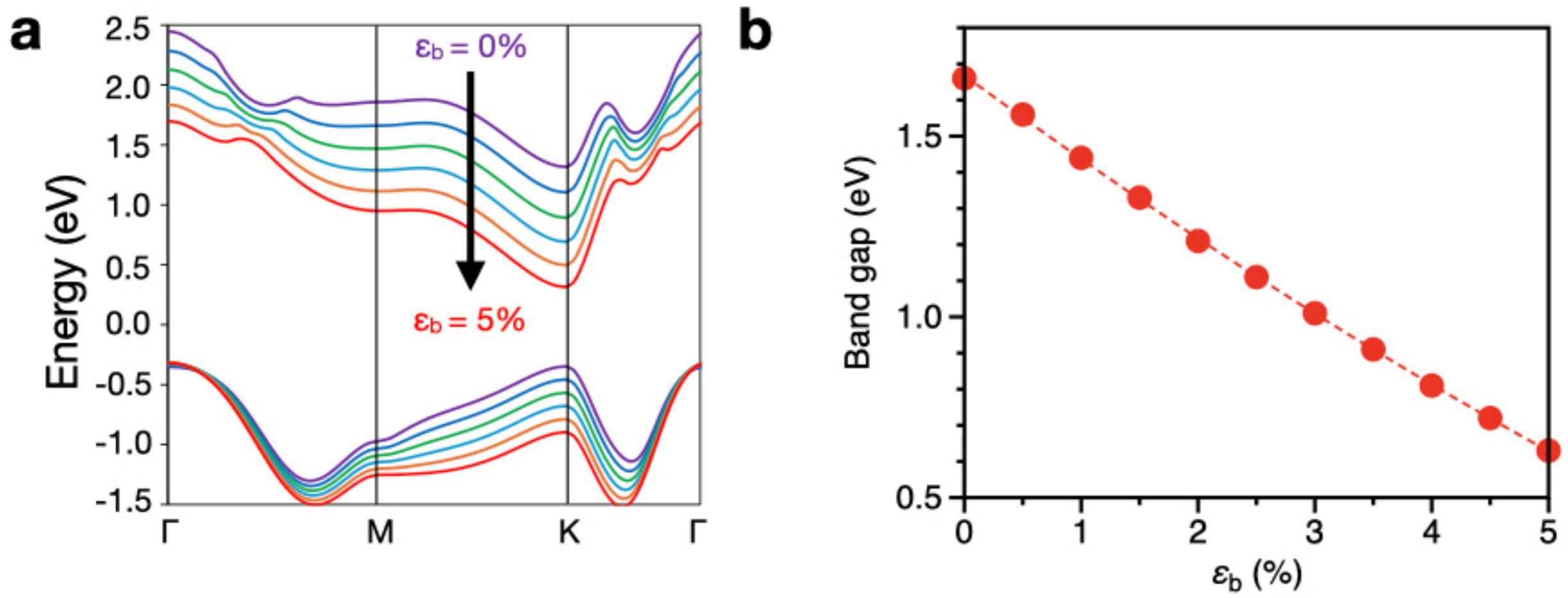

**Supplementary Figure 20 | DFT predictions of electronic band structure. a,** Electronic band structure of monolayer $MoS_2$ under biaxial strain ($\varepsilon_b$). **b,** Extracted band gap of monolayer $MoS_2$ versus $\varepsilon_b$. The dashed line is a polynomial fit.

## Supplementary Note 15. Long-term retention of strain

**Supplementary Figure 21** presents Raman analysis of strained monolayer $MoS_2$ over extended time periods, demonstrating the long-term retention of strain of monolayer $MoS_2$ conformed to valleys with aspect ratios ≈ 0.040, 0.075, and 0.085.

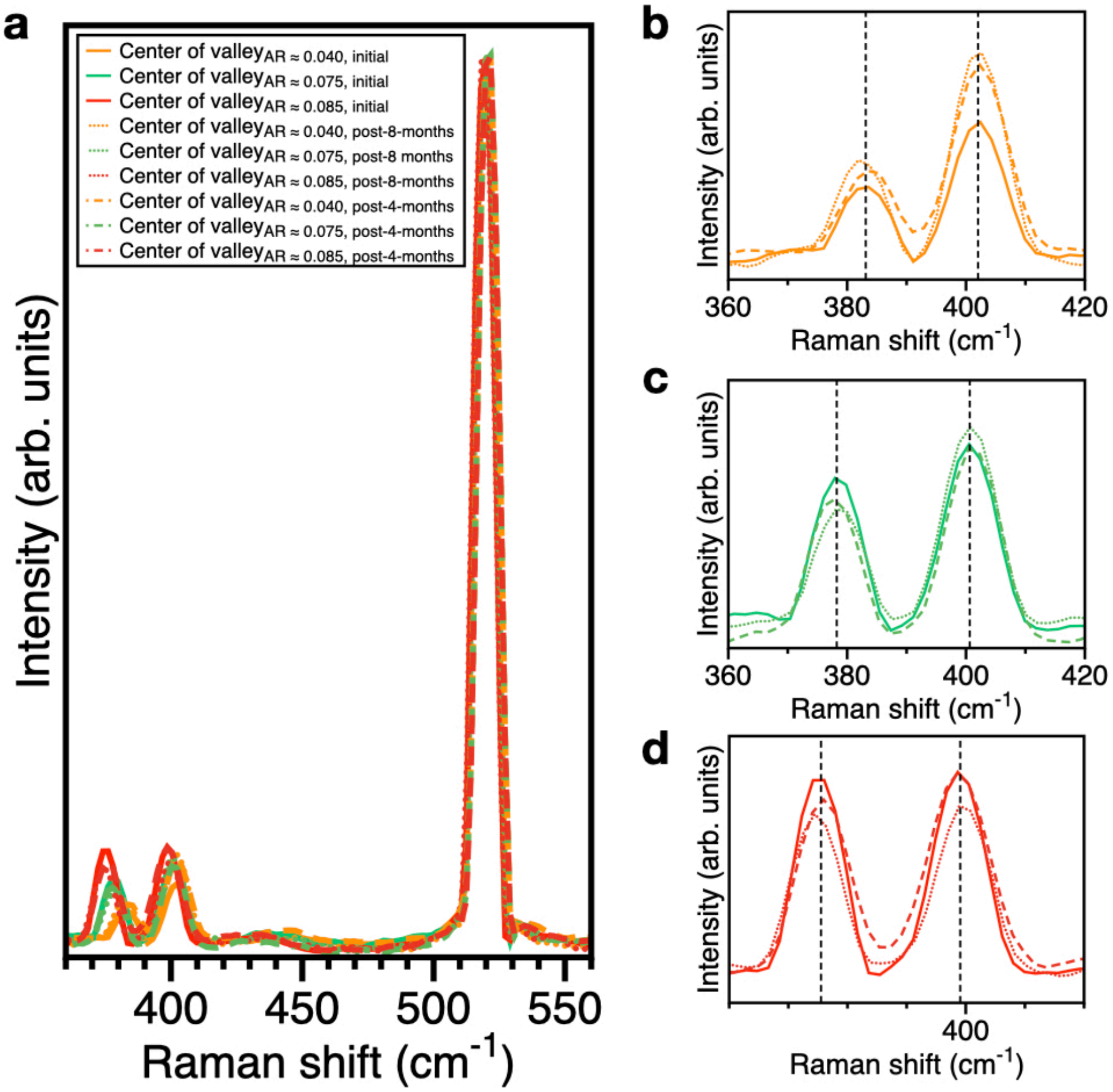

**Supplementary Figure 21 | Long-term retention of strain. a,** Wide-range Raman spectra of monolayer $MoS_2$ collected from the centers of valleys with aspect ratios ≈ 0.040, 0.075, and 0.085 after monolayer conformation, after 4 months, and after 8 months. Minor variations in the relative intensities and positions of the $MoS_2$ and Si peaks across time points arise from changes in the Raman setup's working distance and slight shifts in lateral focus. All spectra are normalized to the Si peak. **b–d,** Tight-range (360–420 $cm^{-1}$) Raman spectra from (a), grouped by aspect ratio: (b) ≈ 0.040, (c) ≈ 0.075, and (d) ≈ 0.085. Vertical dashed lines mark the initial E′ and $A_1$′ peak positions of monolayer $MoS_2$ at each valley center, as determined by Lorentzian fitting.

## Supplementary Note 16. Uniaxial 2D strain engineering

2PL (IP-Visio resin) was used to fabricate patterned substrates containing trench-like structures. The structural fidelity of fabricated a trench with $AR_x \approx 0.081$ and $AR_y \approx 0.027$ was characterized via AFM. In **Supplementary Figure 22a** an AFM topographic image, from a region of the trench that does not include its semicircular ends, is shown. In **Supplementary Figure 22b** the height profiles of the blue and red dashed lines in **Supplementary Figure 22a** are provided.

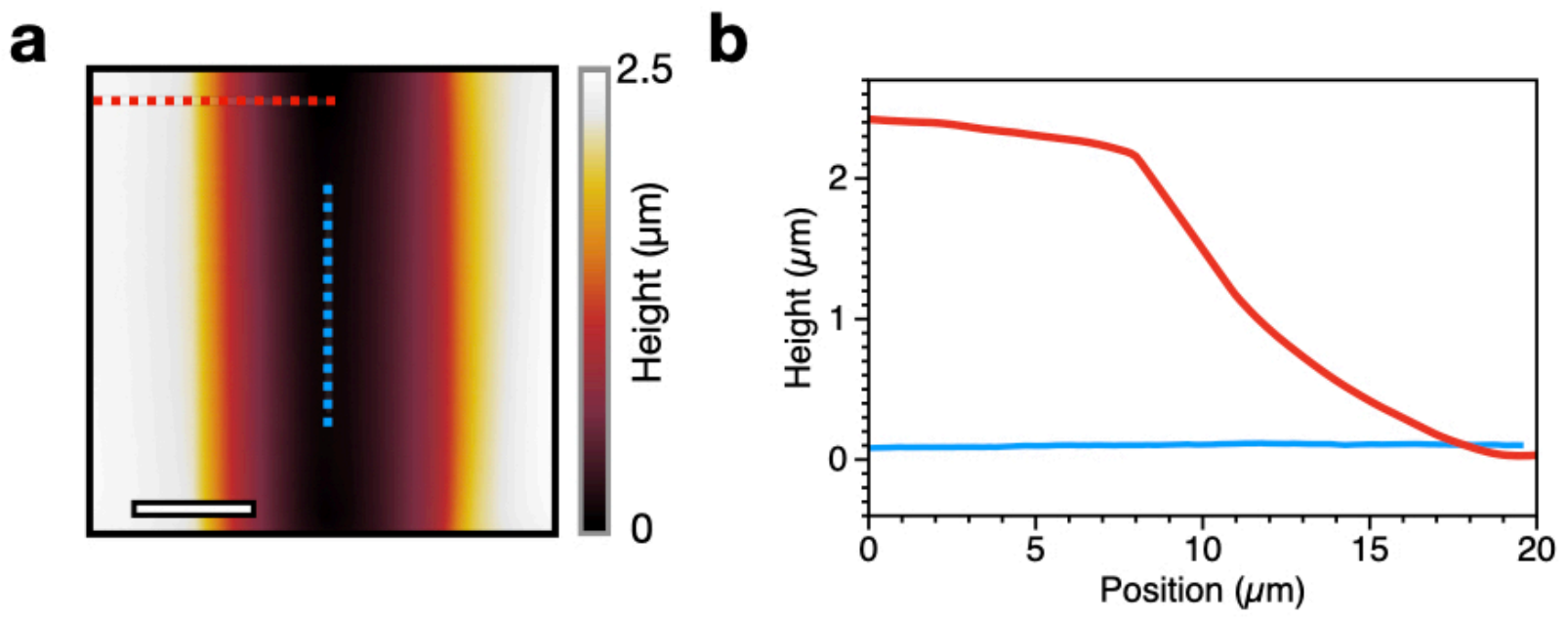

**Supplementary Figure 22 | 2PL-fabricated trench. a,** Topographic atomic force microscope (AFM) image of a 2PL-printed substrate with a trench (image is limited to central area of the trench, excluding the circular edges) with of $AR_x \approx 0.081$. Scale bar, 10 μm. **b,** Line profiles from topographic AFM (blue and red dashed lines) showing the half of the trench profile along x (red) and centreline profile along y (blue).

Finite element analysis predictions of the top-down view of the spatially resolved uniaxial strain distributions ($\varepsilon_{xx}$ and $\varepsilon_{yy}$) in monolayer $MoS_2$ conformed to a trench with $AR_x = 0.15$ and $AR_y = 0.05$ are provided in **Supplementary Fig. 23**.

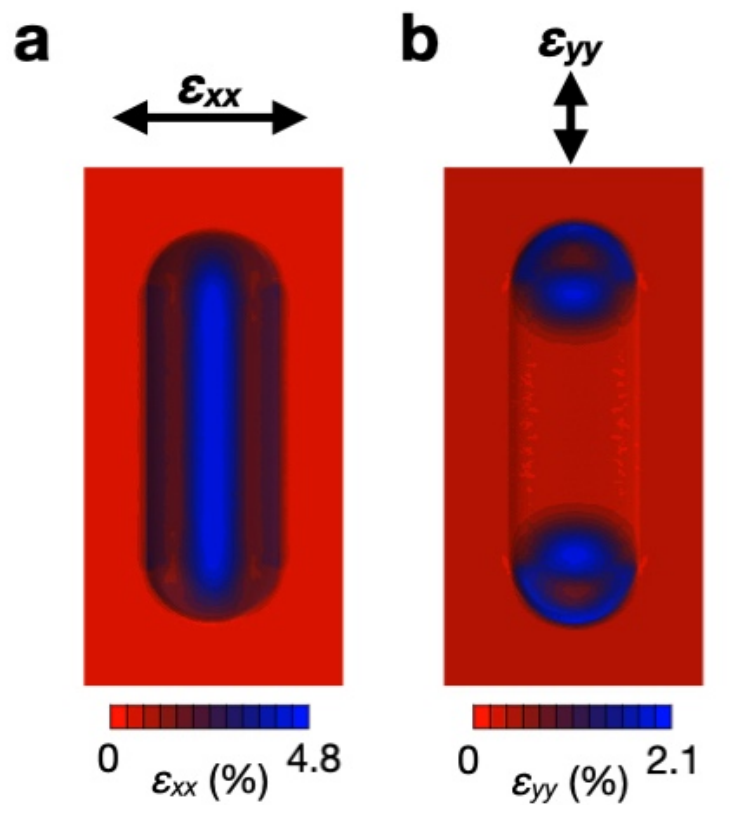

**Supplementary Figure 23 | Simulated uniaxial strain distributions in monolayer $MoS_2$. a & b,** Finite element analysis predictions of the spatially resolved uniaxial strain components $\varepsilon_{xx}$ (a) and $\varepsilon_{yy}$ (b) shown in top-down view for monolayer $MoS_2$ conformed to a trench with aspect ratios $AR_x = 0.15$ and $AR_y = 0.05$.

Wide-range Raman spectra of data presented in Fig. **6g-i** from the main text are presented below **Supplementary Fig. 24**.

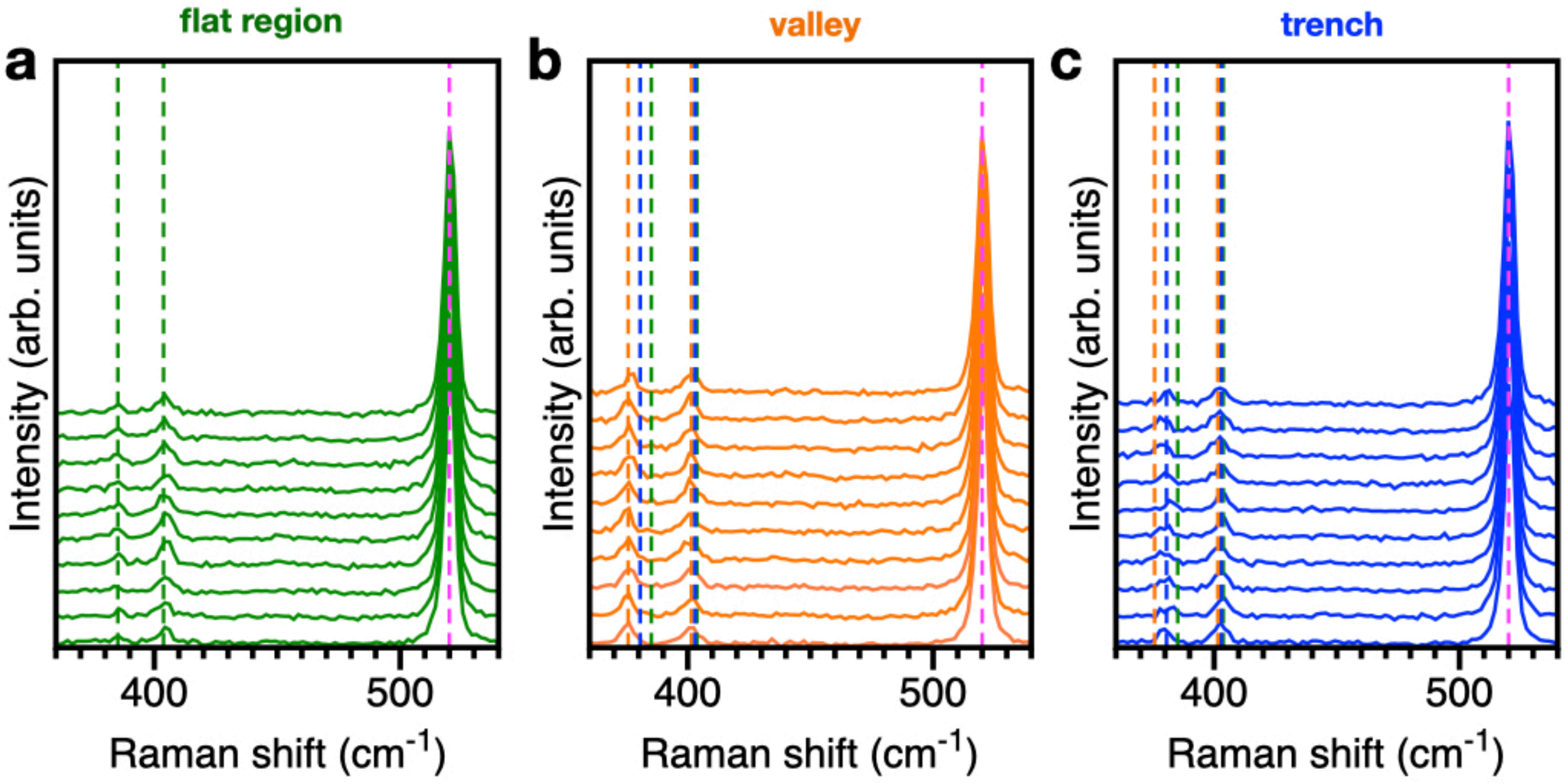

**Supplementary Figure 24 | Raman spectra of monolayer $MoS_2$ on a substrate with a valley and a trench. a-c,** Wide-range Raman spectra of monolayer $MoS_2$ collected from regions conforming to a valley and a trench shown in Fig. **6f** in the main text, with spectra from the flat region (a), the valley center (b), and the trench centerline along the y-direction (c). The green, orange, and blue vertical dashed lines indicate the average E′ and $A_1'$ peak positions, as determined from Lorentzian fits, for monolayer $MoS_2$ on the flat region, at the center of a valley (AR ≈ 0.085), and along the trench y-centerline ($AR_x \approx 0.080$, $AR_y \approx 0.027$), respectively. The pink vertical dashed line indicates the Si peak position (~520.5 $cm^{-1}$).

**Supplementary Note 17. Raw and smoothed Raman spectra of monolayer $MoS_2$**

**Supplementary Figure 25** demonstrates that for unstrained and biaxially strained monolayer $MoS_2$, smoothing does not have an impact on the E′ and $A_1$′ peak locations, which are the only parameters we extract for biaxial strain calculations. In addition, while we discuss the relative intensity of the peaks, we do not report values and only report trends, which are maintained in the presence of smoothing. We also note that the smoothing was carried out using the Smooth function in the Raman manufacturer's (Renishaw) WiRE software.

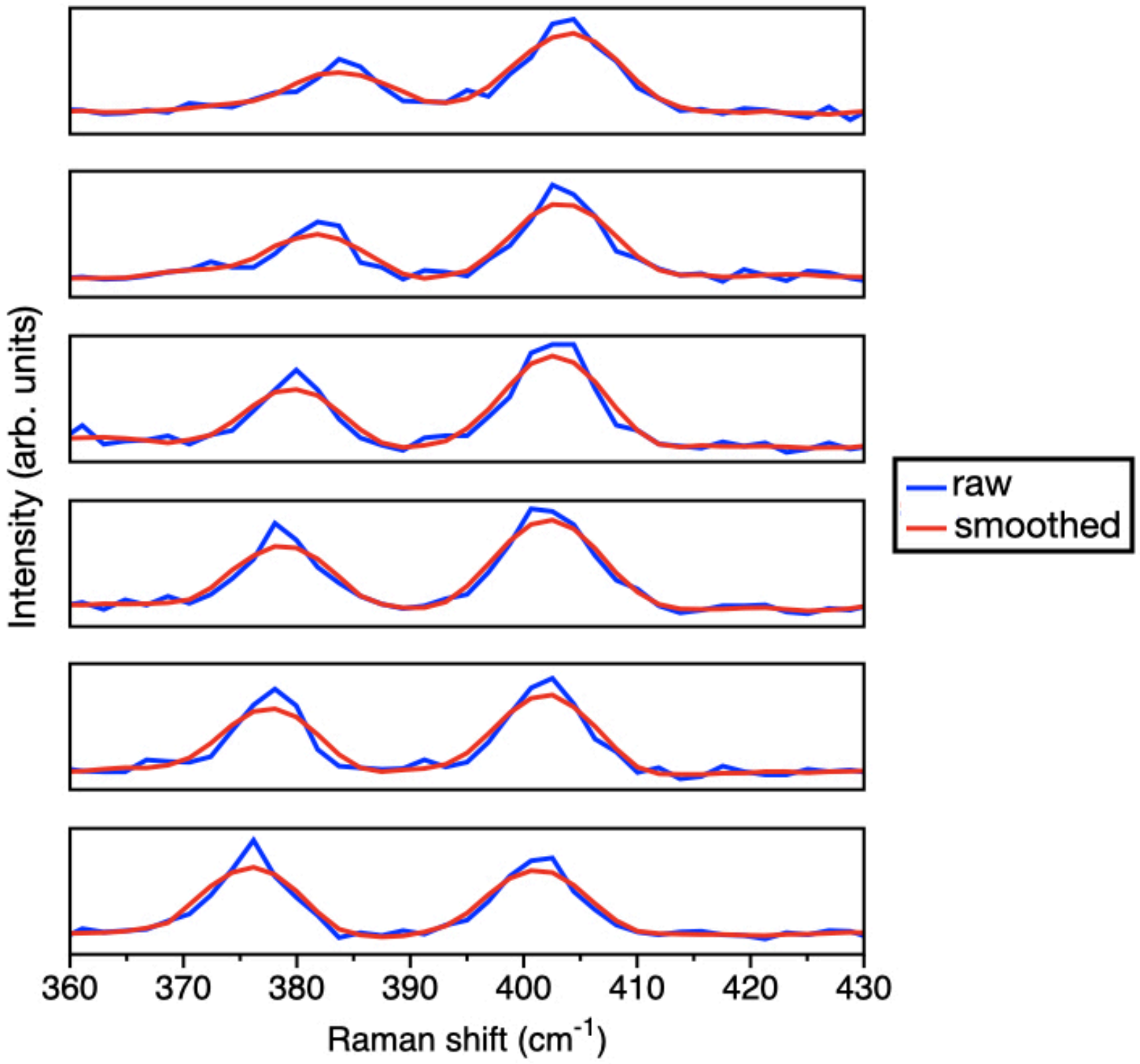

**Supplementary Figure 25 | Raw and smoothed Raman spectra of monolayer $MoS_2$.** Raw and smoothed raman spectra of monolayer $MoS_2$ under increasing biaxial strain. The top spectrum corresponds to monolayer $MoS_2$ on a flat region of a 2PL-fabricated IP-Visio substrate and the bottom spectrum corresponds to monolayer $MoS_2$ at the center of a valley, with intermediate spectra corresponding increasing biaxial strain from top to bottom.

**Supplementary Note 18. Bilayer heterostructure**

**Supplementary Figure 26a** shows an optical microscope image of a prepared $WS_2$–$MoS_2$ heterostructure and **Supplementary Figure 26b** shows an optical microscope image of this bilayer heterostructure after transfer and conforming to a substrate containing a valley.

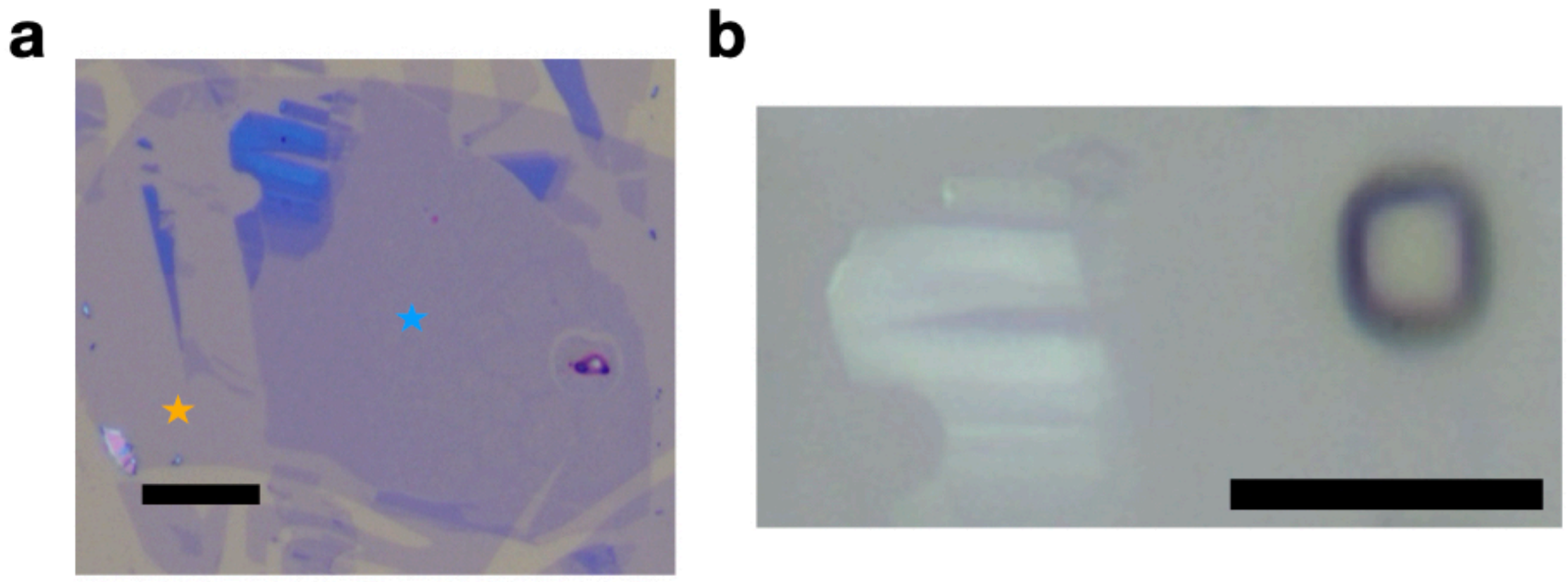

**Supplementary Figure 26 | Bilayer $WS_2$–$MoS_2$ heterostructure. a,** Optical microscope image of a prepared bilayer $WS_2$–$MoS_2$ heterostructure on $SiO_2$. This stack consists of individual monolayers that were initially exfoliated on Au, and subsequently transferred onto $SiO_2$. The orange star marks the bottom monolayer $MoS_2$, and the blue star marks the top monolayer $WS_2$. Scale bar, 50 µm. **b,** Optical microscope image of the prepared bilayer heterostructure after transfer to a patterned substrate. Scale bar, 50 µm

**Supplementary Figure 27** presents the calculated interlayer sliding energy of the bilayer $WS_2$–$MoS_2$ heterostructure along both zigzag and armchair directions.

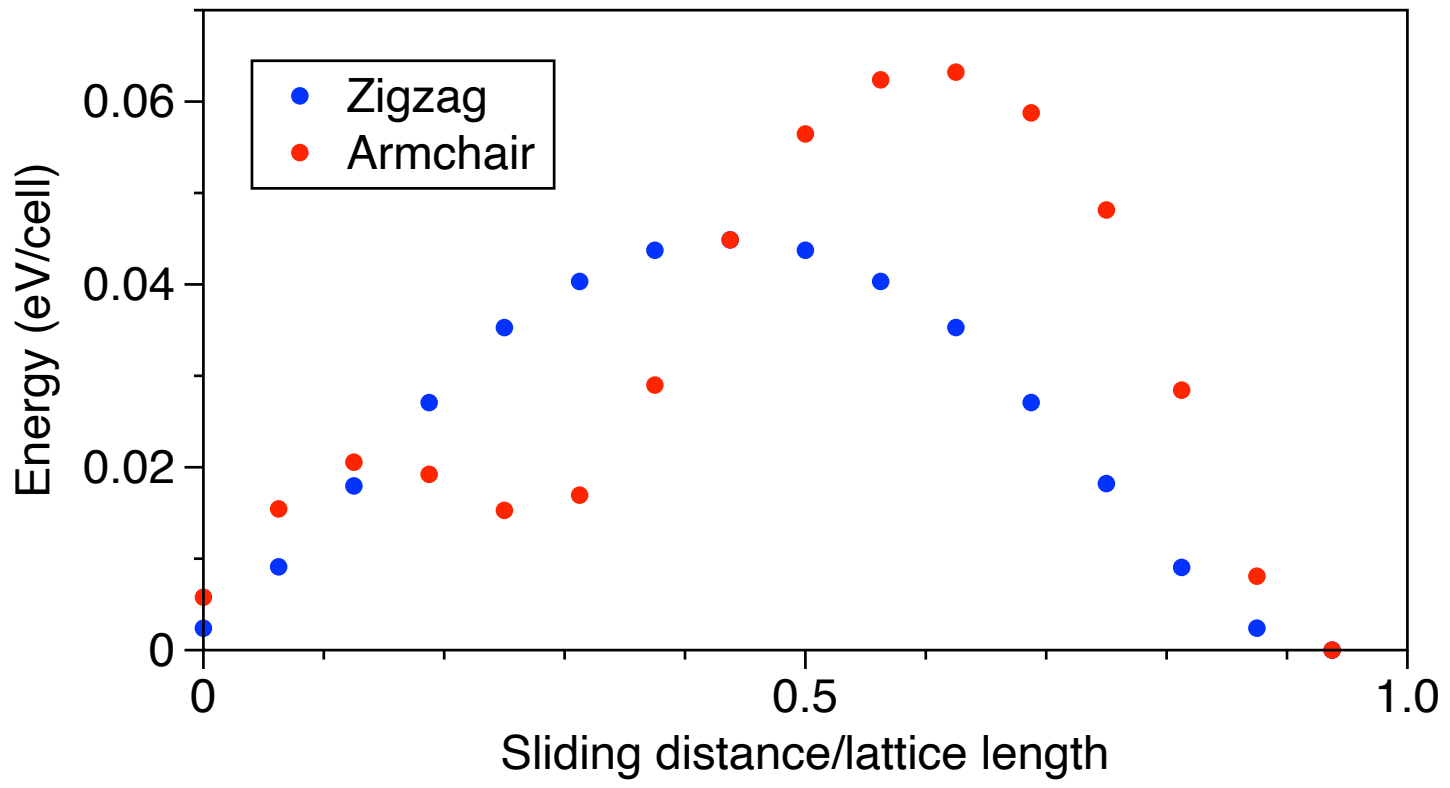

**Supplementary Figure 27 | Interlayer sliding of bilayer $WS_2$–$MoS_2$ heterostructure.** Interlayer sliding energy of bilayer $WS_2$–$MoS_2$ obtained from DFT calculation across zigzag and armchair directions. The

sliding in the $WS_2$–$MoS_2$ bilayer has energy barriers of up to ca. 8 meV/Å$^2$. This corresponds to ~0.07 eV per unit cell, which is close to that of the energy barrier of a $MoS_2$ bilayer[72].

**Supplementary Note 19. Raman spectra of monolayer $WS_2$ on different substrates**

**Supplementary Figure 28** presents the Raman spectra of monolayer $WS_2$ on $SiO_2$ and IP-Visio substrates. For $WS_2$ on $SiO_2$, the peak position difference between the E' and $A_1$' modes is ~61 cm$^{-1}$, consistent with previously reported values[47,73]. It is important to note that the broad feature spanning ~310–370 cm$^{-1}$ encompasses multiple Raman modes, including the E' and 2LA(M) peaks.

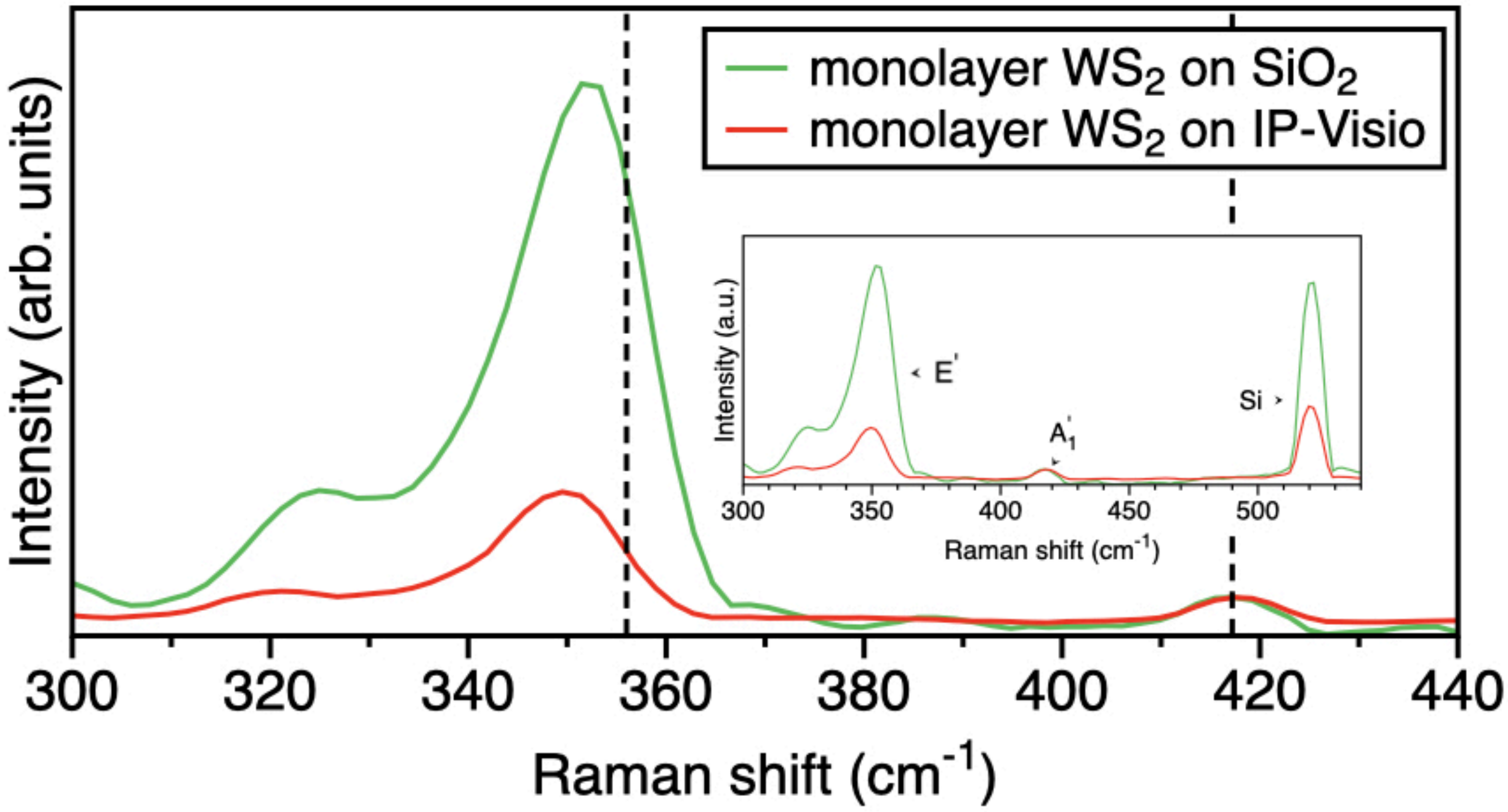

**Supplementary Figure 28 | Raman spectra of monolayer $WS_2$ on different substrates.** Raman spectra of monolayer $WS_2$ on $SiO_2$ and a flat IP-Visio substrate. Spectra are normalized to the $A_1'$ peak intensity. Vertical dashed lines indicate the E′ and $A_1'$ peak positions of monolayer $WS_2$ on $SiO_2$, as determined from Loterntzian fits. The broad feature spanning ~310–370 cm$^{-1}$ encompasses multiple $WS_2$ Raman modes, including the E' and 2LA(M) peaks Inset: wide-range spectrum displaying the Si substrate peak (~520.5 cm$^{-1}$).

**Supplementary Note 20. Extraction of biaxial strain from phonon mode shifts for monolayer $WS_2$**

In monolayer $WS_2$, the positions of the characteristic in-plane E′ and out-of-plane and $A_1'$ modes are sensitive to biaxial strain ($\varepsilon$) and doping ($n$). They are related to strain and doping through the following relation:

$$\begin{pmatrix} \Delta Pos\ E' \\ \Delta Pos\ A_1' \end{pmatrix} = \begin{pmatrix} -2\gamma_{E'} PosE' & k_{n,E'} \\ -2\gamma_{A_1'} PosA_1' & k_{n,A_1'} \end{pmatrix} \begin{pmatrix} \varepsilon \\ n \end{pmatrix} \tag{17}$$

Where $\gamma_{E'}$ and $\gamma_{A_1'}$ are the Grünesian parameters, and $k_{n,E'}$ are the charge doping shift $k_{n,A_{1g}}$ coefficients.

The values of the Grünesian parameters and charge doping shift coefficients are extracted from values are extracted from Michail et al.[74] and Iqbal et al.[75] as $\gamma_{E'} = 0.8, \gamma_{A_1'} = 0.3,\ k_{n,E'} = \frac{3.77}{10^{13}} cm, k_{n,A_1'} = \frac{8.44}{10^{13}} cm.$

## Supplementary References